\documentclass[12pt]{CSM_template}
\usepackage{csmMBFeb07,graphicx,url}

\usepackage[small]{caption}
\usepackage{amssymb,amsmath,amsthm,color,psfrag}
\usepackage{graphicx,epsfig}
\usepackage{algorithm,algpseudocode,xspace}
\usepackage{indentfirst}
\usepackage[resetlabels,labeled]{multibib}
\newcites{S}{\small References}
\usepackage{comment}
\usepackage{soul}
\usepackage{ulem}
\usepackage{bm,times}
\usepackage{framed}
\usepackage{subcaption}
\usepackage{multirow}
\usepackage{afterpage}
\usepackage{algorithm}
\usepackage{algorithmicx}

\usepackage{xr-hyper} 
\usepackage{hyperref} 

\usepackage{vmargin}
\setpapersize{USletter}
\setmarginsrb{1in}{1in}{1in}{1in}%
           {9pt}{9pt}{9pt}{18pt}

\newcommand\oprocendsymbol{\hbox{$\blacksquare$}}
\newcommand\oprocend{\relax\ifmmode\else\unskip\hfill\fi\oprocendsymbol}

\newcounter{sidebar}
\newcounter{sidebarequation}
\newcounter{sidebarproposition}
\newcounter{example}
\renewcommand{\theenumi}{(\roman{enumi})}

\usepackage{tikz,pgfplots}
\usetikzlibrary{shapes,arrows}
\usetikzlibrary{decorations.pathreplacing,decorations.markings}
\usetikzlibrary{positioning,calc}
\pgfplotsset{compat=1.7}

\tikzstyle{block}=[draw opacity=0.7,line width=1.4cm]
\definecolor{CranJ}{cmyk}{0,0.69,0.54,0.04} 
\definecolor{PinkJ}{cmyk}{0,0.71,0.43,0.12} 
\definecolor{Cran}{cmyk}{0,0.73,0.41,0.29} 
\definecolor{VRed}{cmyk}{0,0.75,0.25,0.2} 
\definecolor{ORed}{cmyk}{0,0.75,0.75,0} 
\definecolor{CBlue}{cmyk}{1,0.25,0,0} 

\tikzset{block/.style={%
        inner xsep=1mm,
        inner ysep=1.5mm,
        rectangle,very thick,draw}}
\tikzset{sum/.style={%
        circle,
        minimum size=2mm,inner xsep=1.2mm,inner ysep=1.2mm,
        very thick,draw}}
\tikzset{point/.style={%
        minimum size=0mm,inner xsep=4mm,inner ysep=0mm,draw}}
\tikzset{link/.style={->,very thick,>=stealth}}
\tikzset{undirlink/.style={<->,very thick,>=stealth}}
\tikzset{pole/.style={cross out, draw=black, minimum size=2*(#1-\pgflinewidth), inner sep=0pt, outer sep=0pt},cross/.default={1pt}}

\newcommand{\algorithmoneAbb}{\textsl{FOI-DC}\xspace}

\newcommand{\algorithmone}{\textsl{$1$st-Order-Input Dynamic Consensus}\xspace}

\newcommand*{\1}{\ensuremath{\mathbf{1}}}
\newcommand*{\0}{\ensuremath{\mathbf{0}}}

\newcommand*{\x}{0cm}
\newcommand*{\y}{0cm}
\newcommand{\vectg}[1]{\boldsymbol{{#1}}}
\newcommand{\vect}[1]{\boldsymbol{\mathbf{#1}}}
\usepackage{colonequals}
\newcommand{\defeq}{\colonequals}
\newcommand{\VV}{\mathcal{V}}
\newcommand{\EE}{\mathcal{E}}
\newcommand{\GG}{\mathcal{G}}

\newcommand{\NN}{\mathcal{N}}

\newcommand{\lL}{\vectsf{L}}
\newcommand{\rR}{\vect{\mathsf{R}}}

\newcommand{\pPi}{\vect{\Pi}}
\newcommand{\real}{{\mathbb{R}}} \newcommand{\reals}{{\mathbb{R}}}
 
\newcommand{\realpositive}{{\mathbb{R}}_{>0}}

\newcommand{\realnonnegative}{{\mathbb{R}}_{\ge 0}}
 
\newcommand{\naturals}{{\mathbb{N}}} 
\newcommand{\eps}{\epsilon}

\newcommand{\Hlambda}{\hat{\lambda}}
\newcommand{\argmin}{\operatorname{argmin}}
\newcommand{\rank}{\operatorname{rank}}

\newcommand{\dout}{\mathsf{d}_{\operatorname{out}}}

\newcommand{\re}[1]{\operatorname{Re}(#1)}

\newcommand*{\lr}{\ensuremath{\lambda_r}}

\newcommand{\until}[1]{\{1,\dots,#1\}}

\newcommand{\setdef}[2]{\{#1 \; |\; #2\}}
\newcommand{\argmax}{\operatorname{argmax}}

\newcommand{\vectsf}[1]{\vect{\mathsf{#1}}}
\newcommand{\Bvect}[1]{\bar{\boldsymbol{\mathbf{#1}}}}

\newcommand{\Hvect}[1]{\hat{\boldsymbol{\mathbf{#1}}}}

\newcommand{\dvect}[1]{\dot{\vect{#1}}}
\newcommand{\dvectsf}[1]{\dot{\vectsf{#1}}}
\newcommand{\ddvect}[1]{\ddot{\vect{#1}}}

\newcommand{\Sym}[1]{\operatorname{Sym}(#1)}
\newcommand{\Diag}[1]{\operatorname{Diag}(#1)}

\newcommand{\avrg}[1]{\frac{1}{N}\sum_{j=1}^N#1}

\newcommand{\SUM}[2]{\sum_{#1}^{#2}}

\newcommand*{\R}{\ensuremath{\mathbb{R}}}

\newtheorem{theorem}{\hspace*{1.5cm} Theorem}

\newtheorem{lem}{\hspace*{1.5cm} Lemma}

\begin{document}

\title{Tutorial on Dynamic Average Consensus\\
  {\Large The problem, its applications, and the algorithms}}

\author{S. S. Kia \; B. Van Scoy \; J. Cort\'es \; R. A. Freeman  \; K. M. Lynch \; S. Mart{\'\i}nez
\\
POC: S. S. Kia (solmaz@uci.edu)}

\date{}
\maketitle

\CSMsetup

\begin{figure}[htb]
\centering
     \includegraphics[height=1.8in]{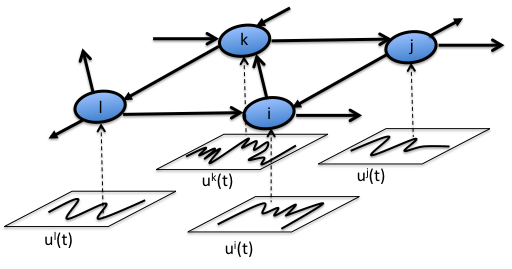}
\caption{A group of communication agents, each endowed with a time-varying reference signal.}\label{fig::DAC-Prob}
\end{figure}

Technological advances in ad-hoc networking and the availability of low-cost reliable computing, data storage and sensing devices have made possible scenarios where the coordination of many subsystems extends the range of human capabilities. Smart grid operations, smart transportation, smart healthcare and sensing networks for environmental monitoring and exploration in hazardous situations are just a few examples of such network operations. In these applications, the ability of a network system to, in a decentralized fashion, fuse information, compute common estimates of unknown quantities, and agree on a common view of the world is critical. These problems can be formulated as agreement problems on linear combinations of dynamically changing reference signals or local parameters. This dynamic agreement problem corresponds to \emph{dynamic average consensus}, which is the problem of interest of this article. The dynamic average consensus problem is for a group of agents to cooperate in order to track the average of locally available time-varying reference signals, where each agent is only capable of local computations and communicating with local neighbors, see Figure~\ref{fig::DAC-Prob}. 

\subsection{Centralized solutions have drawbacks}

The difficulty in the dynamic average consensus problem is that the information is distributed across the network. A  straightforward solution, termed \emph{centralized}, to the dynamic average consensus problem appears to be to gather all of the information in a single place, do the computation (in other words, calculate the average), and then send the solution back through the network to each agent. Although simple, the centralized approach has numerous drawbacks: (1) the algorithm is not robust to failures of the centralized agent (if the centralized agent fails, then the entire computation fails), (2) the method is not scalable since the amount of communication and memory required on each agent scales with the size of the network, (3) each agent must have a unique identifier (so that the centralized agent counts each value only once), (4) the calculated average is delayed by an amount which grows with the size of the network, and (5) the reference signals from each agent are exposed over the entire network which is unacceptable in applications involving sensitive data. The centralized solution is fragile due to existence of a single failure point in the network. This can be overcome by having every agent act as the centralized agent. In this approach, referred to as \emph{flooding}, agents transmit the values of the reference signals across the entire network until each agent knows each reference signal. This may be summarized as ``first do all communications, then do all computations''.  While flooding fixes the issue of robustness to agent failures, it is still subject to many of the drawbacks of the centralized solution. Also, although this approach works reasonably well for small size networks, its communication and storage costs scale poorly in terms of the network size and may incur, depending on how it is implemented, in costs of order $O(N^2)$ per agent (for instance, if each agent has to maintain, for each piece of information, which neighbor it has sent it to and which it has not. This motivates the interest in developing distributed solutions for the dynamics average consensus problem that only involve local interactions and decisions among the agents.

\subsection{Multiple instantiations of static average consensus algorithms are not able to deal with dynamic problems}

It is likely that the reader is familiar with the static version of the dynamic average consensus problem, commonly referred to as static average consensus, where agents seek to agree on a specific combination of fixed quantities. The static problem has been extensively studied  in the literature~\cite{ROS-JAF-RMM:07,wr-rwb:08,WR-YC:11,WR-RWB-EMA:07}, and several simple and efficient distributed algorithms exist with exact convergence guarantees. Given its mature literature, a natural approach to deal with the distributed solution of the dynamic average consensus problem in some literature has been to zero-order sample the reference signals and use a static average consensus algorithm between sampling times (for example, see~\cite{ATK-JAF-AKR:13,DT-JZ-ZS:17}).
If this approach were practical, it would mean that there is no need to worry about designing specific algorithms to solve the dynamic average consensus problem, since we could rely on the algorithmic solutions available for static average consensus.

\begin{figure}[htb]
\centering
\begin{subfigure}{0.5\textwidth}
  \includegraphics[height=1.8in]{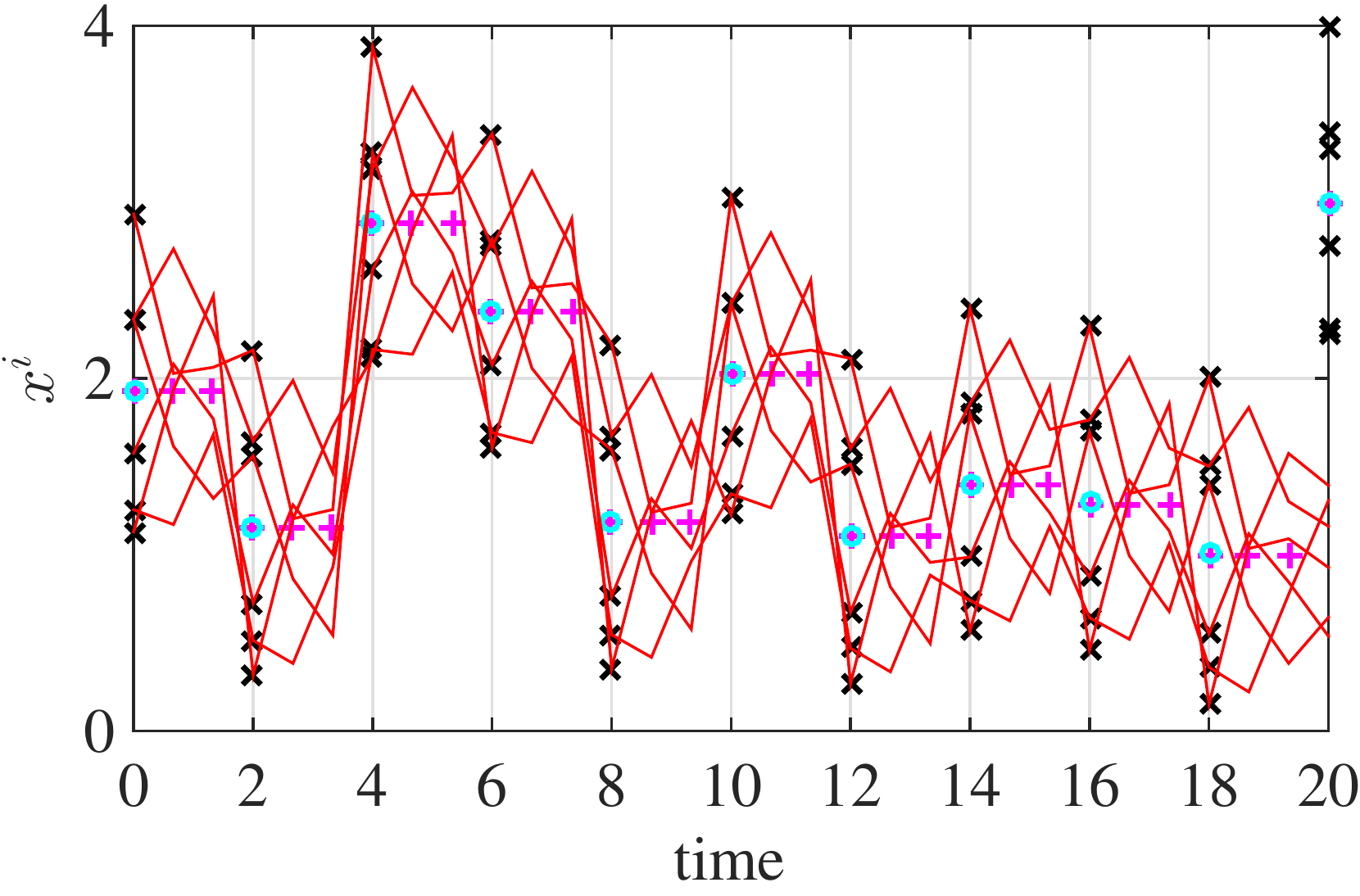}
  \caption{\scriptsize Static algorithm; $3$ communications in $t\in{[}m,m+1{]}$}
\end{subfigure}\hfill
\begin{subfigure}{0.5\textwidth}
  \includegraphics[height=1.8in]{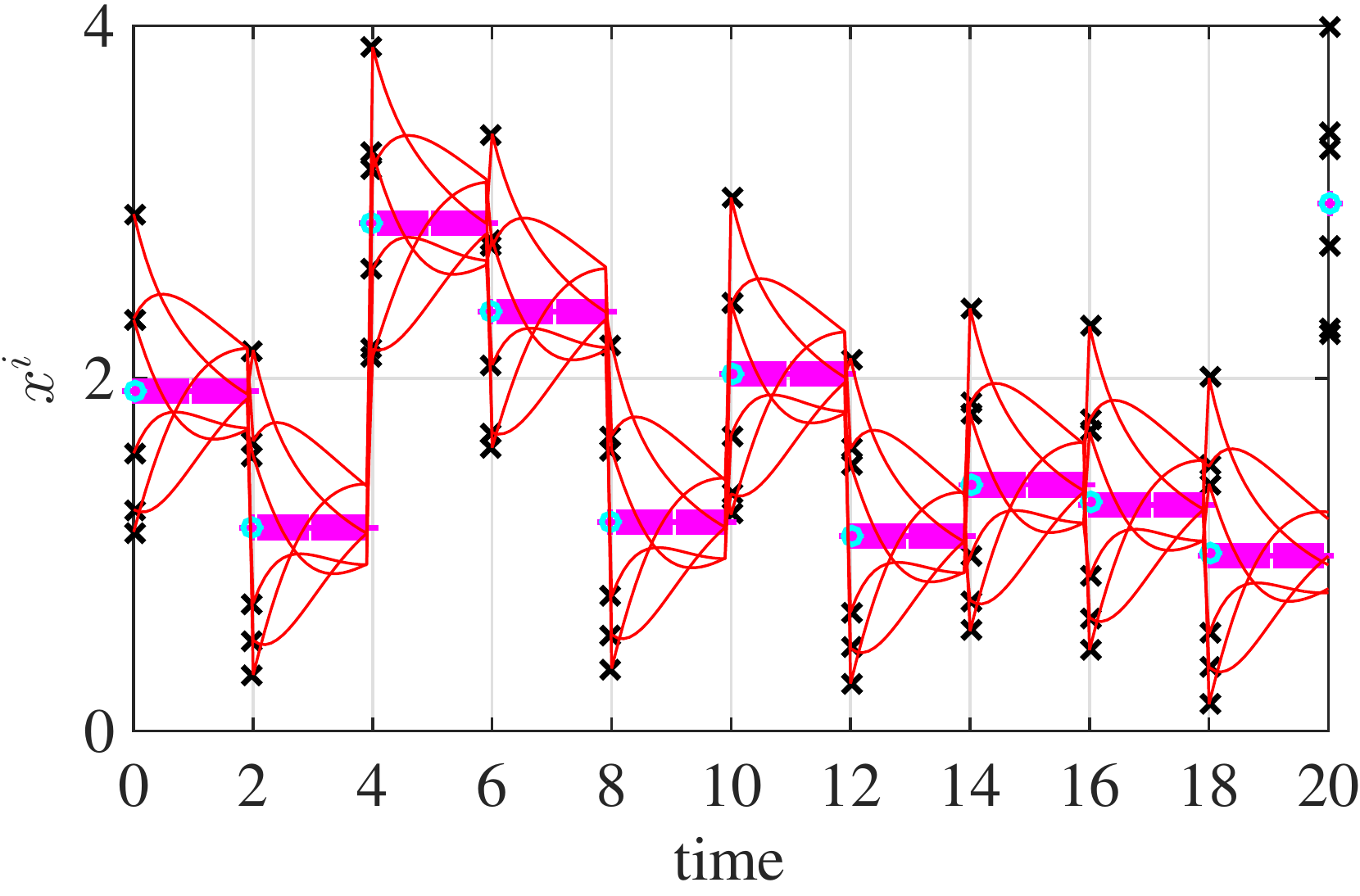}
  \caption{\scriptsize Static algorithm; $20$ communications in $t\in{[}m,m+1{]}$}
\end{subfigure}\\
\begin{subfigure}{0.5\textwidth}
  \includegraphics[height=1.8in]{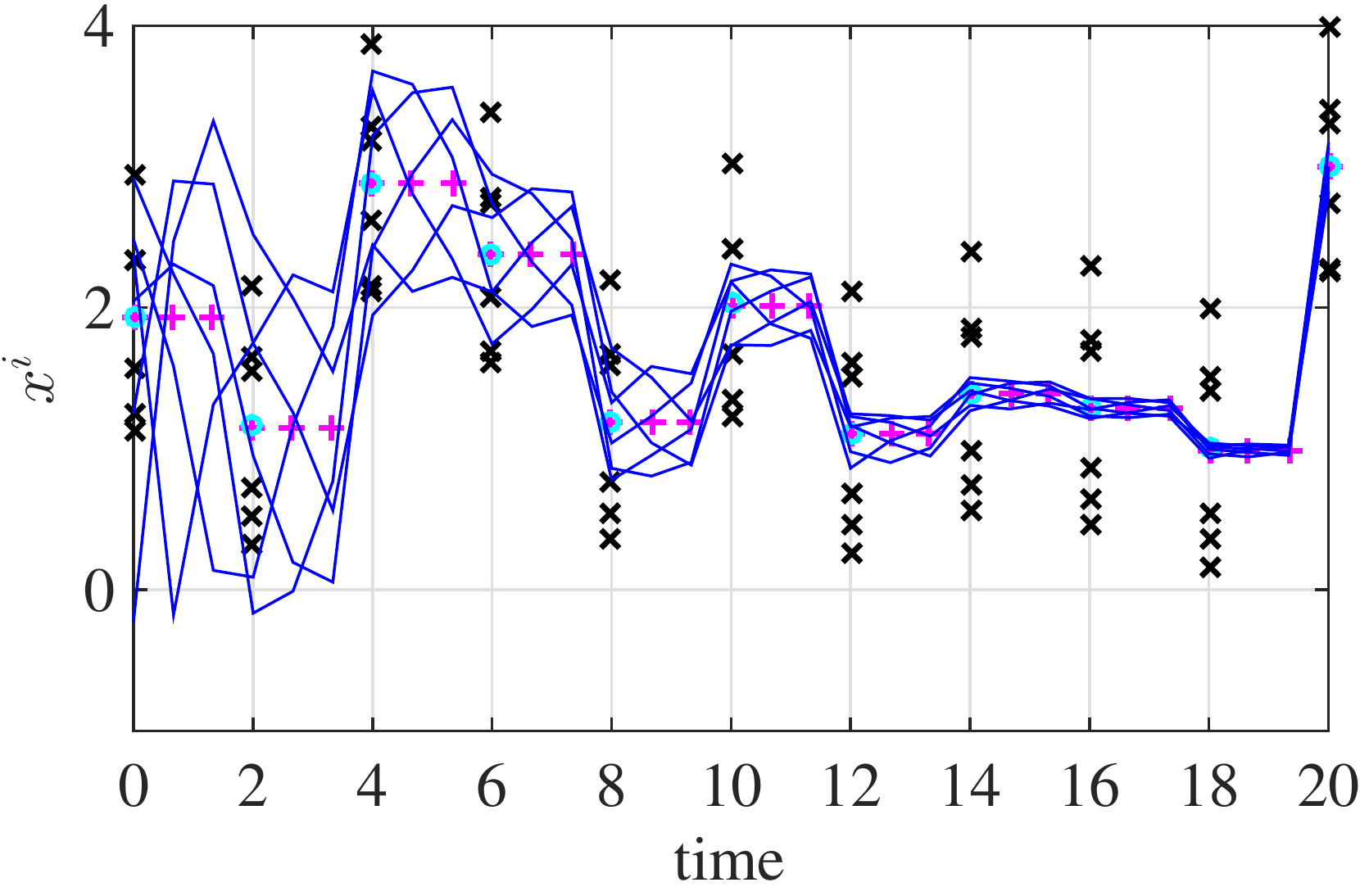}
  \caption{\scriptsize Dynamic algorithm; $3$ communications in $t\in{[}m,m+1{]}$}
\end{subfigure}\hfill
\begin{subfigure}{0.5\textwidth}
  \includegraphics[height=1.8in]{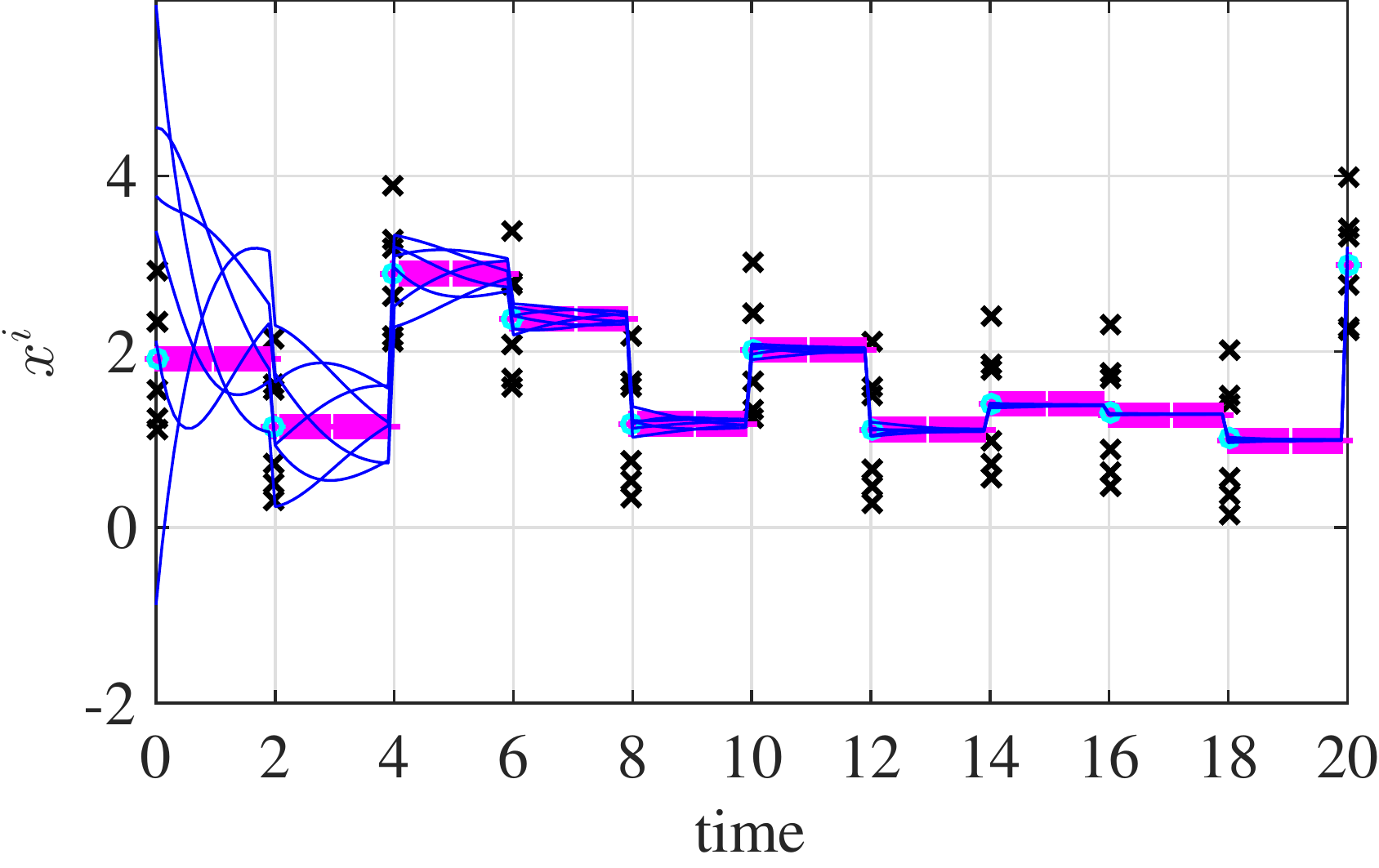}
  \caption{\scriptsize Dynamic algorithm; $20$ communications in $t\in{[}m,m+1{]}$}
\end{subfigure}
\caption{Comparison of performance between a static average consensus algorithm re-initialized at each sampling time vs. a dynamic average consensus algorithm; The solid lines: red curves (resp.  blue curves) represent the time history of the agreement state of each agent generated by the Laplacian static average consensus approach  (resp. the dynamic average consensus of~\eqref{eq::SSK-JC-SM:15-ijrnc-dt-alg});  $\times$: sampling points at $m\,\Delta t$; $\circ$: the average at $m\,\Delta t$; $+$: the average of reference signals at $k\,\delta$. The dynamic consensus algorithm tracks very closely the average as time goes by while the static consensus does not have enough time between sampling times to converge. This trend is preserved even if we increase the frequency of the communication between the agents. In these simulations we used $\alpha=\beta=1$ in~\eqref{eq::SSK-JC-SM:15-ijrnc-dt-alg}.}\label{fig::Ex1sim_disc}
\end{figure}

As the reader might have correctly guessed already, this approach does not work. In order for it to work, we would essentially need a static average consensus algorithm which is able to converge `infinitely' fast. In practice, some time is required for information
to flow across the network, and hence the result of the repeated
application of any static average consensus algorithm operates
with some error whose size depends on its speed of convergence and how
fast inputs change. To illustrate this point better, consider the following numerical example. Consider a process described by a fixed value plus a sine wave whose frequency and phase are changing randomly over time. A group of $6$ agents with the communication topology of directed ring monitors this process by taking synchronous samples, each according to
\begin{align*}
  \mathsf{u}^i(m) = a^i\,(2 + \sin(\omega(m) t(m)+\phi(m))) + b^i, \quad m\in\mathbb{Z}_{\geq0},
\end{align*}
where $a^i$ and $b^i$ are fixed unknown error sources in the measurement of agent $i\in\until{6}$.
To reduce the effect of measurement errors, after each
sampling, every agent wants to obtain the average of the measurements
across the network before the next sampling time. For the numerical simulations, the values $\omega\sim
\mathrm{N}(0,0.25)$, $\phi\sim \mathrm{N}(0,(\pi/2)^2)$, with
$\mathrm{N}(\mu,p)$ indicating a Gaussian distribution with mean $\mu$ and variance $p$ are used. We set the sampling rate at $0.5$ hz ($\Delta t=2$ s). For the simulation under study we use $a^1=1.1$, $a^2=1$, $a^3=0.9$, $a^4=1.05$, $a^5=0.96$, $a^6=1$, $b^1= -0.55$, $b^2=1$, $b^3=0.6$, $b^4=-0.9$, $b^5=-0.6$, and
$b^6=0.4$. To obtain the average, we use the following two approaches: (a) at every sampling time $m$, each agent initializes the standard static discrete-time Laplacian average consensus algorithm
\begin{align*}
  x^i(k+1)=x^i(k)-\delta\sum_{j=1}^N a_{ij}(x^i(k)-x^j(k)),\quad \quad i\in\until{N},
\end{align*}
by the current sampled reference values, $x^i(0)=\mathsf{u}^i(m)$, and implements it with an admissible timestep $\delta$ until just before the next sampling time $m+1$; (b) at time $m=0$, agents start executing a dynamic average consensus algorithm (more specifically, the strategy~\eqref{eq::SSK-JC-SM:15-ijrnc-dt-alg}  which is described in detail later). Between  sampling times $m$ and $m+1$, the reference input $\mathsf{u}^i(k)$ implemented in the algorithm is fixed at $\mathsf{u}^i(m)$, where here $k$ is the communication time index. Figure~\ref{fig::Ex1sim_disc} compares the tracking performance of these two approaches. We can observe that the dynamic average consensus algorithm, by keeping a memory of past actions, produces a better tracking response than the static algorithm initialized at each sampling time with the current values. 
This comparison serves as motivation for the need of specifically designed distributed algorithms that take into account the particular features of the dynamic average consensus problem.

\subsection{Objectives and roadmap of the article}

The objective of this article is to provide an overview of the dynamic average consensus problem that serves as a comprehensive introduction to the problem definition, its applications, and the distributed methods available to solve them. We decided to write this article after realizing that there exist in the literature many works that have dealt with the problem, but that there does not exist a tutorial reference that presents in a unified way the developments that have occurred over the years. "Article Summary" encapsulates the contents of the paper, making emphasis on the value and utility of its algorithms and results. Our primary intention, rather than providing a full account of all the available literature, is to introduce the reader, in a tutorial fashion, to the main ideas behind dynamic average consensus algorithms, the performance trade-offs considered in their design, and the requirements needed for their analysis and convergence guarantees. 

We begin the paper by introducing in  the ``Dynamic Average Consensus: Problem Formulation" section the problem definition and a set of desired properties expected from a dynamic average consensus algorithm. Next, we present various applications of dynamic average consensus in network systems --ranging over distributed formation, distributed state estimation, and distributed optimization problems-- in the ``Applications of Dynamic Average Consensus in Network Systems" section. It is not surprising that the initial synthesis of dynamic average consensus algorithms emerged from a careful look at static average consensus algorithms. In the ``A Look at Static Average Consensus Leading up to the Design of a Dynamic Average Consensus Algorithm" section, we provide a brief review of standard algorithms
for the static average consensus, and then build on this discussion to describe the first dynamic average consensus algorithm presented in the paper. We also elaborate on various features of these initial algorithms and identify their shortcomings. This sets the stage to introduce in the ``Continuous-Time Dynamic Average Consensus Algorithms" section various algorithms that aim to address these shortcomings. The design of continuous-time algorithms for network systems is often motivated by the conceptual ease for design and analysis, rooted in the relatively mature theoretical basis for the control of continuous-time systems. However,
the implementation of these continuous-time algorithms on cyber-physical systems may not be feasible due to practical constraints such as limited inter-agent communication bandwidth. This motivates 
the ``Discrete-Time Dynamic Average Consensus Algorithms" section, where we specifically discuss methods to accelerate the convergence rate and enhance the robustness of the proposed algorithms. Since the information of each agent takes some time to propagate through the network, we can expect that tracking an arbitrarily fast average signal with zero error is not feasible unless agents have some a priori information about the dynamics generating the signals. We address this topic in the ``Perfect Tracking Using A Priori Knowledge of the Input Signals" section, where we take advantage of knowledge of the nature of the reference signals. Many other topics exist that are related to the dynamic average consensus problem that we do not explore in this article. Interested reader can find several intriguing pointers for such topics in ``Further Reading''.
Throughout the paper, unless otherwise noted, we consider network systems whose communication topology is described by strongly connected and weight-balanced directed graphs. Only in a couple of specific cases, we particularize our discussion to the setup of undirected graphs, and we make explicit mention of this to the reader.

\subsection{Required mathematical background and available resources for implementation}

Graph theory plays an essential role in design and performance analysis of dynamic consensus algorithms. ``Basic Notations from Graph Theory" provides a brief overview of the relevant graph theoretic concepts, definitions and notations that we use in the article. Dynamic average consensus algorithms are linear time-invariant (LTI) systems in which the reference signals of the agents enter the system as an external input, in contrast to (Laplacian) static average consensus algorithm where the reference signals enter as initial conditions. Thus, in addition to the internal stability analysis, which is sufficient for the static average consensus algorithm, we need to asses the input-to-state stability (ISS) of the algorithms. The reader can find a brief overview of the ISS analysis of LTI systems in ``Input-to-State Stability of LTI Systems". All of the algorithms described can be implemented using modern computing languages such as C and Matlab. Furthermore, Matlab provides functions for simple construction, modification, and visualization of graphs.

\section{Dynamic Average Consensus: Problem Formulation}\label{sec::WhDynAvCons}

Consider a group of $N$ agents where each agent is capable of (1) sending and receiving information with other agents, (2) storing information, and (3) performing local computations. For example, the agents may be cooperating robots or sensors in a wireless sensor network. The communication topology among the agents is described by a fixed digraph, see ``Basic Notations from Graph Theory" for further details and the graph related notations used throughout the article. Suppose that each agent has a local scalar reference signal, denoted $\mathsf{u}^i(t) : [0,\infty)\to\real$ in continuous time and $\mathsf{u}^i(k): \naturals\to\real$ in discrete time. This signal may be the output of a sensor located on the agent, or it could be the output of another algorithm that the agent is running. The dynamic average consensus problem
then consists of designing an algorithm that allows individual agents to track the time-varying average of the reference signals, given by
\begin{alignat*}{2}
&\text{CT:}\quad & \mathsf{u}^\text{avg}(t) \defeq \frac{1}{N} \sum\nolimits_{i=1}^N \mathsf{u}^i(t) \\
&\text{DT:}\quad & \mathsf{u}^\text{avg}(k)\defeq \frac{1}{N} \sum\nolimits_{i=1}^N \mathsf{u}^i(k)
\end{alignat*}
in continuous time (CT) and discrete time (DT), respectively. For discrete-time signals and algorithms, for any variable $p$ sampled at time $t_k$, we use the shorthand notation $p(k)$ or $p_k$ to refer to~$p(t_k)$. For reasons that we specify below, we are specifically interested in the design of distributed algorithms, meaning that to obtain the average, the policy that each agent implements only depends on its variables (represented by $J^i$, which include its own reference signal) and those of its out-neighbors (represented by $\{I^j\}_{j\in\NN_{\text{out}}^i}$).

In continuous time, we seek a \emph{driving command} $c^i(J^i(t),{\{I^j(t)\}}_{j\in\NN_{\text{out}}^i})\in\real$ for each agent $i\in\until{N}$ such that, with perhaps an appropriate initialization, a local state $x^i(t)$, which we refer to as the \emph{agreement state} of agent $i$, converges to the average $\mathsf{u}^\text{avg}(t)$ asymptotically. Formally, for
\begin{equation}\label{eq::AgentSingInt}
\text{CT}:\quad\quad  \dot{x}^i=c^i(J^i(t),{\{I^j(t)\}}_{j\in\NN_{\text{out}}^i}),~~~i \in \until{N},
\end{equation}
with proper initialization if necessary, we have $x^i(t)\to\mathsf{u}^\text{avg}(t)$ as $t\to\infty$. The driving command $c^i$ can be a memoryless function or an output of a local internal dynamics. Note that, by using the out-neighbors, we are making the convention that information flows in the opposite direction specified by a directed edge (there is no loss of generality in doing it so, and the alternative convention of using in-neighbors instead would also be equally valid).

Dynamic average consensus can also be accomplished using discrete-time dynamics, especially when the time-varying inputs are sampled at discrete times. In such a case, we seek a \emph{driving command} for each agent $i\in\until{N}$ so that 
\begin{equation}\label{eq::AgentSingInt-DT}
\text{DT}:\quad\quad  {x}^i(t_{k+1})=c^i(J^i(t_k),{\{I^j(t_k)\}}_{j\in\NN_{\text{out}}^i}),~~~i \in \until{N},
\end{equation}
under proper initialization if necessary, accomplishes $x^i(t_k)\to \mathsf{u}^\text{avg}(k)$ as $t_k\to\infty$. Algorithm~\ref{alg:Alg} illustrates how a discrete-time dynamic average consensus algorithm can be executed over a network of $N$ communicating agents.
\begin{algorithm}[t]
{\footnotesize
\caption{Execution of a discrete-time dynamic average consensus algorithm at each agent $i\in\until{N}$}\label{alg:Alg}
\begin{algorithmic}[1]
\State \textbf{Input}: $J^i(k)$ and $\{I^j(k)\}_{j\in\NN_{\text{out}}^i}$
\State \textbf{Output}: $x^i(k+1)$, $J^i(k+1)$, and $I^i(k+1)$
\State $x^i(k+1)\leftarrow c^i(J^i(t_k),{\{I^j(t_k)\}}_{j\in\NN_{\text{out}}^i})$
\State Generate $J^i(k+1)$ and $I^i(k+1)$
\State Broadcast $I^i(k+1)$
\end{algorithmic}}
\end{algorithm}

We also consider a third class of dynamic average consensus algorithms in which the dynamics at the agent level is in continuous time but the communication among the agents, because of the restrictions of the wireless communication devices, takes place in discrete time,
\begin{equation}\label{eq::AgentSingInt-CT-DT}
\text{CT-DT}:\quad\quad  \dot{x}^i(t)=c^i(J^i(t),{\{I^j(t^j_{k^j})\}}_{j\in\NN_{\text{out}}^i}),~~~i \in \until{N},
\end{equation}
such that $x^i(t)\to\mathsf{u}^\text{avg}(t)$ as $t\to\infty$. Here $t^j_{k^j}\in\real$ is the $k^j$-th transmission time of agent $j$, which is not necessarily synchronous with the transmission time of other agents in the network.

The consideration of simple dynamics of the form in~\eqref{eq::AgentSingInt},~\eqref{eq::AgentSingInt-DT} and~\eqref{eq::AgentSingInt-CT-DT} is motivated by the fact that the state of the agents does not necessarily correspond to some physical quantity, but instead to some logical variable on which agents perform computation and processing. Agreement on the average is  also of relevance in scenarios where the agreement state is a physical state with more complex dynamics, for example, position of a mobile agent in a robotic team. In such cases, we can leverage the discussion here by, for instance, having agents compute reference signals that are to be tracked by the states with more complex dynamics. Interested readers can consult ``Further Reading" for a list of relevant literature on dynamic average consensus problems for higher-order dynamics.

Given the drawbacks of centralized solutions, here we identify several desirable properties when designing algorithmic solutions to the dynamic average consensus problem. The algorithm is desired to be
\begin{itemize}
    \item \emph{scalable}, so that the amount of computations and resources required on each agent does not grow with the network size, 
    \item
    \emph{robust} to the disturbances present in practical scenarios, such as communication delays and packet drops, agents entering/leaving the network, noisy measurements, and 
    \item
\emph{correct}, meaning that the algorithm converges to the exact average or, alternatively, a formal guarantee can be given about the distance between the estimate and the exact average.
\end{itemize}
Regarding the last property, to achieve agreement, network connectivity must be such that information about the local reference input of each agent reaches other agents frequently enough. Also, as the information of each agent takes some time to propagate through the network, tracking an arbitrarily fast average signal with zero error is not feasible unless agents have some a priori information about the dynamics generating the signals. Therefore, a recurring theme throughout the article is how the convergence guarantees of dynamic average consensus algorithms depend on the network connectivity and on the rate of change of the reference signal of each agent.

\section{Applications of Dynamic Average Consensus in 
  Network Systems}\label{sec::App}

The ability to compute the  average of a set of time-varying reference signals turns out to be useful in numerous applications, and this explains why distributed algorithmic solutions have found their way into 
 many seemingly different problems involving the interconnection
of dynamical systems. This section provides a selected overview of problems to further motivate the reader to learn about dynamic average consensus algorithms and illustrate their range of applicability. 
Other applications of dynamic average consensus can be found in~\cite{PY-RAF-KML:06,KML-IBS-PY-RAF:08,PY-RAF-KML:08,py-raf-gjg-kml-sss-rs:10,RA-JC-CS:12,SD-JMFM:15,fc-wr:17}.

\subsection{Distributed formation control}

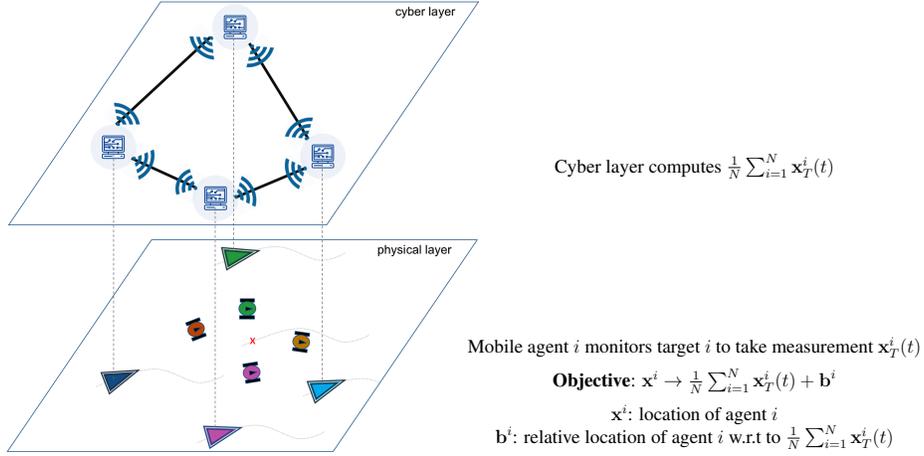
\begin{figure}[htb]
  \centering 
   \includegraphics[width=2.5in]{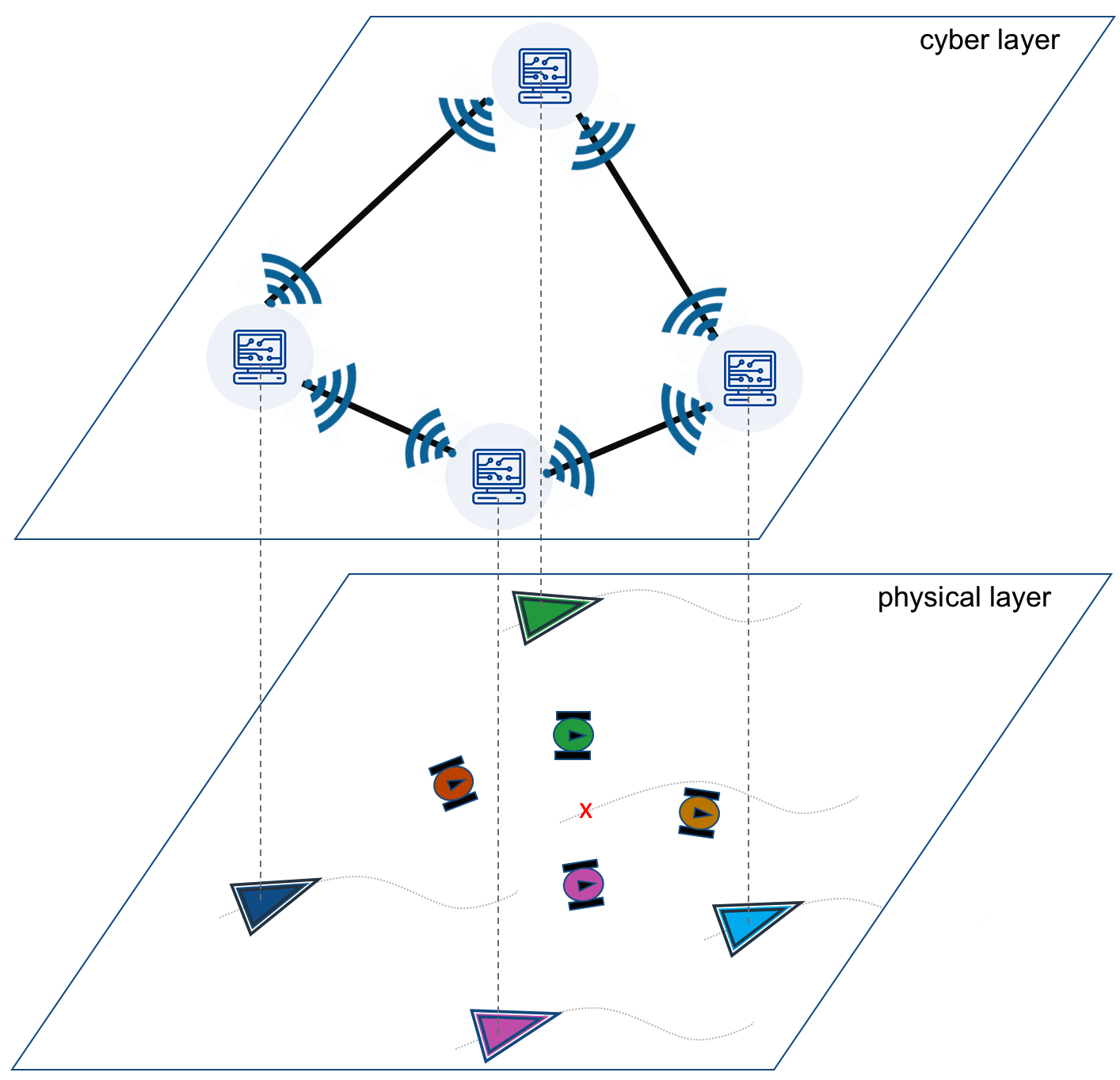}\!\!\!\!\!
\begin{tikzpicture}[auto,thick,scale=0.6, every node/.style={scale=0.6}]
\node (text)[draw=none] at (0,-0.75){$\vect{x}^i$: location of agent $i$};
\node (text)[draw=none] at (0,-1.25){$\vect{b}^i$: relative location of agent $i$ w.r.t to $\frac{1}{N}\sum_{i=1}^N\vect{x}_T^i(t)$};
\node (text)[draw=none] at (0,0.75){Mobile agent $i$ monitors target $i$ to take measurement $\vect{x}^i_T(t)$};
\node (text)[draw=none] at (0,0){\textbf{Objective}: $\vect{x}^i\to \frac{1}{N}\sum_{i=1}^N\vect{x}_T^i(t)+\vect{b}^i$};\node (text)[draw=none] at (0,4.75){Cyber layer computes $\frac{1}{N}\sum_{i=1}^N\vect{x}_T^i(t)$};
\end{tikzpicture}
  \caption{A two-layer consensus-based formation for tracking a team of mobile targets: the larger triangle robots are the mobile agents and the smaller round robots are the mobile moving targets. }\label{fig::two-layer-formation}
\end{figure}

Autonomous networked mobile agents are playing an increasingly important role in coverage, surveillance and patrolling applications in both commercial and military domains. The tasks accomplished by mobile agents often require dynamic motion coordination and formation among team members. Consensus algorithms have been commonly used in the design of formation control strategies~\cite{JAF-RMM:04,WR:07a,KDL-MVM-JA:09}. These algorithms have been used, for instance, to arrive at agreement on the geometric center of formation, so that the formation can be achieved by spreading the agents in the desired geometry about this center, see~\cite{ROS-JAF-RMM:07}. However, most of the existing results are for static formations. Dynamic average consensus algorithms can effectively be used in dynamic formation control, where quantities of interest like the geometric center of the formation change with time.
Figure~\ref{fig::two-layer-formation} depicts an example scenario in which a group of mobile agents tracks a team of mobile targets. Each agent monitors a mobile target with location $\vect{x}_T^i$. The objective is for the agents to follow the team of mobile targets by spreading out in a pre-specified formation, which consists of each agent being positioned at a relative vector $\vect{b}^i$ with respect to the time-varying geometric center of the target team. A two-layer approach  can be used to accomplish the formation and tracking objectives in this scenario: a dynamic consensus algorithm in the cyber layer that computes the geometric center in a distributed manner, and a physical layer that tracks this average plus $\vect{b}^i$.
Note that dynamic average consensus algorithms can also be employed to compute the time-varying variance of the positions of the mobile targets with respect to the geometric center, and this can help the mobile agents adjust the scale of the formation in order to avoid collisions with the target team.
Examples of the use of dynamic consensus algorithms in this two-layer approach with multi-agent systems with first-order, second-order or higher-order dynamics can be found in~\cite{MP-GDR-DJS:07,SG-SR-WR:16,SSK-JC-SM:15-ijrnc}.

\subsection{Distributed state estimation}

\begin{figure}[htb]
\begin{center}
       \includegraphics[height=2.4in]{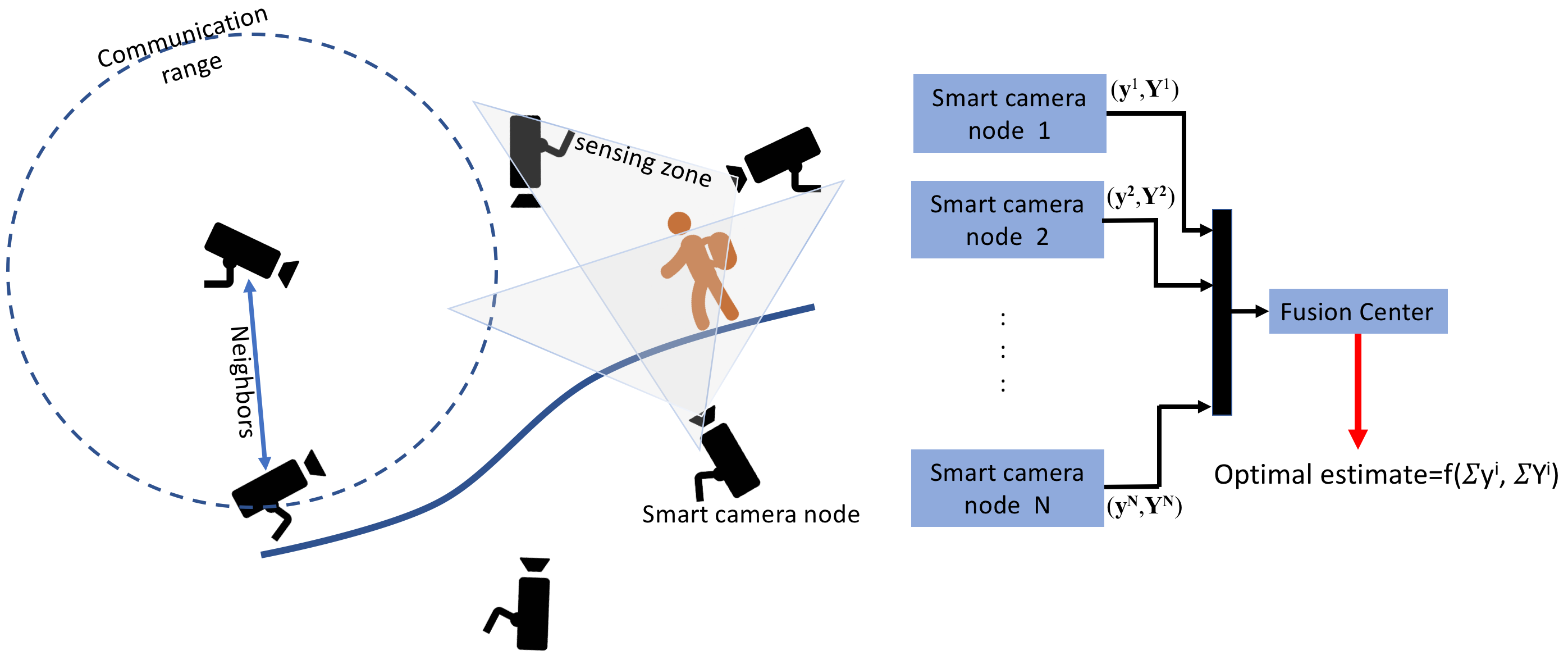}
\caption{A networked smart camera system that monitors and estimates the position of moving targets.}\label{fig::CameraNet}
\end{center}
\end{figure}

Wireless sensors with embedded computing and communication capabilities play a vital role in provisioning real-time monitoring and control
in many applications such as environmental monitoring,
fire detection, object tracking, and body area networks. 
Consider a model of the process of interest given by
\begin{align*}
\vect{x}(k+1)=\vect{A}(k)\,\vect{x}(k)+\vect{B}(k)\,\vectg{\omega}(k),
\end{align*}
where $\vect{x}\in\real^n$ is the state, $\vect{A}(k)\in\real^{n\times n}$ and $\vect{B}(k)\in\real^{n\times m}$ are known system matrices and $\vectg{\omega}\in\real^{m}$ is the white Gaussian process noise with $E[\vectg{\omega}(k)\vectg{\omega}^\top(k)]=\vect{Q}>0$. Let the measurement model at each sensor station $i\in\until{N}$ be
\begin{align*}
\vect{z}^i(k+1)=\vect{H}^i(k+1)\vect{x}(k+1)+\vectg{\nu}^i,
\end{align*}
where $\vect{z}^i\in\real^q$ is the measurement vector, $\vect{H}^i\in\real^{q\times n}$ is the measurement matrix and $\vectg{\nu}^i\in\real^q$ is the white Gaussian measurement noise with $E[\vectg{\nu}^i(k)\vectg{\nu}^i(k)^\top]=\vect{R}^i>0$. If all the measurements are transmitted to a fusion center, a Kalman filter can be used to obtain the minimum variance estimate of the state of the process of interest as follows (see Figure~\ref{fig::CameraNet})
\begin{itemize}
    \item propagation stage:
    \begin{subequations}\label{eq::Kalman-prop}
    \begin{align}
   & \Hvect{x}^{-}(k+1)=\vect{A}(k)\Hvect{x}^{-}(k)\\
   &\vect{P}^{-}(k+1)=\vect{A}(k)\vect{P}^{-}(k+1)\vect{A}(k)^\top+\vect{B}(k)\vect{Q}(k)\vect{B}(k)^\top,\\
  &  \vect{Y}^{-}(k+1)=\vect{P}^{-}(k+1)^{-1},\\
   & \vect{y}^{-}(k+1)= \vect{Y}^{-}(k+1)\,\Hvect{x}^{-}(k+1);
    \end{align}
    \end{subequations}
    \item update stage:
    \begin{align*}
       &\vect{Y}^i(k+1)=\vect{H}^i(k+1)^\top\vect{R}^i(k+1)^{-1}\vect{H}^i(k+1),\\    &\vect{y}^i(k+1)=\vect{H}^i(k+1)^{\top}\vect{R}^i(k+1)^{-1}\vect{z}^i(k+1),\\
       &\vect{P}^{+}(k+1)=\big(\vect{Y}^{-}(k+1)+\sum\nolimits_{i=1}^N\vect{Y}^i(k+1)\big)^{-1},\\
   & \Hvect{x}^{+}(k+1)=\Hvect{x}^{-}(k+1)-\big(\sum\nolimits_{i=1}^N\vect{Y}^i(k+1)-\sum\nolimits_{i=1}^N\vect{y}^i(k+1)\,\Hvect{x}^{-}(k+1)\big).
    \end{align*}
\end{itemize}

Despite its optimality, this implementation is not desirable in many sensor network applications due to existence of a single  point of failure at the fusion center and the high cost of  communication between the sensor stations and the fusion center. An alternative that has previously gained interest~\cite{ROS:05,ROS:09,ATK-JAF-AKR:13,WQ-PZ-ZD:14,GW-NL-YZ:17} is to employ distributed algorithmic solutions that have each sensor station maintain a local filter to process its local measurements and fuse them with the estimates of its neighbors. Some work~\cite{ROS:05,RA-CS-YM:15,WR-UMA:17} employ dynamic average consensus to synthesize distributed implementations of the Kalman filter. For instance, one of the early solutions for distributed minimum variance estimation, has each agent maintain a local copy of the propagation filter~\eqref{eq::Kalman-prop} and employ a dynamic average consensus algorithm to generate the coupling time-varying terms $\frac{1}{N}\sum\nolimits_{i=1}^N\vect{y}^i(k+1)$
and $\frac{1}{N}\sum\nolimits_{i=1}^N\vect{Y}^i(k+1)$. If agents know the size of the network, they can duplicate the update equation locally.

\subsection{Distributed unconstrained convex optimization}
The control literature has introduced numerous distributed algorithmic
solutions~\cite{AN-AO:09,BJ-MR-MJ:09,JW-NE:11,MZ-SM:12,JL-CYT:12,BG-JC:14-tac,SSK-JC-SM:15-auto,GQ-NL:17}
to solve unconstrained convex optimization problems over networked
systems. In a distributed unconstrained convex optimization problem,
a group of $N$ communicating agents, each with access to a local
convex cost function $f^i:\real^n\to\real$, $i\in\until{N}$, seek to
determine the minimizer of the joint global optimization problem
\begin{align}\label{eq::convex_uncon_opti}
  \vect{x}^{\star}=\argmin\,\frac{1}{N}\sum\nolimits_{i=1}^Nf^i(\vect{x}),\end{align}
by local interactions with their neighboring agents.  This problem
appears in network system applications such as multi-agent
coordination, distributed state estimation over sensor networks, or
large scale machine learning problems.  Some of the algorithmic
solutions for this problem are developed using agreement algorithms to
compute global quantities that appear in existing centralized
algorithms. For example, a centralized solution
for~\eqref{eq::convex_uncon_opti} is the Nesterov gradient descent
algorithm~\cite{YN:13} described by
\begin{subequations}\label{eq::Nesterov_central}
\begin{align}
  \vect{x}(k+1)&=\vect{y}(k) -\eta\, (\frac{1}{N}\sum\nolimits_{i=1}^N\nabla f^i(\vect{y}(k))),\\
  \vect{v}(k+1)&=\vect{y}(k) -\frac{\eta}{\alpha_k} (\frac{1}{N}\sum\nolimits_{i=1}^N\nabla f^i(\vect{y}(k))),\\
  \vect{y}(k+1)&=(1-\alpha_{k+1})\,\vect{x}(k + 1) +
  \alpha_{t+1}\vect{v}(k + 1).
\end{align}
\end{subequations}
where $\vect{x}(0),\vect{y}(0),\vect{v}(0)\in\real^n$, and
$\{\alpha_k\}_{k=0}^{\infty}$ is defined by an arbitrarily chosen
$\alpha_0 \in (0, 1)$ and the update equation
$\alpha^2_{k+1}=(1-\alpha_{k+1})\alpha_k^2$, where $\alpha_{k+1}$
always takes the unique solution in $(0,1)$. If all $f^i$,
$i\in\until{N}$, are convex, differentiable and have $L$-Lipschitz
gradients, then every trajectory $k\!  \mapsto\! \vect{x}(k)$
of~\eqref{eq::Nesterov_central} converges to the optimal solution
$\vect{x}^\star$ for any $0<\eta<\frac{1}{L}$.

Note that in~\eqref{eq::Nesterov_central}, the cumulative gradient term $\frac{1}{N}\sum\nolimits_{i=1}^N\nabla f^i(\vect{y}(k))$ is a source of coupling  among the computations performed by each agent. It does not seem reasonable to halt the execution of this algorithm at each step until the agents have figured out the value of this term. Instead, we could employ dynamic average consensus in conjunction with~\eqref{eq::Nesterov_central}: the dynamic average consensus algorithm estimates the coupling term, this estimate is employed in executing~\eqref{eq::Nesterov_central}, which in turn changes the value of the coupling term being estimated. This approach is taken in~\cite{GQ-NL:17} to solve the optimization problem~\eqref{eq::convex_uncon_opti} over connected graphs, and also pursued in other implementations of distributed convex or nonconvex optimization algorithms, see for instance~\cite{RC-GN-LS-DV:15,JX-SZ-YCS-LX:15,DV-FZ-AC-PG-LS:16,PDL-GS:16,AN-AO-WS:17}.

\subsection{Distributed resource allocation}
\begin{figure}[htb]
\begin{center}
       \includegraphics[height=2.4in]{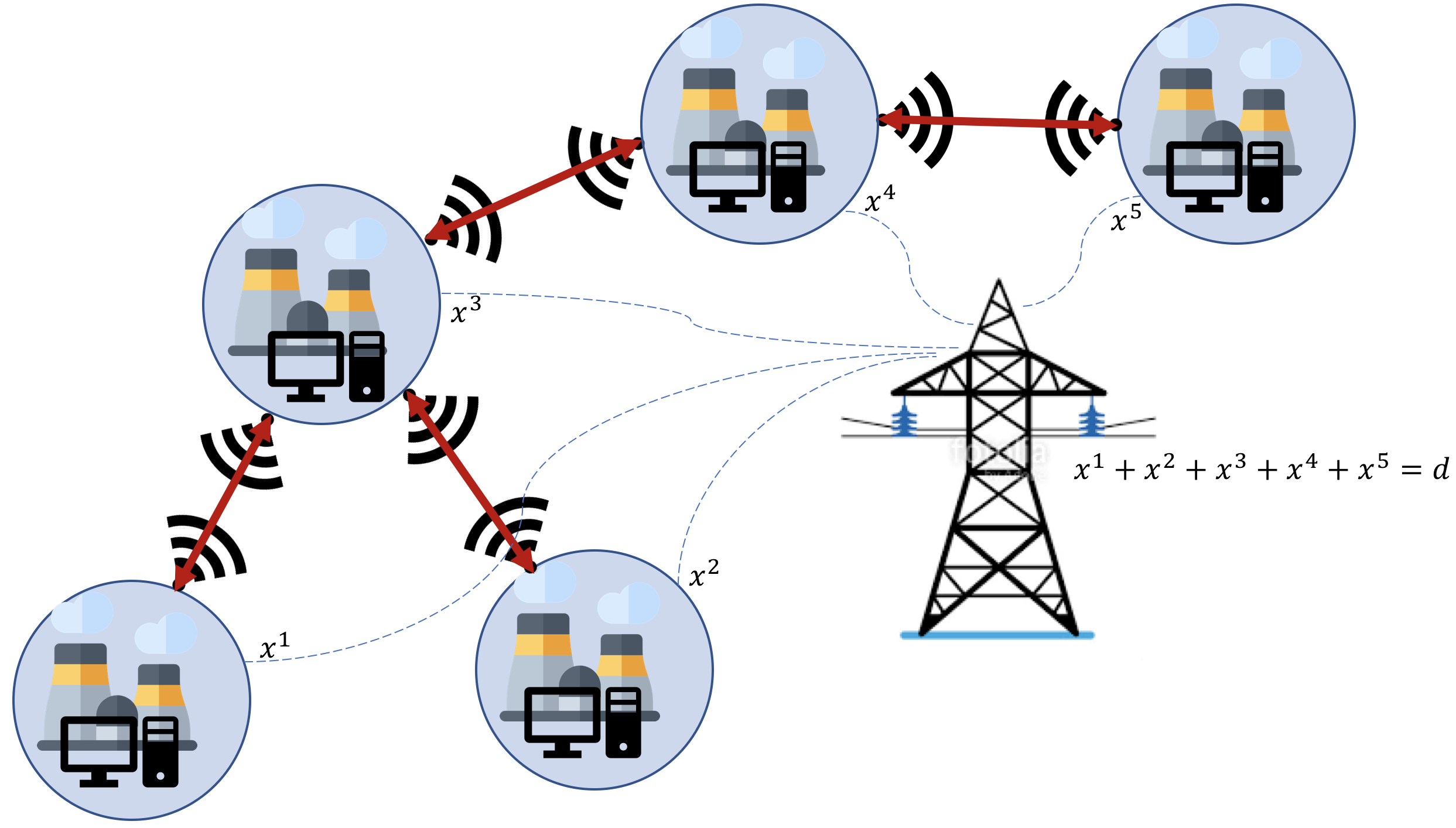}
\caption{A network of $5$ generators with connected undirected topology work together to meet a demand of $x^1+x^2+x^3+x^4+x^5=d$ in a manner that overall cost $\sum_{i=1}^5f^i(x^i)$ for the group is minimized. }\label{fig::ResourceAlloc}
\end{center}
\end{figure}

In optimal resource allocation, a group of agents work cooperatively to meet a demand in an efficient way (see Figure~\ref{fig::ResourceAlloc}). Each agent incurs a cost for the resources it provides.  Let the cost of each agent $i\in \until{N}$ be modeled by a convex and differentiable function  $f^i:\mathbb{R}\to\mathbb{R}$. The objective is to meet the demand ${d}\in\real$ so that the total cost $f(\vect{x})=\Sigma_{i=1}^N f^i(x^i)$ is minimized. Each agent $i\in\until{N}$ therefore seeks to find the $i^\text{th}$ element of $\vect{x}^\star$ given by 
\begin{align*}
  \vect{x}^\star=&\arg\min_{\vect{x}\in
    \reals^N} \,\,\sum\nolimits_{i=1}^N f^i(x^i),~~\text{subject~to~} \\
    &\!\!\!x^1+\cdots+x^N-{d}=0,
\end{align*}
This problem appears in many optimal decision making tasks such as optimal dispatch in power networks~\cite{AJW-FW-GBS:13,AC-JC:15-tcns}, optimal routing~\cite{RM-SL:06}, and economic systems~\cite{GMH:69}. For instance, the group of agents could correspond to a set of flexible loads in a microgrid that receive a request from a grid operator to collectively adjust their power consumption to provide a desired amount of regulation to the bulk grid. In this demand response scenario, $x^i$ corresponds to the amount of deviation from the nominal consumption of load $i$,  the function $f_i$ models the amount of discomfort caused by deviating from it, and ${d}$ is the amount of regulation requested by the grid operator.

A centralized algorithmic solution is given by the popular saddle-point or primal-dual dynamics~\cite{KJA-LH-HU:58,AC-BG-JC:17-sicon} associated to the optimization problem,
\begin{subequations}\label{eq::saddle}
\begin{align}
\dot{\mu}(t)&=x^1(t)+\cdots+x^N(t)-{d},\qquad {\mu}(0)\in\real,\label{eq::saddle-a}\\
\dot{x}^i(t)&=-\nabla f^i(x^i(t))-{\mu}(t),~~ i\in\until{N},~~\, x^i(0)\in\real,
\end{align}
\end{subequations}
If the local cost functions are strictly convex, every trajectory $t\!  \mapsto\! \vect{x}(t)$ converges to the optimal solution $\vect{x}^\star$. The source of coupling in~\eqref{eq::saddle} is the demand mismatch which appears in the right hand side of~\eqref{eq::saddle-a}. We can, however, employ dynamic average consensus to estimate this quantity online and feed it back into the algorithm. This approach is taken in~\cite{AC-JC:16-auto,SSK:17}. This can be accomplished, for instance, by having agent $i$ use the reference signal $x^i(t) - {d}/N $ (this assumes every agent knows the demand and the number of agents in the network, but other reference signals are also possible) in a dynamic consensus algorithm coupled with the execution of~\eqref{eq::saddle}.

\section{A Look at Static Average Consensus Leading up to the Design of a Dynamic Average Consensus Algorithm}\label{sec::DAC}

Consensus algorithms to solve the static average consensus problem have been studied at least as far back as~\cite{JNT:1984}. The commonality in their design is the idea of having agents start their agreement state with their own reference value and adjust it based on some weighted linear feedback which takes into account the difference between their agreement state and their neighbors'. This leads to algorithms of the following form
\begin{subequations}\label{eq::Static_CT_DT}
\begin{alignat}{3}
&\text{CT:}\quad
&\dot{x}^i(t) &= -\sum_{j=1}^N a_{ij} \bigl(x^i(t) - x^j(t)\bigr), 
\label{eq::Static_CT}\\
&\text{DT:}\quad
&x^i({k+1}) &= x^i({k})-\sum_{j=1}^N a_{ij} \bigl(x^i(k) - x^j(k)\bigr),
\label{eq::Static_DT}
\end{alignat}
\end{subequations}
for $i\in\until{N}$, with $x^i(0) = \mathsf{u}^i$ constant for both algorithms. Here $[a_{ij}]_{N\times N}$ is the adjacency matrix of the communication graph (see ``Basic Notions from Graph Theory"). By stacking the agent variables into vectors, the static average consensus algorithms can be written compactly using a graph Laplacian as
\begin{subequations} \label{eq::static_alg}
\begin{alignat}{3}
&\text{CT:}\quad
&\dvect{x}(t) &= -\lL\,\vect{x}(t),
\\
&\text{DT:}\quad
&\vect{x}({k+1}) &= (\vect{I}-\lL)\,\vect{x}(k), 
\end{alignat}
\end{subequations}
with $\vect{x}(0) = \mathsf{u}$.
When the communication graph is fixed, this system is linear time invariant and can be analyzed using standard time domain and frequency-domain techniques in control. Specifically, the frequency-domain representation of the dynamic average consensus algorithm output signal is given by
\begin{subequations}\label{eq::Freq_Static_CT_DT}
\begin{align}
\text{CT:}\quad
\vect{X}(s) &= [s\vect{I}+\lL]^{-1} \vect{x}(0)=[s\vect{I}+\lL]^{-1} \vect{U}(s) ~\label{eq::Freq_Static_CT}\\
\text{DT:}\quad
\vect{X}(z) &= [z\vect{I}-(\vect{I}-\lL)]^{-1} \vect{U}(z), ~\label{eq::Freq_Static_DT}
\end{align}
\end{subequations}
where $\vect{X}(s)$ and $\vect{U}(s)$, respectively, denote the Laplace transform of $\vect{x}(t)$ and $\vectsf{u}$, while $\vect{X}(z)$ and $\vect{U}(z)$, respectively, denote the $z$-transform of $\vect{X}_k$ and $\vectsf{u}$. For a static signals we have $\vect{U}(s)=\vectsf{u}$ and $\vect{U}(z)=\vectsf{u}$.

\begin{figure}[htb]
\begin{subfigure}{0.5\textwidth}
\centering \begin{tikzpicture}[align=center]
 \node (sum1) [sum] {};
 \node (f)    [block,right=of sum1] {$\dfrac{1}{s} \vect{I}_N$};
 \node (sum2) [sum,right=0.5cm of f] {};
 \node (L)    [block,below=of f]  {$\lL$};
 \coordinate (x) at ($(sum2.east)+(5mm,0)$) {};
 \node (y)    [right=0.5cm of x] {$\vect{x}(t)$};
 \node (x0)   [below left=0.5cm and 0.25cm of sum2] {$\vect{x}(0)$};
 \draw [link] (sum1) -- (f);
 \draw [link] (f) -- (sum2);
 \draw [link] (x) |- (L);
 \draw [link] (L) -| node[pos=0.9,xshift=-2.5mm] {\footnotesize{$-$}} (sum1);
 \draw [link] (sum2) -- (y);
 \draw [link] (x0) -| (sum2);
\end{tikzpicture}
\caption{Continuous time}
\label{Fig:BlockDiagramStatic_CT}
\end{subfigure}%
\begin{subfigure}{0.5\textwidth}
\centering \begin{tikzpicture}[align=center]
 \node (sum1) [sum] {};
 \node (f)    [block,right=of sum1] {$\dfrac{1}{z-1} \vect{I}_N$};
 \node (sum2) [sum,right=0.5cm of f] {};
 \node (L)    [block,below=of f]  {$\lL$};
 \coordinate (x) at ($(sum2.east)+(5mm,0)$) {};
 \node (y)    [right=0.5cm of x] {$\vect{x}_k$};
 \node (x0)   [below left=0.5cm and 0.25cm of sum2] {$\vect{x}_0$};
 \draw [link] (sum1) -- (f);
 \draw [link] (f) -- (sum2);
 \draw [link] (x) |- (L);
 \draw [link] (L) -| node[pos=0.9,xshift=-2.5mm] {\footnotesize{$-$}} (sum1);
 \draw [link] (sum2) -- (y);
 \draw [link] (x0) -| (sum2);
\end{tikzpicture}
\caption{Discrete time}
\label{Fig:BlockDiagramStatic_DT}
\end{subfigure}
\caption{Block diagram of the static average consensus algorithms~\eqref{eq::static_alg}. The input signals are assigned to the initial conditions, that is, $\vect{x}(0) = \vectsf{u}$ (in continuous time) or $\vect{x}_0=\vectsf{u}$ (in discrete time). The feedback loop consists of the Laplacian matrix of the communication graph and an integrator ($1/s$ in continuous time and $1/(z-1)$ in discrete time).
}
\label{Fig:BlockDiagramStatic}
\end{figure}
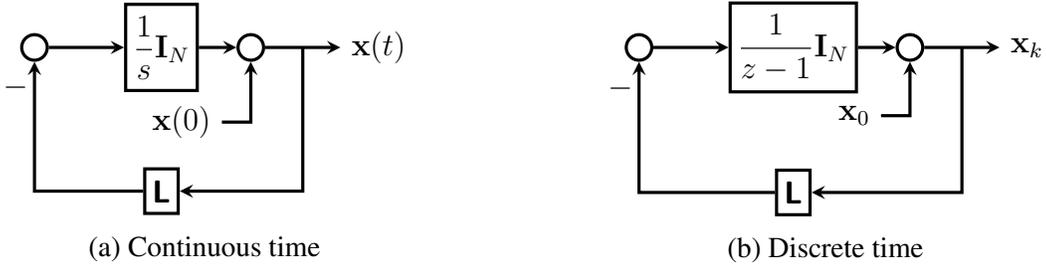
The block diagram of these static average consensus algorithms is shown in Figure~\ref{Fig:BlockDiagramStatic}. The dynamics of these algorithms consists of a negative feedback loop where the feedback term is composed of the Laplacian matrix and an integrator ($1/s$ in continuous time and $1/(z-1)$ in discrete time). For the static average consensus algorithms, the reference signal enters the system as the initial condition of the integrator state. Under certain conditions on the communication graph, we can show that the error of these algorithms converges to zero, as stated next.

\begin{theorem}[Convergence guarantees of the CT and DT static average consensus algorithms~\eqref{eq::Static_CT_DT}~\cite{ROS-JAF-RMM:07}]\label{Thm:static}
Suppose that the communication graph is constant, strongly connected, and weight-balanced digraph, and that the reference signal $\mathsf{u}^i$ at each agent $i\in\until{N}$ is a constant scalar. Then the following convergence results hold for the CT and DT static average consensus algorithms~\eqref{eq::Static_CT_DT}
\begin{description}
\item[CT:] As $t\to\infty$ every agreement state $x^i(t)$, $i\in\until{N}$ of the CT static average consensus algorithm~\eqref{eq::Static_CT} converges to $\mathsf{u}^\text{avg}$ with an exponential rate no worse than $\Hlambda_2$, the smallest nonzero eigenvalue of $\text{Sym}(\lL)$.

\item[DT:] As $k\to\infty$ every agreement state $x_k^i$, $i\in\until{N}$ of the DT static average consensus algorithm~\eqref{eq::Static_DT} converges to $\mathsf{u}^\text{avg}$ with an exponential rate no worse than $\rho\in(0,1)$, provided that the Laplacian matrix satisfies $\rho = \|\vect{I}_N-\lL-\mathbf{1}_N\mathbf{1}_N^\top/N\|_2 < 1$.
\end{description}
\end{theorem}

Note that, given a weighted graph with Laplacian matrix $\lL$, we can scale the graph weights by a nonzero constant $\delta\in\real$ to produce a scaled Laplacian matrix $\delta\lL$, see ``Basic Notions from Graph Theory." This extra scaling parameter can then be used to produce a Laplacian matrix that satisfies the conditions in Theorem~\ref{Thm:static}.

\subsection{A first design for dynamic average consensus}
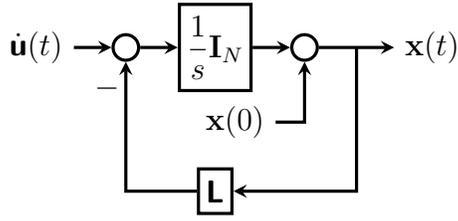
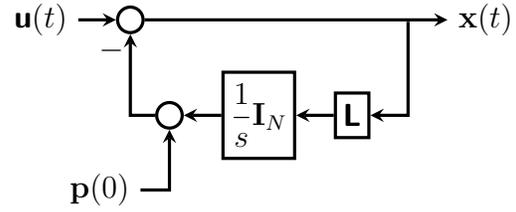
\begin{figure}[htb]
\begin{subfigure}{0.5\textwidth}
\centering \begin{tikzpicture}[align=center]
 \node (u)    {$\dvectsf{u}(t)$};
 \node (sum1) [sum,right=5mm of u] {};
 \node (f)    [block,right=5mm of sum1] {$\dfrac{1}{s} \vect{I}_N$};
 \node (sum2) [sum,right=5mm of f] {};
 \node (L)    [block,below=of f] {$\lL$};
 \coordinate (x) at ($(sum2.east)+(5mm,0)$);
 \node (y)    [right=5mm of x] {$\vect{x}(t)$};
 \node (x0)   [below left=5mm and 2.5mm of sum2] {$\vect{x}(0)$};
 \draw [link] (u) -- (sum1);
 \draw [link] (sum1) -- (f);
 \draw [link] (f) -- (sum2);
 \draw [link] (sum2) -- (y);
 \draw [link] (x) |- (L);
 \draw [link] (L) -| node[pos=0.9,xshift=-2.5mm] {\footnotesize{$-$}} (sum1);
 \draw [link] (x0) -| (sum2);
\end{tikzpicture}
\caption{Dynamic average consensus algorithm~\eqref{eq::alg_DPS-ROS-RMM:05b}}
\label{Fig:BlockDiagramDynamic_CT1}
\end{subfigure}%
\begin{subfigure}{0.5\textwidth}
\centering \begin{tikzpicture}[align=center]
 \node (u)    {$\vectsf{u}(t)$};
 \node (sum1) [sum,right=0.5cm of u] {};
 \node (sum2) [sum,below right=1cm and 0.25cm of sum1] {};
 \node (f)    [block,right=0.5cm of sum2] {$\dfrac{1}{s} \vect{I}_N$};
 \node (L)    [block,right=0.5cm of f]  {$\lL$};
 \node (x)    at (sum1 -| L.east) [xshift=0.5cm,inner xsep=0cm,inner ysep=0cm] {};
 \node (y)    [right=0.5cm of x] {$\vect{x}(t)$};
 \node (x0)   [below left=0.5cm and 0.25cm of sum2] {$\vect{p}(0)$};
 \draw [link] (u) -- (sum1);
 \draw [link] (sum1) -- (y);
 \draw [link] (x) |- (L);
 \draw [link] (L) -- (f);
 \draw [link] (f) -- (sum2);
 \draw [link] (sum2) -| node[pos=0.9,xshift=-2.5mm] {\footnotesize{$-$}} (sum1);
 \draw [link] (x0) -| (sum2);
\end{tikzpicture}
\caption{Dynamic average consensus algorithm~\eqref{eq::alg_DPS-ROS-RMM:05b_no_udot}}
\label{Fig:BlockDiagramDynamic_CT2}
\end{subfigure}%
\caption{Block diagram of the continuous-time dynamic average consensus algorithms~\eqref{eq::alg_DPS-ROS-RMM:05b} and~\eqref{eq::alg_DPS-ROS-RMM:05b_no_udot}. While the reference signals are applied as initial conditions for the static consensus algorithms, the reference signals are here applied as inputs to the system. Although both systems are equivalent, system (a) is in the form~\eqref{eq::AgentSingInt-CT-DT} and explicitly requires the derivative of the reference signals, while system (b) does not require differentiating the reference signals.}
\label{Fig:BlockDiagramDynamic}
\end{figure}

Since the reference signals enter the static average consensus algorithms~\eqref{eq::Static_CT_DT} as initial conditions, they cannot track time-varying signals. Looking at the frequency-domain representation in Figure~\ref{Fig:BlockDiagramStatic} of the static average consensus algorithms~\eqref{eq::Static_CT_DT} we can see that what is needed instead is to continuously inject the signals as inputs into the dynamical system. This allows the system to naturally respond to changes in the signals without any need for re-initialization. This basic observation is made in~\cite{DPS-ROS-RMM:05b}, resulting in the systems shown in Figure~\ref{Fig:BlockDiagramDynamic}.

More precisely,~\cite{DPS-ROS-RMM:05b} argues that considering the static inputs as a dynamic step function, the algorithm 
 \begin{align*}
 & \dvect{x}(t)=-\lL\vect{x}(t)+\dot{\vectsf{u}}(t),\quad
  x^i(0)=\mathsf{u}^i(0),~\mathsf{u}^i(t)=\mathsf{u}^i \text{h}(t),\quad i\in\until{N},
\end{align*}
in which the reference value of the agents enter the dynamics as an external input, results in the same frequency representation~\eqref{eq::Freq_Static_CT} (here $\text{h}(t)$ is the Heaviside step function). Therefore, convergence to the average of reference values is guaranteed. Based on this observation,~\cite{DPS-ROS-RMM:05b} proposes one of the earliest algorithms for dynamic average consensus,
\begin{subequations}\label{eq::alg_DPS-ROS-RMM:05b}
\begin{align}
\dot{x}^i(t)&=-\sum_{j=1}^N{a}_{ij}(x^i(t)-x^j(t))+\dot{\mathsf{u}}^i(t),\quad i\in\until{N},\label{eq::alg_DPS-ROS-RMM:05b-a}\\
x^i(0)&=\mathsf{u}^i(0).\label{eq::alg_DPS-ROS-RMM:05b-init}
\end{align}
\end{subequations}
Using a Laplace domain analysis, \cite{DPS-ROS-RMM:05b} shows that, if each input signal $\mathsf{u}^i$, $i\in\until{N}$, has a Laplace transform with all poles in the left half-plane and at most one zero pole (such signals are asymptotically constant), all the agents implementing algorithm~\eqref{eq::alg_DPS-ROS-RMM:05b} over a connected graph track $\mathsf{u}^\textup{avg}(t)$ with zero error asymptotically. As we show below, the convergence properties of algorithm~\eqref{eq::alg_DPS-ROS-RMM:05b} can be described more comprehensively using time-domain ISS analysis.

Define the tracking error of agent $i$ by
 \begin{align*}
    e^i(t)=x^i(t)-\mathsf{u}^\text{avg}(t), \quad i\in\until{N}.
\end{align*}
To analyze the system, the error is decomposed into the \emph{consensus direction} (that is, the direction $\mathbf{1}_N$) and the \emph{disagreement directions} (that is, the directions orthogonal to $\mathbf{1}_N$). To this end, define the transformation matrix
$\vect{T} = \begin{bmatrix}\frac{1}{\sqrt{N}}\vect{1}_N & \vect{\mathsf{R}}\end{bmatrix}$ where $\vect{\mathsf{R}}\in\R^{N\times(N-1)}$ is such that $\vect{T}^\top\vect{T}=\vect{T}\vect{T}^\top=\vect{I}_N$, and consider the change of variables
\begin{align}\label{eq::error_trans}
    \Bvect{e}=\begin{bmatrix}\bar{e}_1\\\Bvect{e}_{2:N}\end{bmatrix}=\vect{T}^\top\vect{e}.
\end{align}
In the new coordinates,~\eqref{eq::alg_DPS-ROS-RMM:05b} takes the form
\begin{subequations}
\begin{alignat}{2}
\dot{\bar{e}}_1&=0,&  \bar{e}_1(t_0)&=\frac{1}{\sqrt{N}}\sum\nolimits_{j=1}^N(x^i(t_0)-\mathsf{u}^i(t_0)),
\label{eq::bar_e_1}
\\
\dot{\Bvect{e}}_{2:N}&=-\vect{\mathsf{R}}^\top\,\lL\, \vect{\mathsf{R}}\,\Bvect{e}_{2:N}+\vect{\mathsf{R}}^\top\dvect{\mathsf{u}},& \qquad \qquad\Bvect{e}_{2:N}(t_0)&=\vect{\mathsf{R}}^\top\,\vect{x}(t_0),
\end{alignat}
\end{subequations}
where $t_0$ is the initial time. Using~the ISS bound on the trajectories of LTI systems (see ``Input-to-State Stability of LTI Systems"), the tracking error of each agent $i\in\until{N}$ while implementing~\eqref{eq::alg_DPS-ROS-RMM:05b} over a strongly connected and weight-balanced digraph is (here we use $\vect{\Pi}=(\vect{I}_N-\frac{1}{N}\vect{1}_N\vect{1}_N^\top)$)
\begin{align}\label{eq::alg_DPS-ROS-RMM:05b-tracking-error}
|e^i(t)|
&\leq \sqrt{\|\Bvect{e}_{2:N}(t)\|^2+|\bar{e}_1(t)|^2} \notag
\\
&\leq \sqrt{\Big(\text{e}^{-\hat{\lambda}_2 \,(t-t_0)} \|\vect{\Pi} \vectsf{x}(t_0)\|+\frac{\sup_{t_0\leq\tau\leq t}\|\vect{\Pi} \dvect{\mathsf{u}}(\tau)\|}{\hat{\lambda}_2}\Big)^2+\Big(\frac{\sum_{j=1}^N(x^i(t_0)-\mathsf{u}^i(t_0))}{\sqrt{N}}\Big)^2}, 
\end{align}
for all $t\in[t_0,\infty)$,
where $\Hlambda_2$ is the smallest nonzero eigenvalue of $\Sym{\lL}$.

The tracking error bound~\eqref{eq::alg_DPS-ROS-RMM:05b-tracking-error} reveals several interesting facts. First, it highlights the necessity for the special initialization~\eqref{eq::alg_DPS-ROS-RMM:05b-init}. Without it, a fixed offset from perfect tracking is present regardless of the type of reference input signals --instead, we would expect that a proper dynamic consensus algorithm should be capable of perfectly tracking static reference signals. Next,~\eqref{eq::alg_DPS-ROS-RMM:05b-tracking-error} shows that the algorithm~\eqref{eq::alg_DPS-ROS-RMM:05b} renders perfect asymptotic tracking not only for reference input signals with decaying rate, but also for unbounded reference signals whose uncommon parts asymptotically converge to a constant value. This fact is due to the ISS tracking bound depending on $ \|(\vect{I}_N-\frac{1}{N}\vect{1}_N\vect{1}_N^\top) \dvect{\mathsf{u}}(\tau)\|$ rather than on $\|\dvect{\mathsf{u}}(\tau)\|$. Note that if the reference signal of each agent $i\in\until{N}$ can be written as  $\mathsf{u}^i(t)=\underline{\mathsf{u}}(t)+\hat{\mathsf{u}}^i(t)$, where $\underline{\mathsf{u}}(t)$ is the (possibly unbounded) common part  and $\hat{\mathsf{u}}^i(t)$ is the uncommon part of the reference signal, we obtain 
\begin{align*}
\|(\vect{I}_N-\frac{1}{N}\vect{1}_N\vect{1}_N^\top) \dvect{\mathsf{u}}(\tau)\|=\|(\vect{I}_N-\frac{1}{N}\vect{1}_N\vect{1}_N^\top)  (\underline{\dot{\mathsf{u}}}(t)\vect{1}_N+\dvect{\hat{\mathsf{u}}}\,(t))\|=\|(\vect{I}_N-\frac{1}{N}\vect{1}_N\vect{1}_N^\top)  \dvect{\hat{\mathsf{u}}}\,(t)\|.
\end{align*}
This goes to show that the algorithm~\eqref{eq::alg_DPS-ROS-RMM:05b} properly uses the local knowledge about the unbounded but common part of the reference dynamic signals to compensate for the tracking error that would be induced due to the natural lag in diffusion of information across the network for dynamic signals. Finally, the tracking error bound~\eqref{eq::alg_DPS-ROS-RMM:05b-tracking-error} shows that, as long as the uncommon part of reference signals has bounded rate, the algorithm~\eqref{eq::alg_DPS-ROS-RMM:05b} tracks the average with some bounded error. For the reader's convenience, we summarize the convergence guarantees of algorithm~\eqref{eq::alg_DPS-ROS-RMM:05b} (equivalently~\eqref{eq::alg_DPS-ROS-RMM:05b_no_udot}) in the following result. 

\begin{theorem}[Convergence of~\eqref{eq::alg_DPS-ROS-RMM:05b} over a strongly connect and weight-balanced digraph]\label{lem::Alg_ContiTime}
  Let $\GG$ be a strongly connect and weight-balanced digraph. Let $\sup_{\tau\in[t,\infty)}\|(\vect{I}_N-\frac{1}{N}\vect{1}_N\vect{1}_N^\top)\dvectsf{u}(\tau)\| \!=\! \gamma(t)\!<\!\infty$.  Then, the trajectories of algorithm~\eqref{eq::alg_DPS-ROS-RMM:05b} are bounded and~satisfy
  \begin{equation} \label{eq::Alg_Alg1_ultimate_bound}
    \lim_{t\to\infty} \big| x^i(t)-\mathsf{u}^\text{avg}(t)\big| \leq 
    \frac{\gamma(\infty)}{ 
      \Hlambda_2},\qquad i\in\until{N}, 
  \end{equation}
  provided $\sum_{i=1}^Nx^i(t_0)=\sum_{i=1}^N\mathsf{u}^i(t_0)$. The convergence rate to this error bound is no worse than~$\textup{Re}({\lambda}_2)$. Moreover, we have $\sum_{i=1}^Nx^i(t)=\sum_{i=1}^N\mathsf{u}^i(t)$ for $t\in[t_0,\infty)$.
\end{theorem}

The explicit expression~\eqref{eq::Alg_Alg1_ultimate_bound} for the tracking error performance is of value for designers.  The smallest non-zero eigenvalue $\hat{\lambda}_2$ of the graph Laplacian is a measure of connectivity of a graph~\cite{MF:73,NMMDA:07}. For highly connected graphs (those with large $\hat{\lambda}_2$), we expect that the diffusion of information across the graph is faster. Therefore, the tracking performance of a dynamic average consensus algorithm over such graphs should be better. Alternatively, when the graph connectivity is low, we expect the opposite effect. The ultimate tracking bound~\eqref{eq::Alg_Alg1_ultimate_bound} highlights this inverse relationship between graph connectivity and steady-state tracking error. Given this inverse relationship, a designer can decide on the communication range of the agents and the expected tracking performance.  Various upper bounds of $\hat{\lambda}_2$ that are function of other graph invariants such as graph degree or the network size for special families of the graphs~\cite{NMMDA:07,RC-FF-AS-SZ:08} can be exploited to design the agents' interaction topology and yield an acceptable tracking performance.

\subsection{Implementation challenges and solutions}
We next discuss some of the features of the algorithm~\eqref{eq::alg_DPS-ROS-RMM:05b} from an implementation perspective. First, we note that even though algorithm~\eqref{eq::alg_DPS-ROS-RMM:05b} tracks~$\mathsf{u}^\text{avg}(t)$ with a steady state error~\eqref{eq::Alg_Alg1_ultimate_bound}, the error can be made infinitesimally small by introducing a high gain $\beta\in\real_{>0}$ to write $\lL$ as $\beta\,\lL$. By doing this, the tracking error becomes $\lim_{t\to\infty} \big| x^i(t)-\mathsf{u}^\text{avg}(t)\big| \leq 
    \frac{\gamma(\infty)}{\beta 
      \Hlambda_2}$, $i\in\until{N}$.
However, for scenarios where the agents are first-order physical systems $\dot{x}^i=c^i(t)$, the introduction of this high gain results in an increase of the control effort $c^i(t)$. To address this, a balance between the control effort and the tracking error margin can be achieved by introducing a two-stage algorithm in which an internal dynamics creates the average using a high-gain dynamic consensus algorithm and feeds the agreement state of the dynamic consensus algorithm as a reference signal to the physical dynamics. This approach is discussed further in the ``Controlling the rate of convergence'' section.

A concern that may exists with the algorithm~\eqref{eq::alg_DPS-ROS-RMM:05b} is that it requires explicit knowledge of the derivative of the reference signals. In applications where the input signals are measured online, computing the derivative can be costly and prone to error.  The other concern is  the particular initialization condition requiring $\sum_{i=1}^Nx^i(t_0)=\sum_{i=1}^N\mathsf{u}^i(t_0)$. In a distributed setting, to comply with this condition agents need to initialize with $x^i(t_0)=\mathsf{u}^i(t_0)$. If the agents are acquiring their signal $\mathsf{u}^i$ from measurements or the signal is the output of a local process, any perturbation in $\mathsf{u}^i(t_0)$ results in a steady-state error in the tracking process. Moreover, if an agent, say agent $N$, leaves the operation permanently at any time $\bar{t}$, then $\sum_{i=1}^{N-1}x^i$ is no longer equal to $\sum_{i=1}^{N-1}\mathsf{u}^i$ after~$\bar{t}$. Therefore, the remaining agents, without re-initialization, carry over a steady-state error in their tracking signal.

Interestingly, all these concerns except for the one regarding an agent's permanent departure can be resolved by a change of variables, corresponding to an alternative implementation of algorithm~\eqref{eq::alg_DPS-ROS-RMM:05b}. Let $p^i=\mathsf{u}^i-x^i$ for $i\in\until{N}$. Then,~\eqref{eq::alg_DPS-ROS-RMM:05b} may be written in the equivalent form as,
\begin{subequations}\label{eq::alg_DPS-ROS-RMM:05b_no_udot}
  \begin{align}
    \dot{p}^i(t)&=\sum_{j=1}^N{a}_{ij}(x^i(t)-x^j(t)),\quad \sum_{i=1}^Np^i(t_0)=0,\quad i\in\until{N},\label{eq::alg_DPS-ROS-RMM:05b_no_udot-a}
    \\
    x^i(t)&=\mathsf{u}^i(t)-p^i(t).
  \end{align}
\end{subequations}
Doing so eliminates the need to know the derivative of the reference signals and generates the same trajectories $t\mapsto x^i(t)$ as~\eqref{eq::alg_DPS-ROS-RMM:05b}.
We note that the initialization condition $\sum\nolimits_{i=1}^Np^i(t_0)=0$ can be easily satisfied if each agent $i\in\until{N}$ starts at $p^i(0)=0$. Note that  this requirement is mild, because $p^i$ is an internal state for agent $i$ and therefore is not affected by communication errors. This initialization condition however, limits the use of algorithm~\eqref{eq::alg_DPS-ROS-RMM:05b_no_udot} in applications where agents join or permanently leave the network at different points in time. To demonstrate robustness of algorithm~\eqref{eq::alg_DPS-ROS-RMM:05b_no_udot} to measurement disturbances, we note that any bounded perturbation in the reference input does not affect the initialization condition but rather appears as an additive disturbance in the communication channel. In particular, observe the following
\begin{subequations}\label{eq::init_ROS-RMM:05b}
\begin{align}
   & \eqref{eq::alg_DPS-ROS-RMM:05b-a}\Rightarrow\sum_{i=1}^N\dot{x}^i(t)=\sum_{i}^N \dot{\mathsf{u}}^i(t)\Rightarrow \sum_{i=1}^N{x}^i(t)=\sum_{i}^N{\mathsf{u}}^i(t)+(\sum_{i=1}^N{x}^i(t_0)-\sum_{i}^N{\mathsf{u}}^i(t_0)),\label{eq::init_ROS-RMM:05b-orig}\\
   &\eqref{eq::alg_DPS-ROS-RMM:05b_no_udot-a}\Rightarrow \sum_{i=1}^N\dot{p}^i(t)=0\quad\quad\quad\Rightarrow \sum_{i=1}^N{p}^i(t)=\sum_{i=1}^N{p}^i(t_0)\label{eq::init_ROS-RMM:05b-alt}
\end{align}
\end{subequations}
As seen in~\eqref{eq::init_ROS-RMM:05b-orig}, if $\sum_{i=1}^N{x}^i(t_0)\neq\sum_{i}^N{\mathsf{u}}^i(t_0)$, then $\sum_{i=1}^N{x}^i(t)\neq\sum_{i}^N{\mathsf{u}}^i(t)$ persists in time. Therefore, if the perturbation on the reference input measurement is removed, the algorithm~\eqref{eq::alg_DPS-ROS-RMM:05b} still inherits the adverse effect of the initialization error. Instead, as~\eqref{eq::init_ROS-RMM:05b-alt} shows for the case of the alternative algorithm~\eqref{eq::alg_DPS-ROS-RMM:05b_no_udot}, $\sum_{i=1}^N{p}^i(t)=0$ is preserved in time as long as the algorithm is initialized such that $\sum_{i=1}^N{p}^i(t_0)=0$, which can be easily done by setting $ p^i(t_0)=0$ for $i\in\until{N}$. Consequently, when the perturbations are removed, the algorithm~\eqref{eq::alg_DPS-ROS-RMM:05b_no_udot} recovers the convergence guarantee of the perturbation-free case. Following steps similar to the ones leading to the bound~\eqref{eq::alg_DPS-ROS-RMM:05b-tracking-error}, we summarize the effect of the additive bounded reference signal measurement perturbation on the convergence of the algorithm~\eqref{eq::alg_DPS-ROS-RMM:05b_no_udot} in the next result.

\begin{lem}[Convergence of~\eqref{eq::alg_DPS-ROS-RMM:05b_no_udot} over a strongly connect and weight-balanced digraph in the presence of additive reference input perturbations]\label{lem::Alg_ContiTime_alt}
  Let $\GG$ be a strongly connect and weight-balanced digraph. Suppose $\mathsf{w}^i(t)$ is an additive perturbation on the measured  reference input signal $\mathsf{u}^i(t)$. Let $\sup_{\tau\in[t,\infty)}\|(\vect{I}_N-\frac{1}{N}\vect{1}_N\vect{1}_N^\top)\dvectsf{u}(\tau)\| \!=\! \gamma(t)\!<\!\infty$ and $\sup_{\tau\in[t,\infty)}\|(\vect{I}_N-\frac{1}{N}\vect{1}_N\vect{1}_N^\top)\dvectsf{w}(\tau)\| \!=\! \omega(t)\!<\!\infty$ . Then, the trajectories of algorithm~\eqref{eq::alg_DPS-ROS-RMM:05b_no_udot} are bounded and~satisfy
  \begin{equation*} 
    \lim_{t\to\infty} \big| x^i(t)\!-\!
        \mathsf{u}^\text{avg}(t) \big| \leq 
    \frac{\gamma(\infty)+\omega(\infty)}{ 
      \Hlambda_2},\qquad i\in\until{N}, 
  \end{equation*}
  provided $\sum_{i=1}^N p^i(t_0)=0$. The convergence rate to this error bound is no worse than~$\textup{Re}({\lambda}_2)$. Moreover, we have $\sum_{i=1}^Np^i(t)=0$ for $t\in[t_0,\infty)$.
\end{lem}

The perturbation $\mathsf{w}^i$ in Lemma~\ref{lem::Alg_ContiTime_alt} can also be regarded as a bounded communication perturbation. Therefore, we can claim that algorithm~\eqref{eq::alg_DPS-ROS-RMM:05b_no_udot} (and similarly~\eqref{eq::alg_DPS-ROS-RMM:05b}) is naturally robust to bounded  communication error. 

From an implementation perspective, it is also desirable that a distributed algorithm is robust to changes in the communication topology that may arise as a result of unreliable transmissions, limited communication/sensing range, network re-routing, or the presence of obstacles. To analyze this aspect, consider a time-varying
digraph $\mathcal{G}(\mathcal{V},\mathcal{E}(t),\vectsf{A}_\sigma(t))$, where the nonzero entries of the adjacency matrix $\vectsf{A}(t)$ are
uniformly lower and upper bounded (in other words, $\mathsf{a}_{ij}(t)\in[\underline{a},
\bar{a}]$, where $0<\underline{a}\leq\bar{a}$, if $(j,i)\in\EE(t)$,
and $\mathsf{a}_{ij}=0$ otherwise). Here,
$\sigma:[0,\infty)\to \mathcal{P} = \until{m}$ is a piecewise constant signal, meaning that it only has a finite number of discontinuities in any finite time interval and that is constant between consecutive discontinuities. Intuitively, consensus in switching networks occurs if there is occasionally enough flow of information from every node in the network to every other node.

Formally, we refer to an admissible switching set $\mathcal{S}_{\mathrm{admis}}$ as a set of piecewise constant switching signals $\sigma: [0,\infty) \rightarrow \mathcal{P}$ with some dwell time $t_L$ (in other words, $t_{k+1}-t_{k}>t_L>0$, for all $k=0,1,\dots$) such that
\begin{itemize}
  \item $\mathcal{G}(\mathcal{V},\mathcal{E}(t),\vectsf{A}_\sigma(t))$ is weight-balanced for $t\geq t_0$;
  \item the number of contiguous, nonempty, uniformly bounded
    time-intervals $[t_{i_j},t_{i_{j+1}})$, $j=1,2,\dots$, starting
    at $t_{i_1}=t_0$, with the property that
    ${\cup}_{t_{i_j}}^{t_{i_{j+1}}}\mathcal{G}(\mathcal{V},\mathcal{E}(t),\vectsf{A}_\sigma(t))$
    is a jointly strongly connected digraph goes to infinity as
    $t\to\infty$.
\end{itemize}

When the switching signal belongs to the admissible set $\mathcal{S}_{\mathrm{admis}}$,~\cite{SSK-JC-SM:15-ijrnc} shows that there always exists $\underline{\lambda}\in\real_{>0}$ and $\kappa\in\real_{\geq1}$ such that $\|\text{e}^{-\vect{\mathsf{R}}^\top\,\lL_\sigma\, \vect{\mathsf{R}}}\|\leq \kappa \text{e}^{-\underline{\lambda}t}$, $t\in\real_{\geq0}$.
Then, implementing the change of variables~\eqref{eq::error_trans}, we can show that the trajectories of algorithm~\eqref{eq::alg_DPS-ROS-RMM:05b_no_udot} satisfy~\eqref{eq::alg_DPS-ROS-RMM:05b-tracking-error} with $\hat{\lambda}_2$ being replaced by $\underline{\lambda}$ and $\|\vect{\Pi}\vect{x}(t_0)\|$ and $\|\vect{\Pi}\dvectsf{u}(\tau)\|$ being multiplied by $\kappa$. We formalize this statement below.

\begin{lem}[Convergence of~\eqref{eq::alg_DPS-ROS-RMM:05b} over switching graphs]\label{lem::Alg_ContiTime_switch}
  Let the communication topology be $\mathcal{G}(\mathcal{V},\mathcal{E}(t),\vectsf{A}_\sigma(t))$ where $\sigma\in\mathcal{S}_{\mathrm{admis}}$. Let $\sup_{\tau\in[t,\infty)}\|(\vect{I}_N-\frac{1}{N}\vect{1}_N\vect{1}_N^\top)\dvectsf{u}(\tau)\| \!=\! \gamma(t)\!<\!\infty$.  Then, the trajectories of algorithm~\eqref{eq::alg_DPS-ROS-RMM:05b} are bounded and~satisfy
  \begin{equation*}
    \lim_{t\to\infty} \big| x^i(t)-\mathsf{u}^\text{avg}(t) \big| \leq 
    \frac{\kappa\gamma(\infty)}{ 
      \underline{\lambda}},\qquad i\in\until{N}, 
  \end{equation*}
  provided $\sum_{i=1}^Nx^i(t_0)=\sum_{i=1}^N\mathsf{u}^i(t_0)$. The convergence rate to this error bound is no worse than~$\underline{\lambda}$. Moreover, we have $\sum_{i=1}^N x^i(t)=\sum_{i=1}^N\mathsf{u}^i(t)$ for $t\in[t_0,\infty)$.
\end{lem}

\subsection{Example: distributed formation control revisited}
\begin{figure}[h]
\begin{subfigure}{0.32\textwidth}
\centering\includegraphics[scale=0.16]{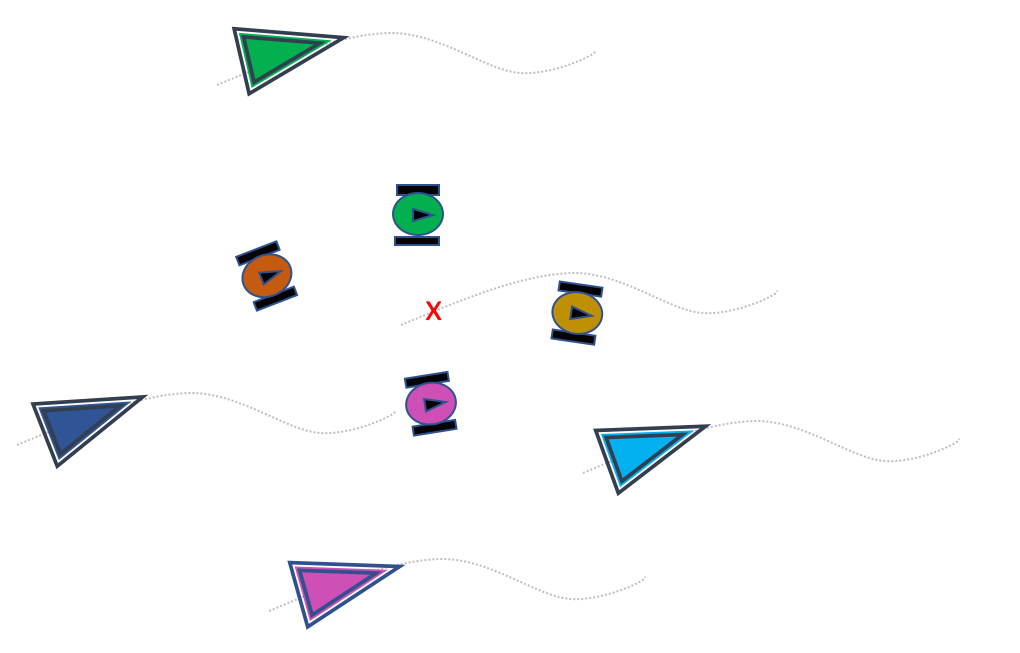}
\caption{}
\end{subfigure}~~
\begin{subfigure}{0.32\textwidth}
\centering\includegraphics[scale=0.16]{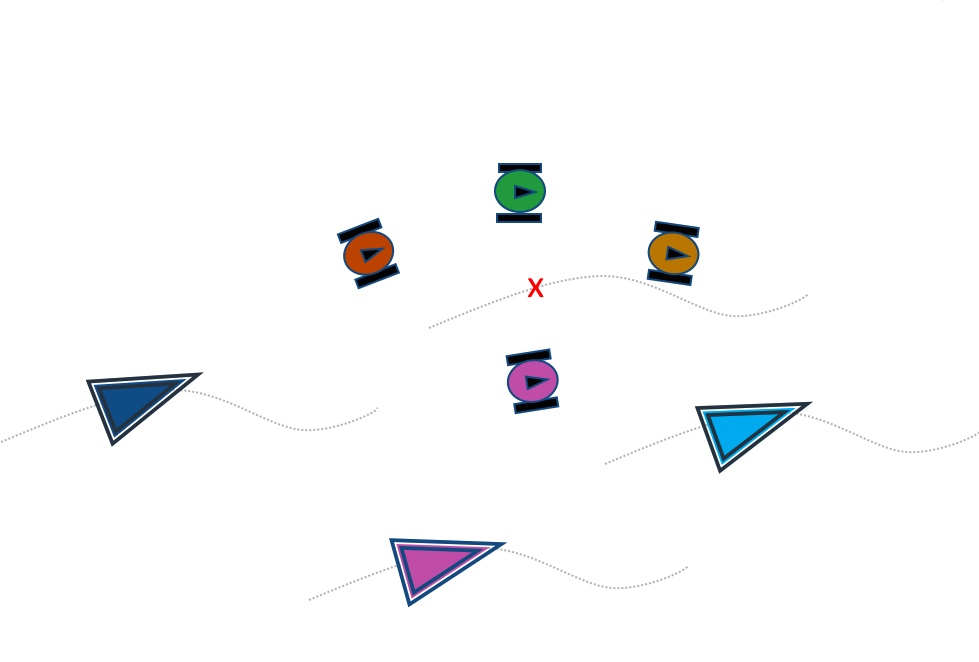}
\caption{}
\end{subfigure}~~
\begin{subfigure}{0.32\textwidth}
\centering\includegraphics[scale=0.16]{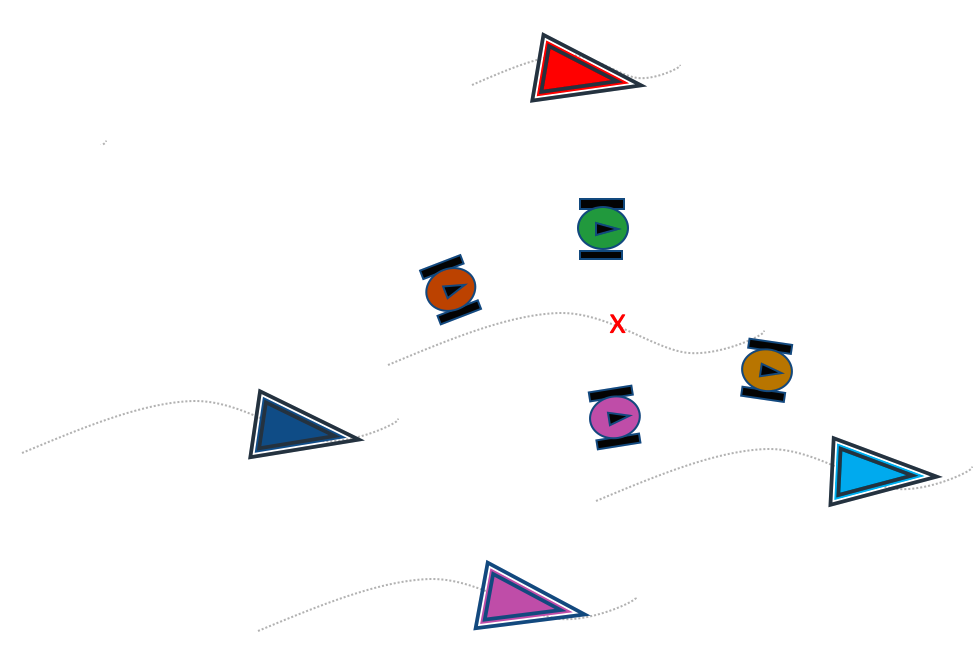}
\caption{}
\end{subfigure}
\caption{A simple dynamic average consensus based containment and tracking of a team of mobile targets: 
The triangle robots cooperatively want to contain the moving round robots by forming a formation around the geometric center of the round robots that they are observing. At plot (b), say after $10$ s from the start of the operation, one of the triangle robots leaves the team. At plot (c), say after $20$ s from the start of the operation, a new triangle robot joins the group to take over tracking the abandoned round robot.}
\label{fig::motiv_examp}
\end{figure}

We come back to one of the scenarios discussed in the "Applications of Dynamic Average Consensus in Network Systems" to illustrate the properties of  algorithm~\eqref{eq::alg_DPS-ROS-RMM:05b} and its alternative implementation~\eqref{eq::alg_DPS-ROS-RMM:05b_no_udot}. Consider a group of four mobile agents (depicted as the triangle robots in Figure~\ref{fig::motiv_examp}) whose communication topology is described by a fixed connected undirected ring.  The objective of these agents is to follow a set of moving targets  (depicted as the round robots in Figure~\ref{fig::motiv_examp}) 
in a containment fashion (that is, making sure that they are surrounded as they move around the environment). 
Let 
\begin{align}\label{eq::ref_example}
x_T^l(t)&=(t/20)^2+0.5\sin((0.35+0.05l)\,t+(5-l)\frac{\pi}{5})+4-2(l-1),\quad l\in\{1,2,3,4\},
\end{align}
be the horizontal position of the set of moving targets (each mobile agent keeps track on one moving target). The term $(t/20)^2$ in the reference signals~\eqref{eq::ref_example} represents the component with unbounded derivative but common across all the agents. 

\begin{figure}[h]
\centering
     \includegraphics[height=1.7in]{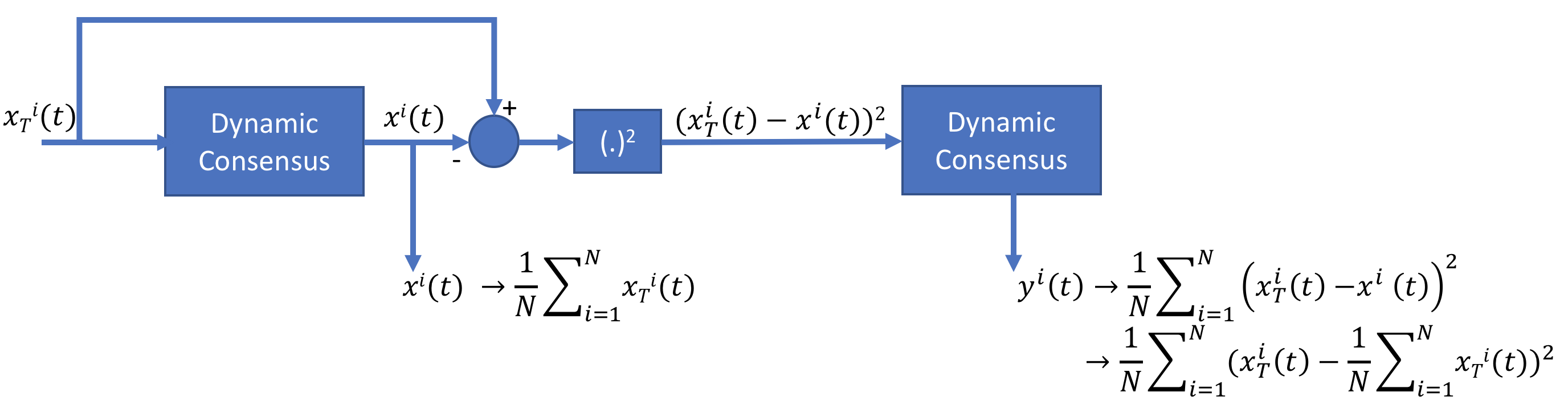}
\caption{A group of $N$ mobile agents use a set of dynamic consensus algorithms to asymptotically track the geometric center $x^\text{avg}_T(t)=\frac{1}{N}\sum_{l=1}^N x_T^l(t)$ and its variance $\frac{1}{N}\sum_{l=1}^N (x^l(t)-x^\text{avg}_T(t))^2$. In this set up each mobile agent $i\in\until{N}$ is monitoring its respective target target's position $x^i_T(t)$. 
 }\label{fig::Example_strategy}
\end{figure}

To achieve their objective, the group of agents seeks to compute on the fly the geometric center $\bar{x}_T(t)=\frac{1}{N}\sum_{l=1}^N x_T^l(t)$ and the associated variance $\frac{1}{N}\sum_{l=1}^N (x^l(t)-\bar{x}_T(t))^2$ determined by the time-varying position of the moving targets. The agents implement two distributed dynamic average consensus algorithms, one for computing the center, the other for computing the variance, as shown in Figure~\ref{fig::Example_strategy}. In this scenario, to illustrate the properties discussed in this section, we have  agent $4$ (the green triangle in Figure~\ref{fig::motiv_examp}) leave the network $10$ s after the beginning of the simulation. Later, $10$ s later, a new agent, labeled $5$  (the red triangle in Figure~\ref{fig::motiv_examp}), joins the network and starts monitoring the target that the old agent $4$ was in charge of. 

\begin{figure}[t]
\centering
\begin{subfigure}{0.45\textwidth}
  \includegraphics[height=1.8in]{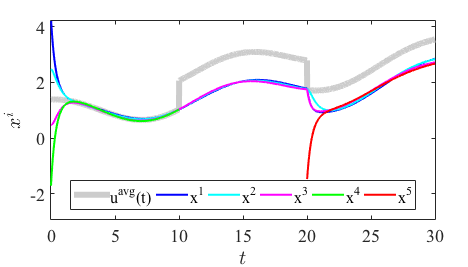}
  \caption{\scriptsize Agent departure and arrival: in this simulation the agents are using algorithm~\eqref{eq::alg_DPS-ROS-RMM:05b_no_udot}. As guaranteed in Lemma~\ref{lem::Alg_ContiTime_alt}, under proper initialization $p^i(0)=0$, $i\in\{1,2,3,4\}$, during the time interval $[0,10]$ s the agents are able to track $x_T^\text{avg}(t)$ with a small error. The challenge presents itself when agent $4$ leaves the operation at $t=10$ s. As seen, because after agent $4$ leaves $\sum_{i=1}^3p^i(10^+)\neq 0$ the remaining agents fail to follow the average of their reference values, which now is $\frac{1}{3}\sum_{l=1}^3\mathsf{u}^l$. Similarly, even with initialization of $p^5(20)=0$ for the new agent $5$, because $p^1(20)+p^2(20)+p^3(20)+p^5(20)\neq 0$, the agents track the average $\frac{1}{4}\sum_{l=1}^4x_T^l(t)$ with a steady state error.}
\end{subfigure}\quad
\begin{subfigure}{0.45\textwidth}
  \includegraphics[height=1.8in]{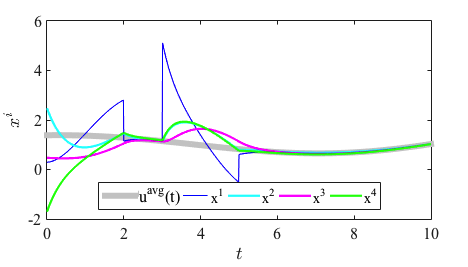}
  \caption{\scriptsize Perturbation of input signals: in this simulation the agents are using algorithm~\eqref{eq::alg_DPS-ROS-RMM:05b_no_udot} and at time interval $[0,10]$ s, agent $1$'s reference input is subject to a measurement perturbation according to $\mathsf{u}^1(t)=x_T^1(t)+\mathsf{w}^1(t)$, where $\mathsf{w}^1(t)=-4\cos(t)$ at $t\in[0,2]$ and $t\in[3,5]$ and $\mathsf{w}^1(t)=0$ in other times.  As guaranteed by Lemma~\ref{lem::Alg_ContiTime_alt}, despite the perturbation, including the initial measurement error of $\mathsf{u}^1(0)=x_T^1(0)-4$, algorithm~\eqref{eq::alg_DPS-ROS-RMM:05b_no_udot} has robustness to the measurement perturbation and recovers its performance after the perturbation is removed. Here, we used a large perturbation error so that its effect is observed more visibly in the simulation plots.~\\}
  \end{subfigure}\\
    \begin{subfigure}{0.45\textwidth}
  \includegraphics[height=1.8in]{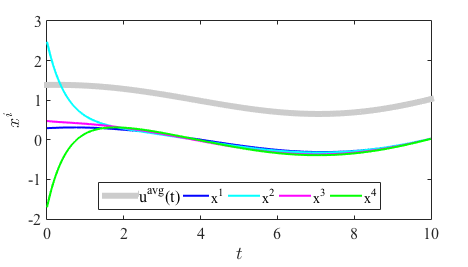}
  \caption{\scriptsize Initialization error: in this simulation the agents are using algorithm~\eqref{eq::alg_DPS-ROS-RMM:05b}. Agent $1$'s reference input has an initial measurement error of $\mathsf{u}^1(0)=x_T^1(0)-4$. Since the measurement error directly effects  the initialization condition of the algorithm, it fails to preserve $\sum_{i=1}^4{x}^l(t)=\sum_{i=1}^4\mathsf{u}^l(t)$. As a result, the effect of initialization error persists and algorithm maintains a significant tracking error.\\~}
\end{subfigure}\quad
  \begin{subfigure}{0.45\textwidth}
  \includegraphics[height=1.8in]{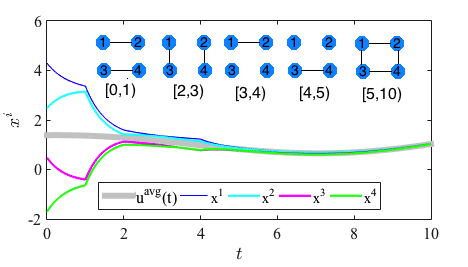}
  \caption{\scriptsize Switching topology: in this simulation the agents are using algorithm~\eqref{eq::alg_DPS-ROS-RMM:05b_no_udot} (similar results also is obtained for algorithm~\eqref{eq::alg_DPS-ROS-RMM:05b}). The network communication topology is a switching graph, where the graph topology at different time intervals is shown on the plot. Since the switching signal $\sigma$ belongs to $\mathcal{S}_{\mathrm{admis}}$, as predicted by Lemma~\ref{lem::Alg_ContiTime_switch}, the trajectories of the algorithm stay bounded and once the topology becomes fully connected, the agents follow their respective dynamic average closely.}
\end{subfigure}
\caption{Performance evaluation of dynamic average consensus algorithms~\eqref{eq::alg_DPS-ROS-RMM:05b} and~\eqref{eq::alg_DPS-ROS-RMM:05b_no_udot} for a group of $4$ agents with reference inputs~\eqref{eq::ref_example} for a tracking scenario described  in Figure~\ref{fig::motiv_examp}.}
\label{fig::motiv_examp_resp}
\end{figure}

For simplicity, we focus our simulation on the calculation of the geometric center. For this computation, agents implement the algorithm~\eqref{eq::alg_DPS-ROS-RMM:05b_no_udot} with reference input $\mathsf{u}^i(t)=x^i_T(t)$, $i\in\{1,2,3,4\}$. Figure~\ref{fig::motiv_examp_resp} shows the algorithm performance for various operational scenarios. As forecast by the discussion of this section, the tracking error vanishes in the presence of  perturbations in the input signals available to the individual agents and switching topologies, and exhibits only partial robustness to agent arrivals, departures, and initialization errors, with a constant bias with respect to the correct average. Additionally, it is worth noticing in  Figure~\ref{fig::motiv_examp_resp} that all the agents exhibit convergence with the same rate.

The introduction of algorithm~\eqref{eq::alg_DPS-ROS-RMM:05b_no_udot} has served as preparation for a more in-depth treatment of the design of dynamic average consensus algorithms, which we tackle next. We discuss how to address the issues of correct initialization (the steady-state error depends on the initial condition $\vect{x}(0)$ or $\vect{x}_0$), adjust convergence rate of the agents, 
and the limitation of tracking with zero steady-state error only constant reference signals (and therefore with small steady-state error for slowly time-varying reference signals).  For clarity of presentation, we discuss continuous-time and discrete-time strategies separately.
\clearpage
\section{Continuous-Time Dynamic Average Consensus Algorithms}

In the following, we discuss various continuous-time dynamic average consensus algorithms and discuss their performance and robustness guarantees. Table~\ref{table::CT-driving_command} summarizes the arguments of the driving command of these algorithms in~\eqref{eq::AgentSingInt} and their special initialization requirements. Some of these algorithms when cast in the form of~\eqref{eq::AgentSingInt} require access to the derivative of the reference signals. Similar to the algorithm~\eqref{eq::alg_DPS-ROS-RMM:05b}, however, this requirement can be eliminated using alternative implementations.
 \begin{table}[htb]
 \footnotesize
    \caption{{\small Arguments of the driving command in~\eqref{eq::AgentSingInt} for the reviewed continuous-time dynamic average consensus algorithms together with their initialization requirements.
      }
    }\label{table::CT-driving_command}\vspace{-0.06in}
  \centering
  \setlength\tabcolsep{2.5pt}
  \begin{tabular}{|c||c|c|c|c|}
    \hline
    Algorithm&\eqref{eq::alg_DPS-ROS-RMM:05b}&\eqref{eq::alg_RAF-PY-KML:06}&\eqref{eq::SSK-JC-SM:13-ecc-A1}&\eqref{eq::alg-SSK-JC-SM:15-ijrnc-ct}\\ \hline
    $J^i(t)$ &
    $\{x^i(t),\dot{\mathsf{u}}(t)\}$&
    $\{x^i(t),v^i(t),{\mathsf{u}}(t)\}$&
    $\left.\begin{array}{c}\{x^i(t),z^i(t),v^i(t),\\{\mathsf{u}}(t),\dot{\mathsf{u}}(t)\}\end{array}\right.$&
    $\left.\begin{array}{c}\{x^i(t),v^i(t),\\
    ~~{\mathsf{u}}(t),\dot{\mathsf{u}}(t)\}\end{array}\right.$
    \\ \hline
   \!\! ${\{I^j(t)\}}_{j\in\NN_{\text{out}}^i}$ &${\{x^j(t)\}}_{j\in\NN_{\text{out}}^i}$ & ${\{x^j(t),v^j(t)\}}_{j\in\NN_{\text{out}}^i}$& ${\{z^j(t),v^j(t)\}}_{j\in\NN_{\text{out}}^i}$&${\{v^i(t)\}}_{j\in\NN_{\text{out}}^i}$\\       \hline
    \parbox{14ex}{Initialization Requirement} & $x^i(0)\!=\!\mathsf{u}^i(0)$& none& none&$\sum_{j=1}^Nv^j(0)=0$\\       \hline
    \end{tabular}
\end{table}

\subsection{Robustness to initialization and permanent agent dropout}

\begin{figure}[!thb]
\begin{subfigure}{0.46\textwidth}
\centering
\begin{tikzpicture}
  \node (alpha) [block] {$\alpha\vect{I}$};
  \node (sum1)  [sum,right=4mm of alpha] {};
  \node (sum2)  [sum,below=8mm of sum1] {};
  \node (u_dot) [above left=4mm and 12mm of sum1] {$\dvectsf{u}(t)$};
  \node (LI1)   [block,below right=8mm and 2mm of sum2] {$\lL_\text{I}^\top$};
  \node (h2)    [block,right=6mm of LI1] {$\dfrac{1}{s}$};
  \node (LI2)   [block,right=6mm of h2] {$\lL_\text{I}$};
  \node (Lp)    at (sum2 -| h2) [block] {$\lL_\text{p}$};
  \node (h1)    at (sum1 -| h2) [block] {$\dfrac{1}{s+\alpha} \vect{I}$};
  \coordinate (x) at ($(sum1 -| LI2.east)+(4mm,0)$) {};
  \draw [link] ($(alpha)-(10mm,0)$) -- node[pos=0.25,yshift=2.5mm] {$\vectsf{u}(t)$} (alpha);
  \draw [link] (alpha) -- (sum1);
  \draw [link] (u_dot) -| (sum1);
  \draw [link] (sum1) -- (h1);
  \draw [link] (Lp) -- (sum2);
  \draw [link] (LI2) -- (h2);
  \draw [link] (h2) -- (LI1);
  \draw [link] (LI1) -| (sum2);
  \draw [link] (sum2) -- node[pos=0.7,xshift=-2.5mm] {\footnotesize{$-$}} (sum1);
  \draw [link] (h1) -- node[pos=1,xshift=-4mm,yshift=2.5mm] {$\vect{x}(t)$} ($(x)+(5mm,0)$);
  \draw [link] (x) |- (LI2);
  \draw [link] (sum2 -| x) -- (Lp);
\end{tikzpicture}
\caption{\footnotesize Continuous-time algorithm in~\eqref{eq::alg_RAF-PY-KML:06} that is robust to initialization. To see why the algorithm is robust, consider multiplying the input signal on the left by $\1^\top$. Then the output of the integrator block ($1/s$) is multiplied by zero (since $\lL_\text{I}\1=0$) and therefore does not affect the output. While the output is affected by the initial state of the $1/(s+\alpha)$ block, this term decays to zero and therefore does not affect the steady-state. Also, the requirement of needing the derivative of the input $\dvectsf{u}(t)$ can be removed by a change of variable.}
\label{Fig:BlockDiagramDynamic_CT_RAF-PY-KML:06}
\end{subfigure}\hfill%
\begin{subfigure}{0.5\textwidth}
\centering \begin{tikzpicture}[align=center]
 \node (sum)   [sum] {};
 \node (alpha) [block,right=4mm of sum] {$\alpha\vect{I}$};
 \node (sum1)  [sum,right=4mm of alpha] {};
 \node (u_dot) [above left=4mm and 2mm of sum] {$\dvectsf{u}(t)$};
 \node (sum2)  [sum,right=4mm of sum1] {};
 \node (sum3)  [sum,right=4mm of sum2] {};
 \node (f)     [block,right=6mm of sum3] {$\dfrac{1}{s} \vect{I}$};
 \node (L)     [block,below=5mm of f] {$\beta\,\lL$};
 \node (LI1)   [block,below=15mm of f] {$\frac{\alpha\beta}{s}\lL$};
 \node (sum4)  [sum,left=8mm of LI1] {};
 \node (q0)    [below left=2.5mm and 2.5mm of sum4] {$\vect{q}(0)$};
 \coordinate (x) at ($(f.east)+(10mm,0)$);
 \draw [link] ($(sum)-(10mm,0)$) -- node[pos=0.25,yshift=2.5mm] {$\vectsf{u}(t)$} (sum);
 \draw [link] (u_dot) -| (sum1);
 \draw [link] (sum) -- (alpha);
 \draw [link] (alpha) -- (sum1);
 \draw [link] (sum1) -- (sum2);
 \draw [link] (sum2) -- (sum3);
 \draw [link] (sum3) -- (f);
 \draw [link] (f) -- node[pos=1,xshift=-5mm,yshift=2.5mm] {$\vect{x}(t)$} ($(f.east)+(15mm,0)$);
 \draw [link] (x) |- (L);
 \draw [link] (x) |- (LI1);
 \draw [link] (L) -| node[pos=1,xshift=-2.5mm,yshift=-3mm] {\footnotesize{$-$}} (sum3);
 \draw [link] (LI1)--(sum4);
 \draw [link] (sum4) -| node[pos=1,xshift=-2.5mm,yshift=-3mm] {\footnotesize{$-$}} node[pos=0.6,xshift=-5mm] {\footnotesize{$\vect{q}(t)$}} (sum2);
 \draw [link] (q0) -| (sum4);
 \draw [link] (x) |- ($(f.south)-(0,32mm)$) -| node[pos=1,xshift=-2.5mm,yshift=-3mm] {\footnotesize{$-$}} (sum);
\end{tikzpicture}
\caption{\footnotesize Continuous-time algorithm in~\eqref{eq::alg-SSK-JC-SM:15-ijrnc-ct}. The parameter $\beta$ may be used to control the tracking error size while $\alpha$ may be used to control the rate of convergence. Furthermore, this algorithm is robust to reference signal measurement perturbations and naturally preserves the privacy of the input signals against adversaries~\cite{SSK-JC-SM:15-ijrnc}.}
\label{Fig:BlockDiagramDynamic_CT_SSK-JC-SM:15}
\end{subfigure}
\caption{Block diagram of continuous-time dynamic average consensus algorithms. These \emph{dynamic} algorithms naturally adapt to changes in the reference signals which are applied as inputs to the system.}
\label{Fig:BlockDiagramDynamic_RAF-SSK}
\end{figure}
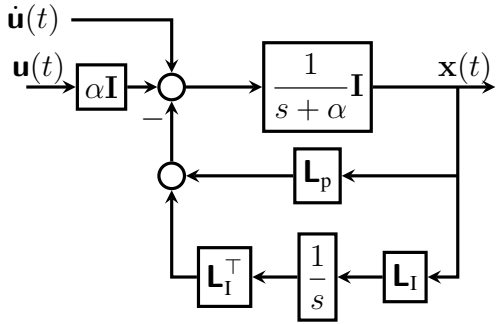
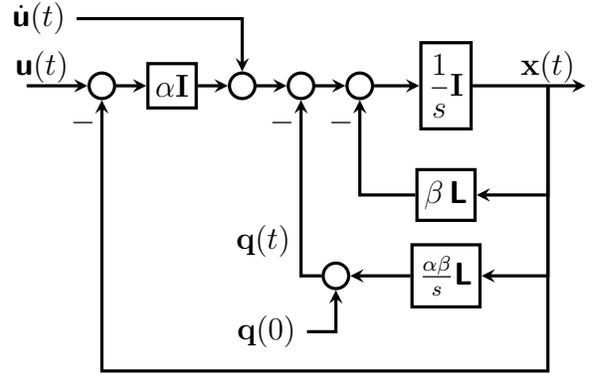

To eliminate the special initialization requirement and to induce robustness with respect to algorithm initialization,~\cite{RAF-PY-KML:06} proposes the following alternative dynamic average consensus algorithm
\begin{subequations}\label{eq::alg_RAF-PY-KML:06}
\begin{align}
&\dot{q}^i(t)=-\sum_{j=1}^N{b}_{ij}\,(x^i-x^j),\\
&\dot{x}^i=-\alpha\,(x^i-\mathsf{u}^i)- \sum_{j=1}^N{a}_{ij}\,(x^i-x^j)+\sum_{j=1}^N{b}_{ji}\,(q^i-q^j)+\dot{\mathsf{u}}^i,\label{eq::alg_RAF-PY-KML:06-x}\\
&q^i(t_0),\, x^i(t_0)\in\real,\quad \quad i\in\until{N},
\end{align}
\end{subequations}
where $\alpha\in\realpositive$. Here, we add $\dot{\mathsf{u}}^i$ to~\eqref{eq::alg_RAF-PY-KML:06-x} to allow agents to track reference inputs whose derivatives have unbounded common components. The necessity of having explicit knowledge of the derivative of reference signals can be removed by using the change of variables ${p}^i=x^i-\mathsf{u}^i$, $i\in\until{N}$.  In algorithm~\eqref{eq::alg_RAF-PY-KML:06}, the agents are allowed to use two different adjacency matrices $[{a}_{ij}]_{N\times N}$ and $[{b}_{ij}]_{N\times N}$, so that they have an extra degree of freedom to  adjust the tracking performance of the algorithm. The Laplacian matrices associated with adjacency matrices $[{a}_{ij}]$ and $[{b}_{ij}]$ are represented by, respectively, $\lL_{\text{p}}$, labeled as \emph{proportional} Laplacian and $\lL_{\text{I}}$, labeled as \emph{integral} Laplacian. Without loss of generality, to simplify the algorithm we can set $\lL_{\text{I}}=\beta\lL_{\text{P}}$, for $\beta\in\real_{>0}$, and use $\beta$ as a design variable to improve the convergence of the algorithm. The compact representation of~\eqref{eq::alg_RAF-PY-KML:06} is as follows 
\begin{subequations}\label{eq::alg_RAF-PY-KML:06-collect}
\begin{align}
&\dvect{q}=-\lL_{\text{I}}\,\vect{x},\\
&\dvect{x}=-\alpha\,(\vect{x}-\vectsf{u})-\lL_{\text{p}}\,\vect{x}+\lL_{\text{I}}^\top\vect{q}+\dvectsf{u},\label{eq::alg_RAF-PY-KML:06-collect-b}
\end{align}
\end{subequations}
which also reads as
\begin{align*}
    \dvect{x}=-\alpha\,(\vect{x}-\vectsf{u})-\lL_{\text{p}}\,\vect{x}-\lL_{\text{I}}^\top\int_{t_0}^t\!\lL_{\text{I}}\,\vect{x}(\tau)\,\text{d}\tau+\lL_{\text{I}}^\top \vect{q}(t_0)+\dvectsf{u}.
\end{align*}
Using a time-domain analysis similar to that employed for algorithm~\eqref{eq::alg_DPS-ROS-RMM:05b}, we characterize the ultimate tracking behavior of the algorithm~\eqref{eq::alg_RAF-PY-KML:06}.
We consider the change of variables~\eqref{eq::error_trans} and
\begin{subequations}\label{eq::v-w-y}
\begin{align}
\vect{w}&=\begin{bmatrix}w_1\\ \vect{w}_{2:N}\end{bmatrix}=\vect{T}^\top\,\vect{q},\label{eq::v-w}\\
\vect{y}&=\vect{w}_{2:N}-\alpha ({\vectsf{R}}^\top\lL_{\text{I}}^\top\vect{\mathsf{R}})^{-1}\vectsf{R}^\top\dvectsf{u},
\end{align}
\end{subequations}
to write~\eqref{eq::alg_RAF-PY-KML:06-collect} in the equivalent form 
\begin{subequations}\label{eq::alg_RAF-PY-KML:06-collectw1-w2-e}
\begin{align}
\dot{w}_1&=0,\\
\begin{bmatrix}
\dvect{y}\\
\dot{\bar{e}}_1\\
\dot{\Bvect{e}}_{2:N}
\end{bmatrix}&=\underbrace{\begin{bmatrix}\vect{0}&\vect{0}&-\vect{\mathsf{R}}^\top\lL_{\text{I}}\vect{\mathsf{R}}\\
0&-\alpha&0\\
\vect{\mathsf{R}}^\top\lL_{\text{I}}^\top\vect{\mathsf{R}}&\vect{0}&-\alpha\vect{I}-\vect{\mathsf{R}}^\top\lL_{\text{p}}\vect{\mathsf{R}}
\end{bmatrix}}_{\vect{A}}\begin{bmatrix}
\vect{y}\\
\bar{e}_1\\
\Bvect{e}_{2:N}
\end{bmatrix}+\underbrace{\begin{bmatrix}-\alpha ({\vectsf{R}}^\top\lL_{\text{I}}^\top\vect{\mathsf{R}})^{-1}\\0 \\  \vect{I} \end{bmatrix}}_{\vect{B}}\vect{\mathsf{R}}^\top\dvectsf{u}.\label{eq::alg_RAF-PY-KML:06-collectw2-e}
\end{align}
\end{subequations}
Let the communication ranges of the agents be such that they can establish adjacency relations $[a_{ij}]$ and $[b_{ij}]$ such that the corresponding $\lL_{\text{I}}$ and $\lL_{\text{P}}$ are Laplacian matrices of strongly connected and weight-balanced digraphs. Then, invoking~\cite[Lemma 9]{RAF-PY-KML:06} we can show that matrix 
$\vect{A}$ in~\eqref{eq::alg_RAF-PY-KML:06-collectw2-e} is Hurwitz. Therefore, using the ISS bound on the trajectories of LTI systems (see ``Input-to-State Stability of LTI Systems"), the tracking error of each agent $i\in\until{N}$ while implementing~\eqref{eq::alg_RAF-PY-KML:06} over a strongly connected and weight-balanced digraph is
\begin{align}\label{eq::errorbound-eq-alg_RAF-PY-KML:06}
|e^i(t)|
\leq &\,\kappa\, \text{e}^{-\underline{\lambda}\,(t-t_0)}\left\|\begin{bmatrix}\vect{w}_{2:N}(t_0)\\\Bvect{e}(t_0)\end{bmatrix}\!\right\| \!+\!
\frac{\kappa\|\vect{B}\|}{\underline{{\lambda}}\,}\sup_{t_0\leq \tau\leq t}\|(\vect{I}_N-\frac{1}{N}\vect{1}_N\vect{1}_N^\top)\dvectsf{u}(\tau)\|,
\end{align}
where ($\kappa,\underline{{\lambda}}$) are given by~\eqref{eq::exp_bound_LTI} for matrix $\vect{A}$ of~\eqref{eq::alg_RAF-PY-KML:06-collectw2-e} and can be computed from~\eqref{eq::LTI_exp_k_lam_tight}. We can show that both $\kappa$ and $\underline{{\lambda}}$ depend on the smallest non-zero eigenvalues of $\Sym{\lL_{\text{I}}}$ and $\Sym{\lL}_{\text{P}}$ as well as~$\alpha$. Therefore, the tracking performance of the algorithm~\eqref{eq::alg_RAF-PY-KML:06} depends on both the magnitude of the derivative of reference signals and also the connectivity of the communication graph.  From this error bound, we observe that for bounded dynamic signals with bounded rate the algorithm~\eqref{eq::alg_RAF-PY-KML:06} is guaranteed to track the dynamic average with an ultimately bounded error. Moreover, we can see that this algorithm does not need any special initialization.
The robustness to initialization can be observed on the block diagram representation of algorithm~\eqref{eq::alg_RAF-PY-KML:06}, shown in 
Figure~\ref{Fig:BlockDiagramDynamic_CT_RAF-PY-KML:06}, as well. For reference, we summarize next the convergence guarantees of algorithm~\eqref{eq::alg_RAF-PY-KML:06}.

\begin{theorem}[Convergence of~\eqref{eq::alg_RAF-PY-KML:06}]\label{thm::alg_RAF-PY-KML:06-thm} Let $\lL_{\textup{P}}$ and $\lL_{\textup{I}}$ be Laplacian matrices corresponding to strongly connected and weight-balanced digraphs. Let $\sup_{\tau\in[t,\infty)}\|(\vect{I}_N-\frac{1}{N}\vect{1}_N\vect{1}_N^\top)\dvectsf{u}(\tau)\| \!=\! \gamma(t)\!<\!\infty$.  Then, starting from any $x^i(t_0),q(t_0)\in\real$, for
  any $\alpha\in\realpositive$ the trajectories of
  algorithm~\eqref{eq::alg-SSK-JC-SM:15-ijrnc-ct} satisfy
  \vspace{-0.1in}
  \begin{equation*}
    \lim_{t\to\infty} \Big| x^i(t)-\mathsf{u}^\text{avg}(t) \Big| \leq
    \frac{\kappa\|\vect{B}\|\gamma(\infty)}{ 
      \underline{\lambda}},\qquad  i\in\until{N},
  \end{equation*}
  where $\kappa,\underline{\lambda}\in\real_{>0}$ satisfy $\|\textup{e}^{\vect{A}t}\|\leq\kappa\,\textup{e}^{-\underline{\lambda} t}$ ($\vect{A}$ and $\vect{B}$ are given in~\eqref{eq::alg_RAF-PY-KML:06-collectw2-e}). Moreover, we have $\sum_{i=1}^N x^i(t)=\sum_{i=1}^N\mathsf{u}^i(t)+\textup{e}^{-\alpha (t-t_0)}(\sum_{i=1}^N x^i(t_0)-\sum_{i=1}^N\mathsf{u}^i(t_0))$ for $t\in[t_0,\infty)$.
\end{theorem}

\begin{figure}[ht]
\centering
  \includegraphics[height=1.8in]{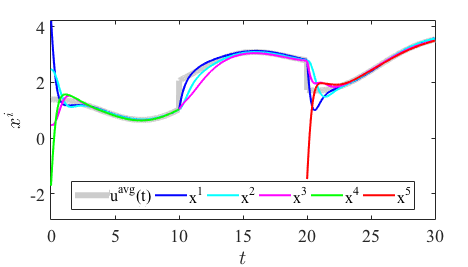}
\caption{Performance of dynamic average consensus algorithm~\eqref{eq::alg_RAF-PY-KML:06} in the distributed formation control scenario of Figure~\ref{fig::motiv_examp}. A group of $4$ mobile agents acquire reference inputs~\eqref{eq::ref_example} corresponding to the time-varying position of a set of moving targets. This algorithm convergence properties are not affected by initialization errors, as stated in Theorem~\ref{thm::alg_RAF-PY-KML:06-thm}. This property also makes it robust to agent arrivals and departures. In fact, in this simulation, agent $4$ leaves the network at time $t=10$ s and a new agent $5$ joins the network at $t=20$ s. Unlike what we observed for  algorithms~\eqref{eq::alg_DPS-ROS-RMM:05b_no_udot} and~\eqref{eq::alg_RAF-PY-KML:06} in Figure~\ref{fig::motiv_examp_resp}, here the execution recovers its tracking performance after a transient.}
\label{fig::motiv_examp_resp_free}
\end{figure}

Figure~\ref{fig::motiv_examp_resp_free} shows the performance of algorithm~\eqref{eq::alg_RAF-PY-KML:06} in the distributed formation control scenario represented in Figure~\ref{fig::motiv_examp}. This plot illustrates how the property of robustness to initialization error of algorithm~\eqref{eq::alg_RAF-PY-KML:06} allows it to accommodate the addition and deletion of agents with satisfactory tracking performance. 

Even though the convergence guarantees of algorithm~\eqref{eq::alg_RAF-PY-KML:06} are valid for strongly connected and weight-balanced digraphs, from an implementation perspective, the use of this strategy over directed graphs may not be feasible. In fact, the presence of the transposed integral Laplacian, $\lL_{\text{I}}^\top$, in~\eqref{eq::alg_RAF-PY-KML:06-collect-b} requires each agent $i\in\until{N}$ to know not only the entries in row $i$ but also the column $i$ of $\lL_{\text{I}}$, and receive information from the corresponding agents.  However, for undirected graph topologies this requirement is satisfied trivially as $\lL_{\text{I}}^\top=\lL_{\text{I}}$.

\subsection{Controlling the rate of convergence}
A common feature of the dynamic average consensus algorithms presented in the "A first design for dynamic average consensus" and "Robustness to initialization and permanent agent dropout" sections
is that the rate of convergence is the same for all agents and is dictated by network topology, as well as some algorithm parameters  (see~\eqref{eq::alg_DPS-ROS-RMM:05b-tracking-error} and~\eqref{eq::errorbound-eq-alg_RAF-PY-KML:06}). However, in some applications, the task is not just to obtain the average of the dynamic inputs but rather to physically track this value, possibly with limited control authority. 
To allow the network to pre-specify its desired worst rate of convergence, $\beta$,~\cite{SSK-JC-SM:13-ecc} proposes dynamic average consensus algorithms whose design incorporates two time scales. The~\algorithmone (\algorithmoneAbb) algorithm is described as follows
\begin{subequations}\label{eq::SSK-JC-SM:13-ecc-A1}
    \begin{align}
      &
      \begin{cases}\epsilon\,\dot{q}^i=-\sum_{j=1}^N{b}_{ij}(z^i-z^j),\\
        \epsilon\,\dot{z}^i=-(z^i+\beta\,
        \mathsf{u}^i+\dot{\mathsf{u}}^i)-\sum_{j=1}^N{a}_{ij}(z^i-z^j)+\sum_{j=1}^N{b}_{ji}(q^i-q^j),
      \end{cases}\label{eq::SSK-JC-SM:13-ecc-A1-a}
      \\
      &\quad\dot{x}^i=-\beta\, x^i-
      z^i,\quad i\in\until{N}\label{eq::SSK-JC-SM:13-ecc-A1-b}
    \end{align}
  \end{subequations}
The fast dynamics here is~\eqref{eq::SSK-JC-SM:13-ecc-A1-a}, and employs a small value for $\eps\in\realpositive$.  The fast dynamics, which builds on the PI algorithm~\eqref{eq::alg_RAF-PY-KML:06}, is intended to generate the average of the sum of the dynamic
input and its first derivative. The slow dynamics~\eqref{eq::SSK-JC-SM:13-ecc-A1-b} then uses the signal generated by the fast dynamics to track the average of the reference signal across the network at a pre-specified smaller rate $\beta\in\realpositive$. The novelty here is that these slow and fast dynamics are running simultaneously and, thus, there is no need to wait for convergence of the fast dynamics and then take slow steps towards the input average.

Similar to the dynamic average consensus algorithm~\eqref{eq::alg_RAF-PY-KML:06},~\eqref{eq::SSK-JC-SM:13-ecc-A1}  does not require any specific initialization. 
The technical approach used in~\cite{SSK-JC-SM:13-ecc} to study the convergence of~\eqref{eq::SSK-JC-SM:13-ecc-A1} is based on singular perturbation theory~\cite[Chapter 11]{HKK:02}, which results in a guaranteed convergence to an $\eps$-neighborhood of $\mathsf{u}^\text{avg}(t)$ for small values of $\eps\in\real_{>0}$.

Using time-domain analysis, we can make more precise the information about the ultimate tracking behavior of~ algorithm\eqref{eq::alg_RAF-PY-KML:06}.
For convenience, we apply the change of variables~\eqref{eq::error_trans},~\eqref{eq::v-w} along
with~$\vect{y}=\vect{w}_{2:N}- ({\vectsf{R}}^\top\lL_{\text{I}}^\top\vect{\mathsf{R}})^{-1}\vectsf{R}^\top(\beta\vectsf{u}+\dvectsf{u}),$ and $\vect{e}_z=\vect{T}^\top(\vect{z}+\beta\avrg{\mathsf{u}}^j\vect{1}_N+\avrg{\dot{\mathsf{u}}}^j\vect{1}_N)$ to write the~\algorithmoneAbb algorithm as follows
\begin{align*}
\dot{w}_1&=0,\\
\begin{bmatrix}
\dvect{y}\\
\dvect{e}_z
\end{bmatrix}&=\!\eps^{-1}\!\underbrace{\begin{bmatrix}\vect{0}&\begin{bmatrix}\vect{0}&-\vect{\mathsf{R}}^\top\lL_{\text{I}}\vect{\mathsf{R}}\end{bmatrix}\\
\begin{bmatrix}
0\\
\vect{\mathsf{R}}^\top\lL_{\text{I}}^\top\vect{\mathsf{R}}\end{bmatrix}&\begin{bmatrix}-1&0\\
\vect{0}&-\vect{I}-\vect{\mathsf{R}}^\top\lL_{\text{p}}\vect{\mathsf{R}}\end{bmatrix}
\end{bmatrix}}_{\bar{\vect{A}}}\!\begin{bmatrix}
\vect{y}\\
\vect{e}_z
\end{bmatrix}+\underbrace{\begin{bmatrix}- ({\vectsf{R}}^\top\lL_{\text{I}}^\top\vect{\mathsf{R}})^{-1}\begin{bmatrix}\vect{0}_{(N-1)\times 1}&\vect{I}_{N-1}\end{bmatrix}\\
\begin{bmatrix}\begin{bmatrix}1&\vect{0}_{1\times N-1}\end{bmatrix}\\\vect{0}_{(N-1)\times N}\end{bmatrix} \end{bmatrix}}_{\bar{\vect{B}}}\vect{f}(t),
\label{eq::FOI_SSK-CJ-SM13-anal-fast-dy}\\
\dvect{e}&=-\beta\,\vect{e}-\vect{e}_z,
\end{align*}
where 
$\vectsf{f}(t)=\vect{T}^\top(\beta\,\dvect{\mathsf{u}}+ \ddvect{\mathsf{u}})$. 
Using the ISS bound on the trajectories of LTI systems, see ``Input-to-State Stability of LTI Systems", the tracking error of each agent $i\in\until{N}$ while implementing~\algorithmoneAbb algorithm with an  $\eps\in\real_{>0}$ is as follows, $i\in\until{N}$,
\begin{align*}
|e^i(t)|&\leq \,\text{e}^{-\beta\,(t-t_0)}|e^i(t_0)|+
\frac{\kappa}{\beta}\,\underset{t_0\leq\tau\leq t}{\sup}\big(\text{e}^{-\eps^{-1}\underline{\lambda}\,(t-t_0) }\Big\|\begin{bmatrix}
\vect{y}(t_0)\\
\vect{e}_z(t_0)
\end{bmatrix}\Big\|+
\frac{\eps\|\bar{\vect{B}}\|}{\underline{\,{\lambda}}\,}\sup_{t_0\leq \tau\leq  t}\|\beta\,\dvect{\mathsf{u}}(\tau)+ \ddvect{\mathsf{u}}(\tau)\|\big),
\end{align*}
where $\|\text{e}^{\bar{\vect{A}}\,t}\|\leq\kappa\text{e}^{-\underline{\lambda}\,t}$. From this error bound, we observe that for dynamic signals with bounded first and second derivatives~\algorithmoneAbb algorithm is guaranteed to track the dynamic average with an ultimately bounded error. This tracking error can be made small using a small $\eps\in\real_{>0}$. Use of small $\eps\in\real_{>0}$ also results in
dynamics~\eqref{eq::FOI_SSK-CJ-SM13-anal-fast-dy} to have a higher decay rate. Therefore, the dominant rate of convergence of \algorithmoneAbb algorithm is determined by $\beta$, which can be pre-specified regardless of the interaction topology. Moreover, $\beta$ can be used to regulate the control effort of the integrator dynamics $\dot{x}^i=c^i(t)$, $i\in\until{N}$ while maintaining a good tracking error via the use of small $\eps\in\real_{>0}$.

\subsection{An alternative algorithm for directed graphs}
As observed, the algorithm~\eqref{eq::alg_RAF-PY-KML:06} is not implementable over directed graphs, since it requires information exchange with both in- and out-neighbors, and these sets are generally different. In~\cite{SSK-JC-SM:15-ijrnc} authors proposed a modified proportional and integral agreement feedback dynamic average consensus algorithm whose implementation does not require the agents to know their respective columns of the Laplacian. This algorithm is 
\begin{subequations}\label{eq::alg-SSK-JC-SM:15-ijrnc-ct}
  \begin{align}
    \dot{q}^i & = \alpha\beta\sum\nolimits_{j=1}^N \mathsf{a}_{ij}(x^i-x^j), 
    \\
    \dot{x}^i & =-\alpha(x^i-\mathsf{u}^i)\!-\!\beta\sum\nolimits_{j=1}^N
    \mathsf{a}_{ij}(x^i-x^j)\!-\!q^i+\dot{\mathsf{u}}^i,\\
     &x^i(t_0),q^i(t_0)\in\real~~s.t.~ \sum\nolimits_{j=1}^Nq^j(t_0)=0,\label{eq::alg-SSK-JC-SM:15-ijrnc-ct-init}
  \end{align}
\end{subequations}
 $i\in\until{N}$, where $\alpha,\beta\in\real_{>0}$. Algorithm~\eqref{eq::alg-SSK-JC-SM:15-ijrnc-ct} in compact form can be equivalently written as
 \begin{align*}
\dvect{x}=-\alpha\,(\vect{x}-\vectsf{u})-\beta\,\lL\,\vect{x}-\alpha\,\beta\,\int_{t_0}^t\!\lL\,\vect{x}(\tau)\,\text{d}\tau-\vect{q}(t_0)+\dvect{\mathsf{u}},
\end{align*}
which demonstrates the proportional and integral agreement feedback structure of this algorithm. As we did for algorithm~\eqref{eq::alg_DPS-ROS-RMM:05b}, we can use a change of variables $p^i=\mathsf{u}^i-x^i$, to write this algorithm in a form whose implementation does not require the knowledge of the derivative of the reference signals. 

We should point out an interesting connection between algorithms~\eqref{eq::alg-SSK-JC-SM:15-ijrnc-ct} and~\eqref{eq::alg_DPS-ROS-RMM:05b_no_udot}. If we write the transfer function from the reference input to the tracking error state~\eqref{eq::alg-SSK-JC-SM:15-ijrnc-ct}, there is a pole-zero cancellation which reduces the algorithm~\eqref{eq::alg-SSK-JC-SM:15-ijrnc-ct} to~\eqref{eq::alg_DPS-ROS-RMM:05b} and \eqref{eq::alg_DPS-ROS-RMM:05b_no_udot}. Despite this close relationship, there are some subtle differences. For example, unlike~\eqref{eq::alg_DPS-ROS-RMM:05b},  algorithm~\eqref{eq::alg-SSK-JC-SM:15-ijrnc-ct} enjoys robustness to reference signal measurement perturbations and naturally preserves the privacy
of the input of each agent against adversaries: specifically~\cite{SSK-JC-SM:15-ijrnc}, an adversary with access to the time history of all network communication messages cannot uniquely reconstruct the reference signal of any agent,  which is not the case for algorithm~\eqref{eq::alg_DPS-ROS-RMM:05b_no_udot}. 

Figure~\ref{Fig:BlockDiagramDynamic_CT_SSK-JC-SM:15} shows the block diagram representation of this algorithm. The next result states the convergence properties of~\eqref{eq::alg-SSK-JC-SM:15-ijrnc-ct}. We refer the reader to~\cite{SSK-JC-SM:15-ijrnc} for the proof of this statement which is established using the time domain analysis we implemented to analyze the algorithms we have reviewed so far.

\begin{theorem}[Convergence of~\eqref{eq::alg-SSK-JC-SM:15-ijrnc-ct} over strongly connected and weight-balanced digraphs for dynamic input signals~\cite{SSK-JC-SM:15-ijrnc}]\label{thm::SSK-JC-SM:15-ijrnc-thm} Let $\GG$ be a strongly connected and weight-balanced digraph. Let $\sup_{\tau\in[t,\infty)}\|(\vect{I}_N-\frac{1}{N}\vect{1}_N\vect{1}_N^\top)\dvectsf{u}(\tau)\| \!=\! \gamma(t)\!<\!\infty$.  Then, for
  any $\alpha,\beta\in\realpositive$, the trajectories of
  algorithm~\eqref{eq::alg-SSK-JC-SM:15-ijrnc-ct} satisfy
  \vspace{-0.1in}
  \begin{equation}
  \label{eq::Alg-SSK-JC-SM_ultimate_bound} 
    \lim_{t\to\infty} \Big| x^i(t)-\mathsf{u}^\text{avg}(t) \Big| \leq
    \frac{\gamma(\infty)}{ 
      \beta\Hlambda_2},\qquad  i\in\until{N}.
  \end{equation}
provided $\sum\nolimits_{j=1}^Nq^j(t_0)=0$.  The convergence rate to the error bound is~$\min\{\alpha,\beta\re{\lambda_2}\}.$
\end{theorem}

The inverse relation between $\beta$ and the tracking error in~\eqref{eq::Alg-SSK-JC-SM_ultimate_bound} indicates that we can use the parameter $\beta$ to control the tracking error size, while $\alpha$ can be used to control the rate of convergence.


\section{Discrete-Time Dynamic Average Consensus Algorithms}\label{sec::discrete-time}

While the continuous-time dynamic average consensus algorithms described in the previous section are amenable to elegant and relatively simple analysis, implementing these algorithms on practical cyber-physical systems requires continuous communication between agents. This requirement is not feasible in practice due to constraints on the communication bandwidth. To address this issue, we study here discrete-time dynamic average consensus algorithms where the communication among agents occurs only at discrete time steps.

The main difference between continuous-time and discrete-time dynamic average consensus algorithms is the rate at which their estimates converge to the average of the reference signals. In continuous time, the parameters may be scaled to achieve any desired convergence rate, while in discrete time the parameters must be carefully chosen to ensure convergence. The problem of optimizing the convergence rate has received significant attention in the literature~\cite{BV-RAF-KML:15c,BV-RAF-KML:15,BV-RAF-KML:15d,BNO-MJC-MGR:10,TE-DZ-EDA-LV:11,EK-PF:09,YY-JL-RMM-JG:12,EM-JIM-CS:13,MLE-RAF-KML:13,MLE-RAF-KML:14}. Here, we give a simple method using root locus techniques for choosing the parameters in order to optimize the convergence rate. We also show how to further accelerate the convergence by introducing extra dynamics into the dynamic average consensus algorithm.

We analyze the convergence rate of four discrete-time dynamic average consensus algorithms in this section, beginning with the discretized version of the continuous-time algorithm~\eqref{eq::alg_DPS-ROS-RMM:05b_no_udot}. We then show how to use extra dynamics to accelerate the convergence rate and/or obtain robustness to initial conditions. Table~\ref{table::DT-driving_command} summarizes the arguments of the driving command of these algorithms in~\eqref{eq::AgentSingInt-DT} and their special initialization requirements.
\begin{table}[htb]
 \footnotesize
    \caption{{\small Arguments of the driving command in~\eqref{eq::AgentSingInt-DT} for the reviewed discrete-time dynamic average consensus algorithms together with their initialization requirements.}
    }\label{table::DT-driving_command}\vspace{-0.06in}
  \centering
  \setlength\tabcolsep{2.5pt}
  \begin{tabular}{|c||c|c|c|c|}
    \hline
    Algorithm & \eqref{Eq:P_DT} & \eqref{Eq:P_DT_accelerated} & \eqref{Eq:PI_DT} & \eqref{Eq:PI_DT_accelerated} \\ \hline
    $J^i(t)$ &
    $\{\mathsf{u}^i_k,p^i_k\}$ &
    $\{\mathsf{u}^i_k,p^i_k,p^i_{k-1}\}$ &
    $\{\mathsf{u}^i_k,p^i_k,q^i_k\}$ &
    $\{\mathsf{u}^i_k,p^i_k,p^i_{k-1},q^i_k,q^i_{k-1}\}$ \\[1mm] \hline
    ${\{I^j(t)\}}_{j\in\NN_{\text{out}}^i}$ &
    ${\{x^j_k\}}_{j\in\NN_{\text{out}}^i}$ &
    ${\{x^j_k\}}_{j\in\NN_{\text{out}}^i,}$ &
    ${\{x^j_k,p^j_k\}}_{j\in\NN_{\text{out}}^i}$ &
    ${\{x^j_k,p^j_k\}}_{j\in\NN_{\text{out}}^i}$ \\[1mm]
    \hline
    Initialization Requirement &
    $\sum_{j=1}^N p^j_0 = 0$ & $\sum_{j=1}^N p^j_0 = 0$ & none & none \\ \hline
    \end{tabular}
\end{table}

For simplicity of exposition, we assume that the communication graph is constant, connected, and undirected. The Laplacian matrix is then symmetric and therefore has real eigenvalues. Since the graph is connected, the smallest eigenvalue is $\lambda_1=0$ and all the other eigenvalues are strictly positive, in other words, $\lambda_2>0$. Furthermore, we assume that the smallest and largest nonzero eigenvalues are known (if the exact eigenvalues are unknown, it also suffices to have lower and upper bounds, respectively, on $\lambda_2$ and $\lambda_N$).

\subsection{Non-robust dynamic average consensus algorithms}

\begin{figure}[htb]
\renewcommand{\x}{3mm}
\renewcommand{\y}{3mm}
\begin{subfigure}{0.4\textwidth}
\centering
\begin{tikzpicture}
 \node (sum1) [sum] {};
 \node (sum2) [sum,below right=10mm and 2mm of sum1] {};
 \node (f)    [block,right=5mm of sum2] {$\small\dfrac{k_I}{z-1} \vect{I}$};
 \node (L)    [block,right=5mm of f]  {$\small\lL$};
 \node (x)    at (sum1 -| L.east) [xshift=5mm,inner xsep=0cm,inner ysep=0cm] {};
 \node (x0)   [below left=5mm and 2mm of sum2] {$\small\vect{p}_0$};
 \draw [link] ($(sum1)-(8mm,0)$) -- node[pos=0.25,yshift=2.5mm] {$\small\vectsf{u}_k$} (sum1);
 \draw [link] (sum1) -- node[pos=1,xshift=-3mm,yshift=2.5mm] {$\small\vect{x}_k$} ($(x)+(5mm,0)$);
 \draw [link] (x) |- (L);
 \draw [link] (L) -- (f);
 \draw [link] (f) -- (sum2);
 \draw [link] (sum2) -| node[pos=0.9,xshift=-2.5mm] {\footnotesize{$-$}} (sum1);
 \draw [link] (x0) -| (sum2);
\end{tikzpicture}
\caption{\footnotesize Non-robust non-accelerated dynamic average consensus algorithm~\eqref{Eq:P_DT}.}
\label{Fig:BlockDiagramP}
\end{subfigure}\hfill%
\begin{subfigure}{0.56\textwidth}
\centering
\begin{tikzpicture}
  \node (sum1)  [sum] {};
  \node (h1)    [block,below right=\y and 2mm of sum1] {$\small\dfrac{k_p}{z-\rho} \vect{I}$};
  \node (L1)    [block,right=5mm of h1] {$\small\lL$};
  \node (sum2)  [sum,right=5mm of L1] {};
  \node (h2)    [block,below right=\y and 2mm of sum2] {$\small\dfrac{k_I}{z-1} \vect{I}$};
  \node (L2)    [block,right=5mm of h2] {$\small\lL$};
  \coordinate (x) at ($(u -| L2.east)+(5mm,0)$) {};
  \draw [link] ($(sum1)-(8mm,0)$) -- node[pos=0.25,yshift=2.5mm] {$\small\vectsf{u}_k$} (sum1);
  \draw [link] (h1) -| node[pos=0.9,xshift=-2.5mm] {\footnotesize{$-$}} (sum1);
  \draw [link] (L1) -- (h1);
  \draw [link] (sum2) -- (L1);
  \draw [link] (h2) -| (sum2);
  \draw [link] (L2) -- (h2);
  \draw [link] (sum1) -- node[pos=1,xshift=-3mm,yshift=2.5mm] {$\small\vect{x}_k$} ($(x)+(5mm,0)$);
  \draw [link] (x) |- (L2);
  \draw [link] (sum2 -| x) -- (sum2);
\end{tikzpicture}
\caption{\footnotesize Robust non-accelerated proportional-integral dynamic average consensus algorithm~\eqref{Eq:PI_DT}.}
\label{Fig:BlockDiagramPI}
\end{subfigure}\\
\begin{subfigure}{0.4\textwidth}
\centering
\begin{tikzpicture}
 \node (sum1) [sum] {};
 \node (sum2) [sum,below right=10mm and 1mm of sum1] {};
 \node (f)    [block,right=\x of sum2] {$\small\dfrac{k_I\,z}{(z-\rho^2)(z-1)} \vect{I}$};
 \node (L)    [block,right=\x of f]  {$\small\lL$};
 \node (x)    at (sum1 -| L.east) [xshift=3mm,inner xsep=0cm,inner ysep=0cm] {};
 \node (x0)   [below left=5mm and 1mm of sum2] {$\small\vect{p}_0$};
 \draw [link] ($(sum1)-(8mm,0)$) -- node[pos=0.25,yshift=2.5mm] {$\small\vectsf{u}_k$} (sum1);
 \draw [link] (sum1) -- node[pos=1,xshift=-3mm,yshift=2.5mm] {$\small\vect{x}_k$} ($(x)+(5mm,0)$);
 \draw [link] (x) |- (L);
 \draw [link] (L) -- (f);
 \draw [link] (f) -- (sum2);
 \draw [link] (sum2) -| node[pos=0.9,xshift=-2.5mm] {\footnotesize{$-$}} (sum1);
 \draw [link] (x0) -| (sum2);
\end{tikzpicture}
\caption{\footnotesize Non-robust accelerated dynamic average consensus algorithm~\eqref{Eq:P_DT_accelerated}.}
\label{Fig:BlockDiagramAccelerated_P}
\end{subfigure}\hfill%
\begin{subfigure}{0.56\textwidth}
\centering
\begin{tikzpicture}
  \node (sum1)  [sum] {};
  \node (h1)    [block,below right=\y and 1mm of sum1] {$\small\dfrac{k_p\,z}{(z-\rho)^2} \vect{I}$};
  \node (L1)    [block,right=\x of h1] {$\small\lL$};
  \node (sum2)  [sum,right=\x of L1] {};
  \node (h2)    [block,below right=\y and 1mm of sum2] {$\small\dfrac{k_I\,z}{(z-\rho^2)(z-1)} \vect{I}$};
  \node (L2)    [block,right=\x of h2] {$\small\lL$};
  \coordinate (x) at ($(u -| L2.east)+(\x,0)$) {};
  \draw [link] ($(sum1)-(8mm,0)$) -- node[pos=0.25,yshift=2.5mm] {$\small\vectsf{u}_k$} (sum1);
  \draw [link] (h1) -| node[pos=0.9,xshift=-2.5mm] {\footnotesize{$-$}} (sum1);
  \draw [link] (L1) -- (h1);
  \draw [link] (sum2) -- (L1);
  \draw [link] (h2) -| (sum2);
  \draw [link] (L2) -- (h2);
  \draw [link] (sum1) -- node[pos=1,xshift=-3mm,yshift=2.5mm] {$\small\vect{x}_k$} ($(x)+(5mm,0)$);
  \draw [link] (x) |- (L2);
  \draw [link] (sum2 -| x) -- (sum2);
\end{tikzpicture}
\caption{\footnotesize Robust accelerated proportional-integral dynamic average consensus algorithm~\eqref{Eq:PI_DT_accelerated}.}
\label{Fig:BlockDiagramAccelerated_PI}
\end{subfigure}
\caption{Block diagram of discrete-time dynamic average consensus algorithms. The right two algorithms use proportional-integral dynamics to obtain robustness to initial conditions, while the bottom two algorithms use extra dynamics to accelerate the convergence rate. When the graph is connected and balanced, and upper and lower bounds on the nonzero eigenvalues of the graph Laplacian are known, closed-form solutions for the parameters which optimize the convergence rate are known; see Theorem~\ref{Thm:DT_rates}.}
\label{Fig:BlockDiagramDT}
\end{figure}
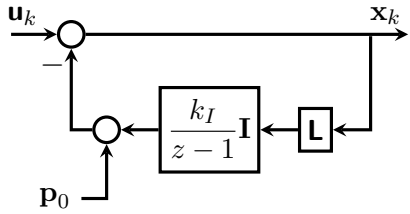
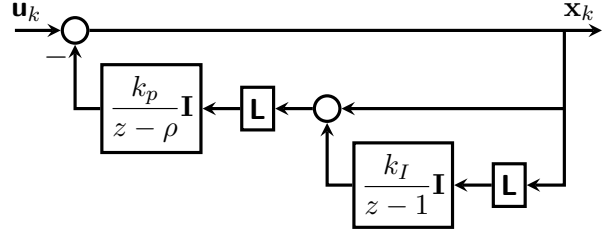
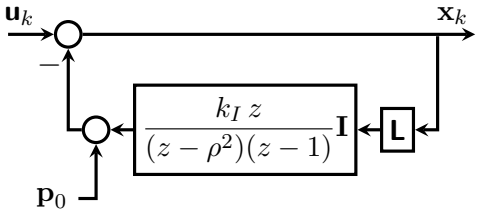
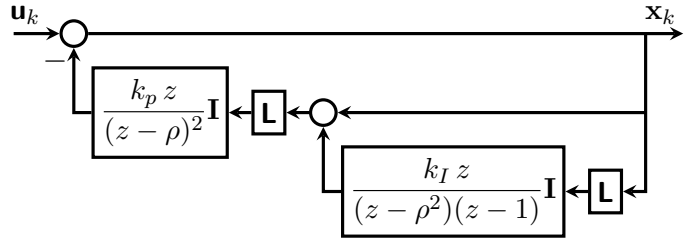

We first consider the discretized version of the continuous-time dynamic average consensus algorithm in~\eqref{eq::alg_DPS-ROS-RMM:05b_no_udot} (``Euler Discretizations of Continuous-Time Dynamic Average Consensus Algorithms" elaborates on the method for discretization and the associated range of admissible stepsizes). This algorithm has the iterations
\begin{subequations}\label{Eq:P_DT}
\begin{align}
{p}^i_{k+1} &= p_k^i+k_I\sum_{j=1}^N{a}_{ij}(x^i_k-x^j_k),\quad p^i_0\in\reals,\quad i\in\until{N},\\
x_k^i&=\mathsf{u}^i_k-p^i_k,
\end{align}
\end{subequations}
where $k_I\in\real$ is the stepsize. We provide the block diagram in Figure~\ref{Fig:BlockDiagramP}.

\begin{figure}[htb]
\begin{subfigure}{0.45\textwidth}\centering
\begin{tikzpicture}[align=center,pole/.style={cross out,draw=black,minimum size=2.5mm,very thick,inner sep=0pt,outer sep=0pt}]
  \begin{axis}[
    xmin=-1.5,xmax=2,
    ymin=-0.8,ymax=0.9,
    minor tick num=0,
    xlabel=\empty,
    ylabel=\empty,
    xtick={-1,1},
    xticklabels={$-1$,$1$},
    ytick=\empty,
    axis equal,
    axis lines=center]
    \node [pole,red] at (axis cs:1,0) {};
    \draw [black,dashed,radius=0.7] (axis cs:0,0) circle;
    \node at (axis cs: 0.63,0.63) {$\rho\mathbb{T}$};
    \draw [draw=blue,very thick] (axis cs: 1,0) -- (axis cs: -1.5,0);
    \draw [draw=blue,->,very thick,decoration={markings,mark=at position 1 with {\arrow[blue]{>}}},postaction={decorate}] (axis cs: 1,0) -- (axis cs: -1.3,0);
    \draw [draw=blue,->,very thick,decoration={markings,mark=at position 1 with {\arrow[blue]{>}}},postaction={decorate}] (axis cs: 1,0) -- (axis cs: -0.4667,0);
    \draw [draw=blue,->,very thick,decoration={markings,mark=at position 1 with {\arrow[blue]{>}}},postaction={decorate}] (axis cs: 1,0) -- (axis cs:  0.2667,0);
    \node [anchor=north west] at (axis cs:1.3,-0.01) {$\text{Re}(z)$};
    \node [anchor=south west] at (axis cs:0.02,0.9) {$\text{Im}(z)$};
  \end{axis}
\end{tikzpicture}
\caption{\footnotesize Non-accelerated dynamic average consensus algorithm in Figure~\ref{Fig:BlockDiagramP}.}
\label{Fig:RootLocus_standard}
\end{subfigure}\hfill%
\begin{subfigure}{0.45\textwidth}\centering
\begin{tikzpicture}[align=center,pole/.style={cross out,draw=black,minimum size=2.5mm,very thick,inner sep=0pt,outer sep=0pt}]
  \begin{axis}[
    xmin=-1.5,xmax=2,
    ymin=-0.8,ymax=0.9,
    minor tick num=0,
    xlabel=\empty,
    ylabel=\empty,
    xtick={-1,0.49,1},
    xticklabels={$-1$,$\rho^2$,$1$},
    ytick=\empty,
    axis equal,
    axis lines=center]
    \node [pole,red] at (axis cs:1,0) {};
    \node [pole,red] at (axis cs:0.49,0) {};
    \draw [very thick,radius=1.5mm,red] (axis cs:0,0) circle;
    \draw [black,dashed,radius=0.7] (axis cs:0,0) circle;
    \draw [draw=blue,very thick,decoration={markings,mark=at position 0.75 with {\arrow[blue]{>}}},postaction={decorate}] (axis cs: 1,0) -- (axis cs: 0.7,0);
    \draw [draw=blue,very thick,decoration={markings,mark=at position 0.75 with {\arrow[blue]{>}}},postaction={decorate}] (axis cs: 0.49,0) -- (axis cs: 0.7,0);
    \draw [draw=blue,very thick,decoration={markings,mark=at position 0.5 with {\arrow[blue]{>}}},postaction={decorate}] (axis cs: -0.7,0) -- (axis cs: 0,0);
    \draw [draw=blue,very thick,decoration={markings,mark=at position 0.5 with {\arrow[blue]{>}}},postaction={decorate}] (axis cs: -0.7,0) -- (axis cs: -1.5,0);
    \draw [draw=blue,very thick,decoration={markings,mark=at position 0.125 with {\arrow[blue]{>}}},postaction={decorate},radius=0.7] (axis cs:0,0) circle;
    \draw [draw=blue,very thick,decoration={markings,mark=at position 0.375 with {\arrow[blue]{>}}},postaction={decorate},radius=0.7] (axis cs:0,0) circle;
    \draw [draw=blue,very thick,decoration={markings,mark=at position 0.625 with {\arrow[blue]{<}}},postaction={decorate},radius=0.7] (axis cs:0,0) circle;
    \draw [draw=blue,very thick,decoration={markings,mark=at position 0.875 with {\arrow[blue]{<}}},postaction={decorate},radius=0.7] (axis cs:0,0) circle;
    \node at (axis cs: 0.63,0.63) {$\rho\mathbb{T}$};
    \node [anchor=north west] at (axis cs:1.3,-0.01) {$\text{Re}(z)$};
    \node [anchor=south west] at (axis cs:0.02,1) {$\text{Im}(z)$};
  \end{axis}
\end{tikzpicture}
\caption{\footnotesize Accelerated dynamic average consensus algorithm in Figure~\ref{Fig:BlockDiagramAccelerated_P}.}
\label{Fig:RootLocus_accelerated}
\end{subfigure}
\caption{Root locus design of dynamic average consensus algorithms. The dynamic average consensus algorithm poles are the points on the root locus at gains $\lambda_i$ for $i\in\until{N}$ where $\lambda_i$ are the eigenvalues of the graph Laplacian. To optimize the convergence rate, the parameters are chosen to minimize $\rho$ such that all poles corresponding to eigenvalues $\lambda_i$ for $i\in\{2,\ldots,N\}$ are inside the circle centered at the origin of radius $\rho$. Then the dynamic average consensus algorithm converges linearly with rate $\rho$.}
\label{Fig:RootLocus}
\end{figure}
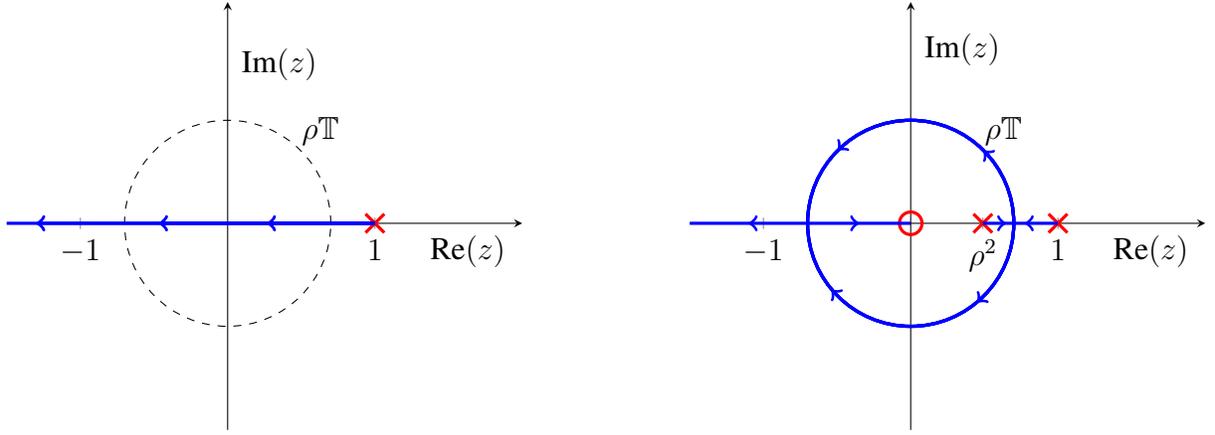

For discrete-time LTI systems, the convergence rate is given by the maximum magnitude of the system poles. The poles are the roots of the characteristic equation, which for the dynamic average consensus algorithm in Figure~\ref{Fig:BlockDiagramP} is
\begin{align*}
0 = z\vect{I} - (\vect{I} - k_I \lL).
\end{align*}
If the Laplacian matrix can be diagonalized, then the system can be separated according to the eigenvalues of $\lL$ and each subsystem analyzed separately. The characteristic equation corresponding to the eigenvalue $\lambda$ of $\lL$ is then
\begin{align}
0 = 1 + \lambda \frac{k_I}{z-1}. \label{Eq:RootLocus}
\end{align}
To observe how the pole moves as a function of the Laplacian eigenvalue, root locus techniques from LTI systems theory can be used. Figure~\ref{Fig:RootLocus_standard} shows the root locus of~\eqref{Eq:RootLocus} as a function of $\lambda$. The dynamic average consensus algorithm poles are then the points on the root locus at gains $\lambda_i$ for $i\in\until{N}$ where $\lambda_i$ are the eigenvalues of the graph Laplacian. To optimize the convergence rate, the system is designed to minimize $\rho$ where all poles corresponding to disagreement directions (that is, those orthogonal to the consensus direction $\mathbf{1}_N$) are inside the circle centered at the origin of radius $\rho$. Since the pole starts at $z=1$ and moves left as $\lambda$ increases, the convergence rate is optimized when there is a pole at $z=\rho$ when $\lambda=\lambda_2$ and at $z=-\rho$ when $\lambda=\lambda_N$, that is,
\begin{align*}
0 = 1 + \lambda_2\,\frac{k_I}{\rho-1}  \quad \text{and} \quad
0 = 1 + \lambda_N\,\frac{k_I}{-\rho-1}.
\end{align*}
Solving these conditions for $k_I$ and $\rho$ gives
\begin{align*}
k_I = \frac{2}{\lambda_2 + \lambda_N}  \quad  \text{and} \quad
\rho = \frac{\lambda_N-\lambda_2}{\lambda_N+\lambda_2}.
\end{align*}

While the previous choice of parameters optimizes the convergence rate, even faster convergence can be achieved by introducing extra dynamics into the dynamic average consensus algorithm. Consider the accelerated dynamic average consensus algorithm in Figure~\ref{Fig:BlockDiagramAccelerated_P}, given by
\begin{subequations}\label{Eq:P_DT_accelerated}
\begin{align}
p^i_{k+1} &= (1+\rho^2) p^i_k - \rho^2 p^i_{k-1} + k_I \sum_{j=1}^N a_{ij} (x^i_k-x^j_k),\qquad p^i_0\in\reals, \qquad i\in\until{N},\\
x^i_k &= u^i_k - p^i_k.
\end{align}
\end{subequations}
Instead of a simple integrator, the transfer function in the feedback loop now has two poles (one of which is still at $z=1$). To implement the dynamic average consensus algorithm, each agent must keep track of two internal state variables ($p_k^i$ and $p_{k-1}^i$). This small increase in memory, however, can result in a significant improvement in the rate of convergence, as discussed below.

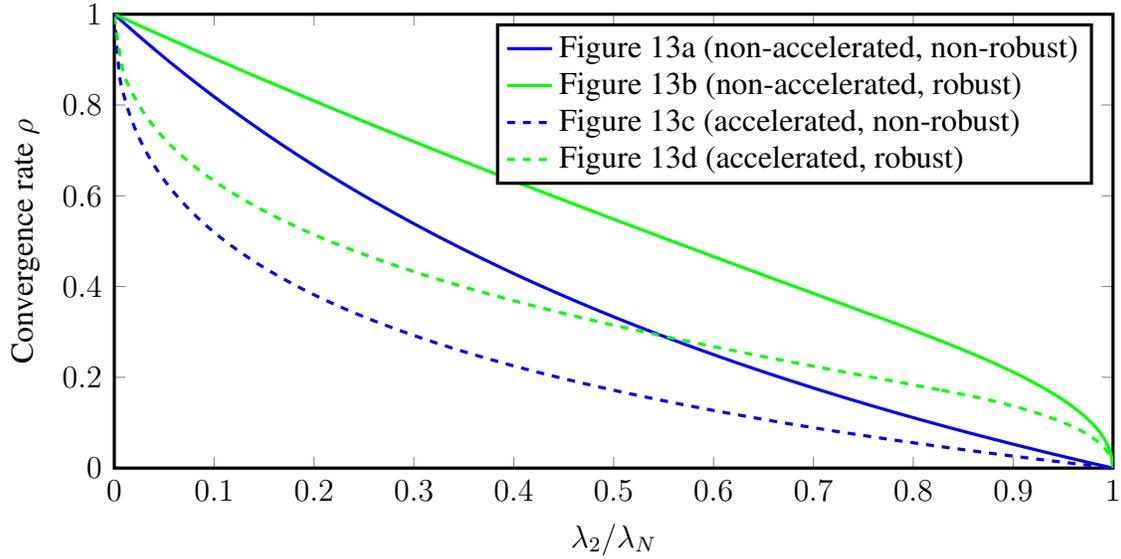
\begin{figure}[htb]
\centering
\begin{tikzpicture}
  \begin{axis}[
    width=0.9\textwidth,
    height=3in,
    xmin=0,xmax=1,
    ymin=0,ymax=1,
    ylabel={Convergence rate $\rho$},
    xlabel={$\lambda_2/\lambda_N$},
    legend cell align=left,
    smooth,very thick,no markers,solid]
    
    \addplot [blue,domain=0:1,samples=201] {(1-x)/(1+x)};  
    \addlegendentry{Figure~\ref{Fig:BlockDiagramP} (non-accelerated, non-robust)}
    
    \addplot [green,domain=0:3-sqrt(5),samples=101] {(8-8*x+x^2)/(8-x^2)};  
    \addlegendentry{Figure~\ref{Fig:BlockDiagramPI} (non-accelerated, robust)}
    
    \addplot [blue,dashed,domain=0:1,samples=201] {(1-sqrt(x))/(1+sqrt(x))};  
    \addlegendentry{Figure~\ref{Fig:BlockDiagramAccelerated_P} (accelerated, non-robust)}
    
    \addplot [green,dashed,domain=0:2*(sqrt(2)-1),samples=101]
      {((2-(2*sqrt(1-x)-x)) + 4*(1-sqrt(2-(2*sqrt(1-x)-x))))/(2+(2*sqrt(1-x)-x))};  
    \addlegendentry{Figure~\ref{Fig:BlockDiagramAccelerated_PI} (accelerated, robust)}
    
    \addplot [green, domain=3-sqrt(5):1,samples=101]
      {(-x*(1-x)+sqrt((1-x)*(4+5*x^2-x^3)))/(2*(1+x^2))};  
    
    \addplot [green,dashed,domain=2*(sqrt(2)-1):1,samples=101]
      {((1+(2*sqrt(1-x)-x)) + 2*(1-sqrt(2+(2*sqrt(1-x)-x))))/(1+(2*sqrt(1-x)-x))};  
  \end{axis}
\end{tikzpicture}
\caption{Convergence rate $\rho$ as a function of $\lambda_2/\lambda_N$ for the dynamic average consensus algorithms in Figure~\ref{Fig:BlockDiagramDT}. The accelerated dynamic average consensus algorithms (dashed lines) use extra dynamics to enhance the convergence rate as opposed to the non-accelerated algorithms (solid lines). Also, the robust algorithms (green) use the proportional-integral structure to obtain robustness to initial conditions as opposed to the non-robust algorithms (blue). The graph is assumed to be constant, connected, and undirected with Laplacian eigenvalues $\lambda_i$ for $i\in\until{N}$. Closed-form expressions for the rates and algorithm parameters are provided in Theorem~\ref{Thm:DT_rates}.}
\label{Fig:rho}
\end{figure}

Once again, root locus techniques can be used to design the parameters to optimize the convergence rate. Figure~\ref{Fig:RootLocus_accelerated} shows the root locus of the accelerated dynamic average consensus algorithm~\eqref{Eq:P_DT_accelerated}. By adding an open-loop pole at $z=\rho^2$ and zero at $z=0$, the root locus now goes around the $\rho$-circle. Similar to the previous case, the convergence rate is optimized when there is a repeated pole at $z=\rho$ when $\lambda=\lambda_2$ and a repeated pole at $z=-\rho$ when $\lambda=\lambda_N$. This gives the optimal parameter $k_I$ and convergence rate $\rho$ given by
\begin{align*}
k_I = \frac{4}{(\sqrt{\lambda_2} + \sqrt{\lambda_N})^2} \quad\text{and}\quad
\rho = \frac{\sqrt{\lambda_N}-\sqrt{\lambda_2}}{\sqrt{\lambda_N}+\sqrt{\lambda_2}}.
\end{align*}
The convergence rate of both the standard (Eq.~\eqref{Eq:P_DT}) and accelerated (Eq.~\eqref{Eq:P_DT_accelerated}) versions of the dynamic average consensus algorithm are plotted in Figure~\ref{Fig:rho} as a function of the ratio $\lambda_2/\lambda_N$.

\subsection{Robust dynamic average consensus algorithms}

While the previous dynamic average consensus algorithms are not robust to initial conditions, we can also use root locus techniques to optimize the convergence rate of dynamic average consensus algorithms that are robust to initial conditions. Consider the discrete-time version of the proportional-integral estimator from~\eqref{eq::alg_RAF-PY-KML:06} whose iterations are given by
\begin{subequations}\label{Eq:PI_DT}
\begin{align}
q^i_{k+1} &= \rho\,q^i_k + k_p \sum_{j=1}^N a_{ij} \bigl( (x^i_k-x^j_k) + (p^i_k-p^j_k) \bigr), \\
p^i_{k+1} &= p^i_k + k_I \sum_{j=1}^N a_{ij} (x^i_k-x^j_k), \\
x^i_k &= u^i_k - q^i_k, \qquad
 p^i_0,q^i_0\in\reals, \qquad i\in\until{N}
\end{align}
\end{subequations}
with parameters $\rho,k_p,k_I\in\reals$. The block diagram of this algorithm is shown in Figure~\ref{Fig:BlockDiagramPI}.

Since the dynamic average consensus algorithms \eqref{Eq:P_DT} and~\eqref{Eq:P_DT_accelerated} have only one Laplacian block in the block diagram, the resulting root loci are linear in the Laplacian eigenvalues. For the proportional-integral dynamic average consensus algorithm, however, the block diagram contains two Laplacian blocks resulting in a quadratic dependence on the eigenvalues. Instead of a linear root locus, the design involves a quadratic root locus. Although this complicates the design process, closed-form solutions for the algorithm parameters can still be found~\cite{BV-RAF-KML:15}, even for the accelerated version using extra dynamics, given by
\begin{subequations}\label{Eq:PI_DT_accelerated}
\begin{align}
q^i_{k+1} &= 2\rho\,q^i_k - \rho^2 q^i_{k-1} + k_p \sum_{j=1}^N a_{ij} \bigl((x^i_k-x^j_k) + (p^i_k-p^j_k)\bigr),\\
p^i_{k+1} &= (1+\rho^2) p^i_k - \rho^2 p^i_{k-1} + k_I \sum_{j=1}^N a_{ij} (x^i_k-x^j_k),\\
x^i_k &= u^i_k - q^i_k,\qquad
 p^i_0,q^i_0\in\reals, \qquad i\in\until{N},
\end{align}
\end{subequations}
whose block diagram is in Figure~\ref{Fig:BlockDiagramAccelerated_PI}. The resulting convergence rate is plotted in Figure~\ref{Fig:rho}. While the convergence rates of the standard and accelerated proportional-integral dynamic average consensus algorithms are slower than those of~\eqref{Eq:P_DT} and~\eqref{Eq:P_DT_accelerated}, respectively, they have the additional advantage of being robust to initial conditions.

The following result summarizes the parameter choices which optimize the convergence rate for each of the DT dynamic average consensus algorithms in Figure~\ref{Fig:BlockDiagramDT}. The results for the first two algorithms follow from the previous discussion, while details of the results for the last two algorithms can be found in~\cite{BV-RAF-KML:15}.

\begin{theorem}[Optimal convergence rates of DT dynamic average consensus algorithms] \label{Thm:DT_rates}
Let $\GG$ be a connected, undirected graph. Suppose the reference signal $\vectsf{u}^i$ at each agent $i\in\until{N}$ is a constant scalar. Consider the dynamic average consensus algorithms in Figure~\ref{Fig:BlockDiagramDT} with the  parameters chosen according to Table~\ref{tab:parameters} (the algorithms in Figures~\ref{Fig:BlockDiagramP} and~\ref{Fig:BlockDiagramAccelerated_P} are initialized such that the average of the initial integrator states is zero, that is, $\sum_{i=1}^N p^i_0 = 0$). Then, the agreement states $x^i_k$, $i\in\until{N}$ converge to $\vectsf{u}^\textup{avg}$ exponentially with rate $\rho$.
\end{theorem}

\begin{table}[htb]
\caption{Parameter selection for the dynamic average consensus algorithms of Figure~\ref{Fig:BlockDiagramDT} as a function of the minimum and maximum nonzero Laplacian eigenvalues $\lambda_2$ and $\lambda_N$, respectively, with $\lr\defeq\lambda_2/\lambda_N$.}\label{tab:parameters}
    \centering
\begin{tabular}{l|l|c|c}
 & $\rho$ & $k_I$ & $k_p$ \\[1mm] \hline
Figure~\ref{Fig:BlockDiagramP} &
  $\frac{\lambda_N-\lambda_2}{\lambda_N+\lambda_2}$ &
  $\frac{2}{\lambda_2 + \lambda_N}$ & \textup{N/A} \\[5mm]
Figure~\ref{Fig:BlockDiagramPI} &
  $\begin{cases}
    \frac{8-8\lr+\lr^2}{8-\lr^2}, & {\scriptstyle 0 < \lr \leq 3-\sqrt{5}} \\[1mm]
    \frac{\sqrt{(1-\lr)(4+\lr^2(5-\lr))} - \lr(1-\lr)}{2(1+\lr^2)}, & {\scriptstyle 3-\sqrt{5} < \lr \leq 1} \end{cases}$ &
  $\frac{1-\rho}{\lambda_2}$ &
  $\frac{1}{\lambda_N} \frac{\rho (1-\rho) \lr}{\rho + \lr - 1}$ \\[8mm]
Figure~\ref{Fig:BlockDiagramAccelerated_P} &
  $\frac{\sqrt{\lambda_N}-\sqrt{\lambda_2}}{\sqrt{\lambda_N}+\sqrt{\lambda_2}}$ &
  $\frac{4}{(\sqrt{\lambda_2} + \sqrt{\lambda_N})^2}$ & \textup{N/A} \\[5mm]
Figure~\ref{Fig:BlockDiagramAccelerated_PI} &
  $\begin{cases}
    \frac{6 - 2\sqrt{1-\lr} + \lr - 4 \sqrt{2 - 2\sqrt{1-\lr} + \lr)}}{2 + 2\sqrt{1-\lr} - \lr}, & {\scriptstyle0 < \lr \leq 2(\sqrt{2}-1)} \\[1mm]
    \frac{-3 - 2\sqrt{1-\lr} + \lr + 2 \sqrt{2 + 2\sqrt{1-\lr} - \lr}}{-1-2\sqrt{1-\lr} + \lr}, & {\scriptstyle 2(\sqrt{2}-1) < \lr \leq 1} \end{cases}$ &
  $\frac{(1-\rho)^2}{\lambda_2}$ &
  ${\scriptstyle (2 + 2\sqrt{1-\lr} - \lr) k_I}$
\end{tabular}
\end{table}

\section{Perfect Tracking Using A Priori Knowledge of the Input Signals}\label{sec::TimeVarying}

The design of the dynamic average consensus algorithms described in the discussion so far does not require prior knowledge of the reference signals and is therefore broadly applicable.  This also comes at a cost: the convergence guarantees of  these algorithms are strong only when the reference signals are constant or slowly varying. The error of such algorithms can be large, however, when the reference signals change quickly in time. In this section, we describe dynamic average consensus algorithms which are capable of tracking fast time-varying signals with either zero or small steady-state error. In each case, their design assumes some specific information about the nature of the reference signals. In particular, we consider reference signals which either (1) have a known model, (2) are bandlimited, or (3) have bounded derivatives.

\subsection{Signals with a known model (discrete time)}

The discrete-time dynamic average consensus algorithms discussed previously are designed with the idea of tracking constant reference signals with zero steady-state error. To do this, the algorithms contain an integrator in the feedback loop. This concept generalizes to time-varying signals with a known model using the \emph{internal model principle}. Consider reference signals whose $z$-transform has the form $\mathsf{u}^i(z)=n^i(z)/d(z)$ where $n^i(z)$ and $d(z)$ are polynomials in $z$ for $i\in\until{N}$. Dynamic average consensus algorithms can be designed to have zero steady-state error for such signals by placing the model of the input signals (that is, $d(z)$) in the feedback loop. Some common examples of models are
\begin{align*}
d(z) &=
\begin{cases}
(z-1)^m, & \text{polynomial of degree }m-1 \\
 z^2-2z\cos(\omega)+1, & \text{sinusoid with frequency }\omega.
\end{cases}
\end{align*}
In this section, we focus on dynamic average consensus algorithms which track degree $m-1$ polynomial reference signals with zero steady-state error.

\begin{figure}[htb]
\begin{subfigure}{\columnwidth}
\centering \begin{tikzpicture}[align=center]
  \renewcommand{\x}{0.6cm};
  \renewcommand{\y}{0.5cm};
  \node (u1)    {$\vectsf{u}_k$};
  \node (delta) [block,right=\x of u1]  {$\Delta^{(m)}$};
  \node (sum1)  [sum,right=\x of delta] {};
  \node (f1)    [block,right=\x of sum1] {$\dfrac{1}{z-1} \vect{I}_N$};
  \node (L1)    [block,below=\y of f1]   {$\lL$};
  \coordinate (x1) at ($(f1.east)+(\x,0)$) {};
  \draw [link] (u1) -- (delta);
  \draw [link] (delta) -- (sum1);
  \draw [link] (sum1) -- (f1);
  \draw [link] (f1) -- ($(x1)+(\x,0)$);
  \draw [link] (L1) -| node[pos=0.9,xshift=-2.5mm] {\footnotesize{$-$}} (sum1);
  \draw [link] (x1) |- (L1);
  \node (e)    [right=1.2cm of x1] {$\ldots$};
  \node (u2)   [right=\x of e,inner sep=0mm] {};
  \node (sum2) [sum,right=\x of u2] {};
  \node (f2)   [block,right=\x of sum2] {$\dfrac{1}{z-1} \vect{I}_N$};
  \node (L2)   [block,below=\y of f2] {$\lL$};
  \coordinate (x2) at ($(f2.east)+(\x,0)$) {};
  \node (y)    [right=\x of x2] {$\vect{x}_k$};
  \draw [link] (u2.west) -- (sum2);
  \draw [link] (sum2) -- (f2);
  \draw [link] (f2) -- (y);
  \draw [link] (L2) -| node[pos=0.9,xshift=-2.5mm] {\footnotesize{$-$}} (sum2);
  \draw [link] (x2) |- (L2);
  \draw [very thick,decorate,decoration={brace,mirror,amplitude=3mm}] ([yshift=-20mm,xshift=-1mm]sum1.center) -- node[below,yshift=-3mm] %
        {\footnotesize{$m$ times}}([yshift=-20mm,xshift=1mm]x1.east);
\end{tikzpicture}
\caption{Dynamic average consensus algorithm~\eqref{Eq:DT_poly_ZM} in \cite{MZ-SM:10-auto} where $\Delta^{(m)} = (1-z^{-1})^m$ is the $m^\text{th}$ divided difference (see also~\cite{EM-JIM-CS-SM:14} for a stepsize analysis). The performance does not degrade when the graph is time-varying, but the estimate is delayed by $m$ iterations. Furthermore, the algorithm is numerically unstable when $m$ is large and eventually diverges from tracking the average when implemented using finite precision arithmetic.}
\label{Fig:BlockDiagram_poly_ZM}
\end{subfigure}\\
\begin{subfigure}{\columnwidth}
\centering \begin{tikzpicture}[align=center]
  \renewcommand{\x}{0.6cm};
  \renewcommand{\y}{0.5cm};
  \node (u1)   {$\vectsf{u}_k$};
  \node (sum1) [sum,right=\x of u1] {};
  \node (f1)   [block,below right=\y and 0.3cm of sum1] {$\dfrac{1}{z-1} \vect{I}_N$};
  \node (L1)   [block,right=\x of f1]   {$\lL$};
  \coordinate (x1) at ($(u1 -| L1.east) +(\x,0)$) {};
  \draw [link] (u1) -- (sum1);
  \draw [link] (f1) -| node[pos=0.9,xshift=-2.5mm] {\footnotesize{$-$}} (sum1);
  \draw [link] (L1) -- (f1);
  \draw [link] (x1) |- (L1);
  \draw [link] (sum1) -- ($(x1)+(\x,0)$);
  \node (e)    [right=1.2cm of x1] {$\ldots$};
  \node (u2)   [right=\x of e,inner sep=0mm] {};
  \node (sum2) [sum,right=\x of u2] {};
  \node (f2)   [block,below right=\y and 0.3cm of sum2] {$\dfrac{1}{z-1} \vect{I}_N$};
  \node (L2)   [block,right=\x of f2] {$\lL$};
  \coordinate (x2) at ($(u2 -| L2.east) + (\x,0)$) {};
  \node (y)    [right=\x of x2] {$\vectsf{x}_k$};
  \draw [link] (u2.west) -- (sum2);
  \draw [link] (sum2) -- (y);
  \draw [link] (f2) -| node[pos=0.9,xshift=-2.5mm] {\footnotesize{$-$}} (sum2);
  \draw [link] (L2) -- (f2);
  \draw [link] (x2) |- (L2);
  \draw [very thick,decorate,decoration={brace,mirror,amplitude=3mm}] ([yshift=-22mm,xshift=-1mm]sum1.center) -- node[below,yshift=-3mm] %
        {\footnotesize{$m$ times}}([yshift=-22mm,xshift=1mm]x1.east);
\end{tikzpicture}
\caption{Dynamic average consensus algorithm~\eqref{Eq:DT_poly_P}, which is the algorithm in~\cite{RAF-PY-KML:06} cascaded in series $m$ times. The estimate of the average is not delayed and the algorithm is numerically stable, but the tracking performance degrades when the communication graph is time-varying.}
\label{Fig:BlockDiagram_poly_P}
\end{subfigure}
\caption{Block diagram of dynamic average consensus algorithms which track polynomial signals of degree $m-1$ with zero steady-state error when initialized correctly (neither algorithm is robust to initial conditions). The indicated section is repeated in series $m$ times.}
\label{Fig:BlockDiagram_poly}
\end{figure}
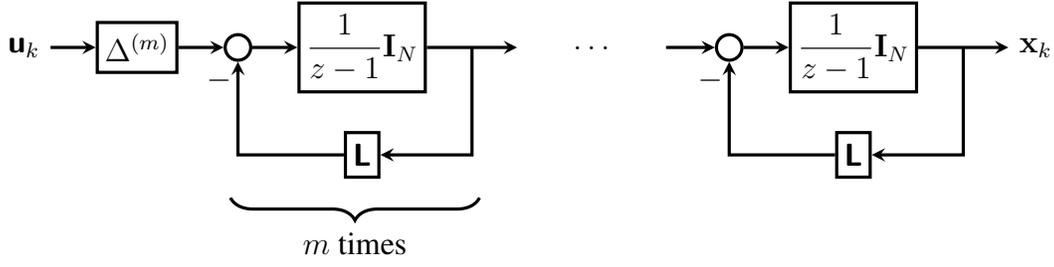
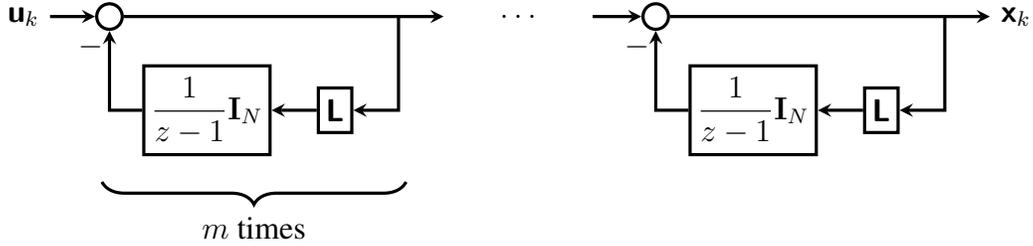

Consider the dynamic average consensus algorithms in Figure~\ref{Fig:BlockDiagram_poly}. The transfer function of each algorithm has $m-1$ zeros at $z=1$, so the algorithms track degree $m-1$ polynomial references signals with zero steady-state error. The time-domain equations for the dynamic average consensus algorithm in Figure~\ref{Fig:BlockDiagram_poly_ZM} are
\begin{subequations}\label{Eq:DT_poly_ZM}
\begin{align}
x^i_{1,k+1} &= x^i_{1,k} - \sum_{j=1}^N a_{ij} (x^i_{1,k}-x^j_{1,k}) + \Delta^{(m)} \mathsf{u}_k \\
x^i_{2,k+1} &= x^i_{2,k} - \sum_{j=1}^N a_{ij} (x^i_{2,k}-x^j_{2,k}) + x^i_{1,k} \\
 &~~\vdots \nonumber \\
x^i_{m,k+1} &= x^i_{m,k} - \sum_{j=1}^N a_{ij} (x^i_{m,k}-x^j_{m,k}) + x^i_{m-1,k} \\
x^i_k &= x^i_{m,k}, \quad x^i_{\ell,0}\in\reals, \quad \ell\in\until{m}, \quad i\in\until{N}
\end{align}
\end{subequations}
where the $m^\text{th}$ divided difference is defined recursively as $\Delta^{(m)} \mathsf{u}^i_k = \Delta^{(m-1)} \mathsf{u}^i_k - \Delta^{(m-1)} \mathsf{u}^i_{k-1}$ for $m\geq 2$ with $\Delta^{(1)} \mathsf{u}^i_k = \mathsf{u}^i_k-\mathsf{u}^i_{k-1}$. The estimate of the average, however, is delayed by $m$ iterations due to the transfer function having a factor of $z^{-m}$ between the input and output. This problem is fixed by the dynamic average consensus algorithm in Figure~\ref{Fig:BlockDiagram_poly_P}, given by
\begin{subequations}\label{Eq:DT_poly_P}
\begin{align}
p^i_{1,k+1} &= p^i_{1,k} + \sum_{j=1}^N a_{ij} \bigl( (\mathsf{u}^i_k-\mathsf{u}^j_k) - (p^i_{1,k}-p^j_{1,k}) \bigr) \\
p^i_{2,k+1} &= p^i_{2,k} + \sum_{j=1}^N a_{ij} \bigl( (\mathsf{u}^i_k-\mathsf{u}^j_k) - (p^i_{1,k}-p^j_{1,k}) - (p^i_{2,k}-p^j_{2,k}) \bigr) \\
 &~~\vdots \nonumber \\
p^i_{m,k+1} &= p^i_{m,k} + \sum_{j=1}^N a_{ij} \Bigl( (\mathsf{u}^i_k-\mathsf{u}^j_k) - \sum_{\ell=1}^m (p^i_{\ell,k}-p^j_{\ell,k}) \Bigr) \\
x^i_k &= \mathsf{u}^i_k - \sum_{\ell=1}^m p^i_{\ell,k}, \quad p^i_{\ell,0}\in\reals, \quad \ell\in\until{m}, \quad i\in\until{N},
\end{align}
\end{subequations}
which tracks degree $m-1$ polynomial reference signals with zero steady-state error without delay. Note, however, that the communication graph is assumed to be constant in order to use frequency domain arguments; while the output of the dynamic average consensus algorithm in Figure~\ref{Fig:BlockDiagram_poly_ZM} is delayed, it also has nice tracking properties when the communication graph is time-varying whereas the dynamic average consensus algorithm in Figure~\ref{Fig:BlockDiagram_poly_P} does not.

To track degree $m-1$ polynomial reference signals, each dynamic average consensus algorithm in Figure~\ref{Fig:BlockDiagram_poly} cascades $m$ dynamic average consensus algorithms, each with a pole at $z=1$ in the feedback loop. The dynamic average consensus algorithm~\eqref{Eq:P_DT} is cascaded in Figure~\ref{Fig:BlockDiagram_poly_P}, but any of the dynamic average consensus algorithms from the previous section could also be used. For example, the proportional-integral dynamic average consensus algorithm could be cascaded $m$ times to track degree $m-1$ polynomial reference signals with zero steady-state error independent of the initial conditions.

In general, reference signals with model $d(z)$ can be tracked with zero steady-state error by cascading simple dynamic average consensus algorithms, each of which tracks a factor of $d(z)$. In particular, suppose $d(z) = d_1(z)\,d_2(z)\ldots d_m(z)$. Then $m$ dynamic average consensus algorithms can be cascaded where the $i^\textup{th}$ component contains the model $d_i(z)$ for $i=1,\ldots,m$. Alternatively, a single dynamic average consensus algorithm can be designed which contains the entire model $d(z)$. This approach using an internal model version of the proportional-integral dynamic average consensus algorithm is designed in~\cite{HB-RAF-KML:10} in both continuous and discrete time.

In many practical applications, the exact model of the reference signals is unknown. However, it is shown in~\cite{HB:15} that a frequency estimator can be used in conjunction with an internal model dynamic average consensus algorithm to still achieve zero steady-state error. In particular, the frequency of the reference signals is estimated such that the estimate converges to the actual frequency. This time-varying estimate of the frequency is then used in place of the true frequency to design the feedback dynamic average consensus algorithm~\cite{HB:15}.

\subsection{Bandlimited signals (discrete time)}
In order to use algorithms designed using the model of the reference signals, the signals must be composed of a finite number of known frequencies. When either the frequencies are unknown or there are infinitely many frequencies, dynamic average consensus algorithms can still be designed if the reference signals are bandlimited. In this case, feedforward dynamic average consensus algorithm designs can be used to achieve arbitrarily small steady-state error.

For our discussion here, we assume that the reference signals are bandlimited with known cutoff frequency. In particular, let $\mathsf{U}^i(z)$ be the $z$-transform of the $i^\text{th}$ reference signal $\{\mathsf{u}^i_k\}$. Then $\mathsf{U}^i(z)$ is bandlimited with cutoff frequency $\theta_c$ if $|\mathsf{U}^i(\exp(j\theta))|=0$ for all $\theta\in(\theta_c,\pi]$.

\begin{figure}
\centering\begin{tikzpicture}[align=center]
  \renewcommand{\x}{0.6cm};
  \renewcommand{\y}{0.8cm};
  \node (u1)   {$\vectsf{u}_k$};
  \node (hpre) [block,right=\x of u1] {$h(z) \vect{I}_N$};
  \coordinate (x1) at ($(hpre.east)+(\x,0)$) {};
  \node (L1)   [block,below right=\y and \x of x1] {$\lL$};
  \node (sum1) at ($(L1.east |- u1)+(\x,0)$) [sum] {};
  \node (f1)   [block,right=\x of sum1] {$\dfrac{1}{z} \vect{I}_N$};
  \draw [link] (u1) -- (hpre);
  \draw [link] (hpre) -- (sum1);
  \draw [link] (L1) -| node[pos=0.9,xshift=-2.5mm] {\footnotesize{$-$}} (sum1);
  \draw [link] (x1) |- (L1);
  \draw [link] (sum1) -- (f1);
  \draw [link] (f1.east) -- ($(f1.east)+(\x,0)$);
  \node (e)    [right=1.2cm of f1.east,inner sep=0mm] {$\ldots$};
  \node (u2)   [right=\x of e,inner sep=0mm] {};
  \coordinate (x2) at ($(u2)+(\x,0)$) {};
  \node (L2)   [block,below right=\y and \x of x2] {$\lL$};
  \node (sum2) at ($(L2.east |- u2)+(\x,0)$) [sum] {};
  \node (f2)   [block,right=\x of sum2] {$\dfrac{1}{z} \vect{I}_N$};
  \node (y)    [right=\x of f2] {$\vect{x}_k$};
  \draw [link] (u2) -- (sum2);
  \draw [link] (sum2) -- (f2);
  \draw [link] (L2) -| node[pos=0.9,xshift=-2.5mm] {\footnotesize{$-$}} (sum2);
  \draw [link] (x2) |- (L2);
  \draw [link] (f2) -- (y);
  \draw [very thick,decorate,decoration={brace,mirror,amplitude=3mm}] ([yshift=-16mm,xshift=-1mm]x1) -- node[below,yshift=-3mm] %
        {\footnotesize{$m$ times}}([yshift=-16mm,xshift=1mm]f1.east);
\end{tikzpicture}
\caption{Feedforward dynamic average consensus algorithm for tracking the average of bandlimited reference signals. The prefilter $h(z)$ is applied to the reference signals \emph{before} passing through the graph Laplacian. For an appropriately designed prefilter, the dynamic average consensus algorithm can track bandlimited reference signals with arbitrarily small steady-state error when using exact arithmetic (and small error for finite precision)~\cite{BV-RAF-KML:16}.}
\label{Fig:BlockDiagramFeedforward}
\end{figure}
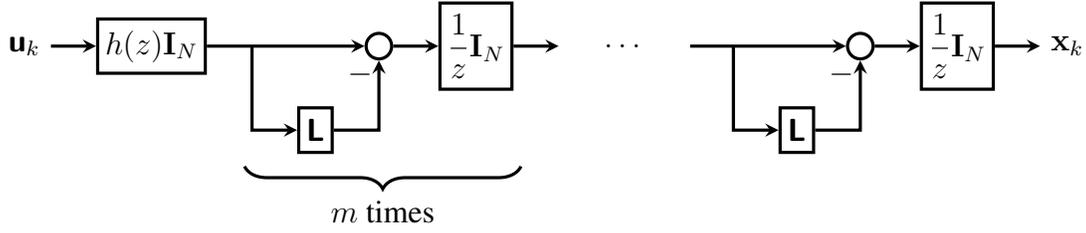

Consider the dynamic average consensus algorithm in Figure~\ref{Fig:BlockDiagramFeedforward}. The reference signals are passed through a prefilter $h(z)$ and then multiplied $m$ times by the consensus matrix $\vect{I}-\lL$ with a delay between each (to allow time for communication). The transfer function from the input $\vectsf{U}(z)$ to the output $\vect{X}(z)$ is
\begin{equation*}
    H(z,\lL) = h(z) \frac{1}{z^m} (\vect{I}-\lL)^m.
\end{equation*}
In order for the tracking error to be small, $h(z)$ must approximate $z^m$ for all $\theta\in[0,\theta_c]$ where $z=\exp(j\theta)$ and $\theta_c$ is the cutoff frequency. In this case the transfer function in the passband is approximately
\begin{equation*}
    H(z,\lL) \approx (\vect{I}-\lL)^m,
\end{equation*}
so the error can be made small by making $m$ large enough (so long as $\lL$ is scaled such that $\|\vect{I}-\lL-\1\1^\top/N\|_2<1$).

In particular, the prefilter is designed such that $h(z)$ is proper and $h(z)\approx z^m$ for $z=\exp(j\theta)$ for all $\theta\in[0,\theta_c]$ (note that $h(z)=z^m$ cannot be used since it is not causal). An $m$-step filter can be obtained by cascading a one-step filter $m$ times in series. In other words, let $h(z) = [z f(z)]^m$ where $f(z)$ is strictly proper and approximates unity in the passband. Since $f(z)$ must approximate unity in both magnitude \emph{and} phase, a standard lowpass filter cannot be used. Instead, set
\begin{align*}
f(z) &= 1 - g(z)/\lim_{z\to\infty} g(z)
\end{align*}
where $g(z)$ is a proper highpass filter with cutoff frequency $\theta_c$. Then $f(z)$ is strictly proper (due to the normalizing constant in the denominator) and approximates unity in the band $[0,\theta_c]$ (since $g(z)$ is highpass). Therefore, a prefilter $h(z)$ which approximates $z^m$ in the passband can be designed using a standard highpass filter $g(z)$.

Using such a prefilter,~\cite{BV-RAF-KML:16} makes the error of the dynamic average consensus algorithm in Figure~\ref{Fig:BlockDiagramFeedforward} arbitrarily small if (1) the graph is connected and balanced at each time step (in particular, it need not be constant), (2) $\lL$ is scaled such that $\|\vect{I}-\lL-\1\1^\top/N\|_2<1$, (3) the number of stages $m$ is made large enough, (4) the prefilter can approximate $z^m$ arbitrarily closely in the passband, and (5) exact arithmetic is used. Note that exact arithmetic is required for arbitrarily small error since rounding errors cause high frequency components in the reference signals.

\subsection{Signals with bounded derivatives (continuous time)}
Stronger tracking results can be obtained using algorithms implemented in continuous time. Here, we present a set of continuous-time dynamic average consensus algorithms which are capable of tracking time-varying reference signals whose derivatives are bounded with zero error in \emph{finite time}. For simplicity, we assume that the communication graph is constant, connected, and undirected. Also, the reference signals are assumed to be differentiable with bounded derivatives.

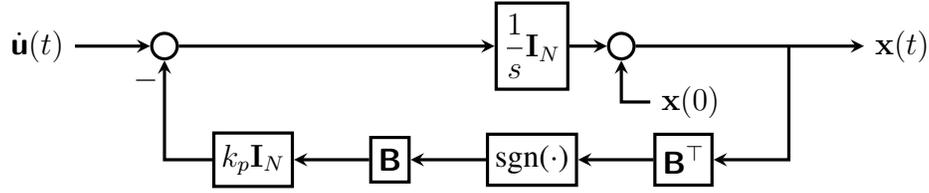
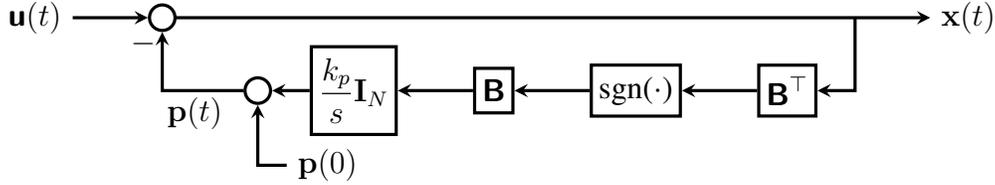
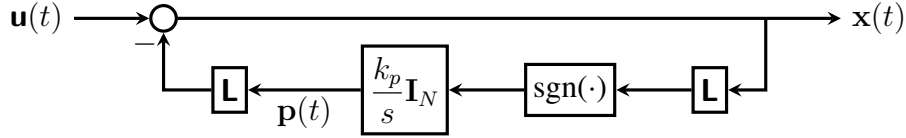
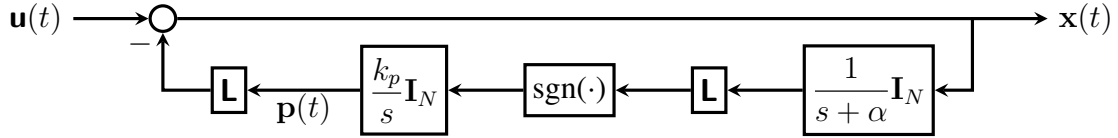
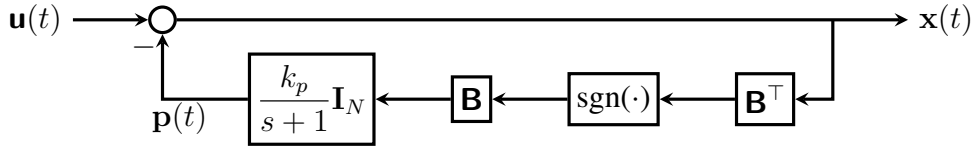
\begin{figure}[!htb]
\begin{subfigure}{\textwidth}
\centering \begin{tikzpicture}
  \node (u)    {$\dvectsf{u}(t)$};
  \node (sum1) [sum,right=of u] {};
  \node (alpha) [block,below right=1cm and 0.5cm of sum1] {$k_p\vect{I}_N$};
  \node (B1)   [block,right=of alpha]  {$\vectsf{B}$};
  \node (sgn)  [block,right=of B1] {\textrm{sgn}($\cdot$)};
  \node (B2)   [block,right=of sgn] {$\vectsf{B}^\top$};
  \node (f)    [block] at (u -| sgn) {$\dfrac{1}{s} \vect{I}_N$};
  \node (sum2) [sum,right=0.5cm of f] {};
  \coordinate (x) at ($(f -| B2.east)+(1cm,0)$) {};
  \node (y)    [right=of x] {$\vect{x}(t)$};
  \node (x0)   [below right=0.25cm and 0.25cm of sum2] {$\vect{x}(0)$};
  \draw [link] (u) -- (sum1);
  \draw [link] (sum1) -- (f);
  \draw [link] (f) -- (sum2);
  \draw [link] (sum2) -- (y);
  \draw [link] (x) |- (B2);
  \draw [link] (B2) -- (sgn);
  \draw [link] (sgn) -- (B1);
  \draw [link] (alpha) -| node[pos=0.9,xshift=-2.5mm] {\footnotesize{$-$}} (sum1);
  \draw [link] (B1) -- (alpha);
  \draw [link] (x0) -| (sum2);
\end{tikzpicture}
\caption{Dynamic average consensus algorithm~\eqref{eq::alg_FC-YC-WR:12} which achieves perfect tracking in finite time and uses one-hop communication, but is not robust to initial conditions (that is, the steady-state error is zero only if $\1^\top \vect{x}(0) = \1^\top \vectsf{u}(0)$). Furthermore, the derivative of the reference signals is required; see~\cite{FC-YC-WR:12}.}
\label{Fig:BlockDiagram_CT1}
\end{subfigure}\\
\begin{subfigure}{\textwidth}
\centering \begin{tikzpicture}
  \node (u)    {$\vectsf{u}(t)$};
  \node (sum1) [sum,right=of u] {};
  \node (sum2) [sum,below right=0.7cm and 1cm of sum1] {};
  \node (f)    [block,right=0.5cm of sum2] {$\dfrac{k_p}{s}\vect{I}_N$};
  \node (L1)   [block,right=of f] {$\vectsf{B}$};
  \node (sgn)  [block,right=of L1] {\textrm{sgn}($\cdot$)};
  \node (L2)   [block,right=of sgn] {$\vectsf{B}^\top$};
  \coordinate (x) at ($(u -| L2.east)+(0.5cm,0)$) {};
  \node (y)    [right=of x] {$\vect{x}(t)$};
  \node (x0)   [below right=0.5cm and 0.25cm of sum2] {$\vect{p}(0)$};
  \draw [link] (u) -- (sum1);
  \draw [link] (sum1) -- (y);
  \draw [link] (x) |- (L2);
  \draw [link] (L2) -- (sgn);
  \draw [link] (sgn) -- (L1);
  \draw [link] (L1) -- (f);
  \draw [link] (f) -- (sum2);
  \draw [link] (sum2) -| node[pos=0.9,xshift=-2.5mm] {\footnotesize{$-$}} node[pos=0.3,yshift=-3mm] {$\vect{p}(t)$} (sum1);
  \draw [link] (x0) -| (sum2);
\end{tikzpicture}
\caption{Dynamic average consensus algorithm~\eqref{eq::alg_FC-YC-WR:12-2} which is equivalent to the algorithm in (a), although this form does not require the derivative of the reference signals. In this case, the requirement on the initial conditions is $\mathbf{1}^\top \vect{p}(0) = 0$.}
\label{Fig:BlockDiagram_CT2}
\end{subfigure}\\
\begin{subfigure}{\textwidth}
\centering \begin{tikzpicture}
  \node (u)    {$\vectsf{u}(t)$};
  \node (sum1) [sum,right=of u] {};
  \node (L1)   [block,below right=0.5cm and 0.5cm of sum1] {$\lL$};
  \node (f1)   [block,right=1.5cm of L1] {$\dfrac{k_p}{s} \vect{I}_N$};
  \node (sgn)  [block,right=of f1] {\textrm{sgn}($\cdot$)};
  \node (L2)   [block,right=of sgn] {$\lL$};
  \coordinate (x) at ($(u -| L2.east)+(0.5cm,0)$) {};
  \node (y)    [right=of x] {$\vect{x}(t)$};
  \draw [link] (u) -- (sum1);
  \draw [link] (sum1) -- (y);
  \draw [link] (x) |- (L2);
  \draw [link] (L2) -- (sgn);
  \draw [link] (sgn) -- (f1);
  \draw [link] (f1) -- node[pos=0.5,yshift=-3mm] {$\vect{p}(t)$} (L1);
  \draw [link] (L1) -| node[pos=0.9,xshift=-2.5mm] {\footnotesize{$-$}} (sum1);
\end{tikzpicture}
\caption{Dynamic average consensus algorithm~\eqref{eq::alg_CT3} which converges to zero error in finite time and is robust to initial conditions, but requires two-hop communication (in other words, two rounds of communication are performed at each time instant); see~\cite{JG-RAF-KML:17}.}
\label{Fig:BlockDiagram_CT3}
\end{subfigure}\\
\begin{subfigure}{\textwidth}
\centering \begin{tikzpicture}
  \node (u)    {$\vectsf{u}(t)$};
  \node (sum1) [sum,right=of u] {};
  \node (L1)   [block,below right=0.5cm and 0.5cm of sum1]  {$\lL$};
  \node (f1)   [block,right=1.5cm of L1] {$\dfrac{k_p}{s} \vect{I}_N$};
  \node (sgn)  [block,right=of f1] {\textrm{sgn}($\cdot$)};
  \node (L2)   [block,right=of sgn] {$\lL$};
  \node (f2)   [block,right=of L2] {$\dfrac{1}{s+\alpha} \vect{I}_N$};
  \coordinate (x) at ($(u -| f2.east)+(0.5cm,0)$) {};
  \node (y)    [right=of x] {$\vect{x}(t)$};
  \draw [link] (u) -- (sum1);
  \draw [link] (sum1) -- (y);
  \draw [link] (x) |- (f2);
  \draw [link] (f2) -- (L2);
  \draw [link] (L2) -- (sgn);
  \draw [link] (sgn) -- (f1);
  \draw [link] (f1) -- node[pos=0.5,yshift=-2.5mm] {$\vect{p}(t)$} (L1);
  \draw [link] (L1) -| node[pos=0.9,xshift=-3mm] {\footnotesize{$-$}} (sum1);
\end{tikzpicture}
\caption{Dynamic average consensus algorithm~\eqref{eq::alg_CT4} which is robust to initial conditions and uses one-hop communication, but converges to zero error exponentially instead of in finite time; see~\cite{JG-RAF-KML:17}.}
\label{Fig:BlockDiagram_CT4}
\end{subfigure}\\
\begin{subfigure}{\textwidth}
\centering \begin{tikzpicture}
  \node (u)     {$\vectsf{u}(t)$};
  \node (sum1)  [sum,right=of u] {};
  \node (alpha) [block,below right=0.3cm and 1cm of sum1] {$\dfrac{k_p}{s+1}\vect{I}_N$};
  \node (B1)  [block,right=of alpha] {$\vectsf{B}$};
  \node (sgn) [block,right=of B1] {\textrm{sgn}($\cdot$)};
  \node (B2)  [block,right=of sgn] {$\vectsf{B}^\top$};
  \coordinate (x) at ($(u -| B2.east)+(0.5cm,0)$) {};
  \node (y) [right=of x] {$\vect{x}(t)$};
  \draw [link] (u) -- (sum1);
  \draw [link] (sum1) -- (y);
  \draw [link] (x) |- (B2);
  \draw [link] (B2) -- (sgn);
  \draw [link] (sgn) -- (B1);
  \draw [link] (B1) -- (alpha);
  \draw [link] (alpha) -| node[pos=0.9,xshift=-2.5mm] {\footnotesize{$-$}} node[pos=0.4,yshift=-2.5mm] {$\vect{p}(t)$} (sum1);
\end{tikzpicture}
\caption{Dynamic average consensus algorithm~\eqref{eq::alg_FC-YC-WR:X-eq} that is robust to initial conditions and uses one-hop communication, although the error converges to zero exponentially instead of in finite time; see~\cite{SR-WR:17}.}\label{Fig:BlockDiagram_CT5}
\end{subfigure}
\caption{Block diagram of discontinuous dynamic average consensus algorithms in continuous time. In each case, the communication graph is assumed to be constant, connected, and balanced with Laplacian matrix $\lL = \vectsf{B}\vectsf{B}^\top$. Furthermore, the reference signals are assumed to have bounded derivatives.}
\label{Fig:BlockDiagram_CT}
\end{figure}

In discrete time, zero steady-state error is obtained by placing the internal model of the reference signals in the feedback loop. This provides infinite loop gain at the frequencies contained in the reference signals. In continuous time, however, the discontinuous signum function \textrm{sgn} can be used in the feedback loop to provide `infinite' loop gain over all frequencies, so no model of the reference signals is required. Furthermore, such continuous-time dynamic average consensus algorithms are capable of achieving zero error tracking in finite time as opposed to the exponential convergence achieved by discrete-time dynamic average consensus algorithms. One such algorithm is described in~\cite{FC-YC-WR:12} as
\begin{subequations}\label{eq::alg_FC-YC-WR:12}
\begin{align}
\dot{x}^i &= \dot{\mathsf{u}}^i - k_p\sum_{j\in\mathcal{N}_\text{out}^i} \text{sgn}(x^i(t)-x^j(t)), \qquad i\in\until{N},\\
 &\sum\nolimits_{i=1}^N x^i(0)=\sum\nolimits_{i=1}^N \mathsf{u}^i(0).
\end{align}
\end{subequations}
The block diagram representation in Figure~\ref{Fig:BlockDiagram_CT1} indicates that this algorithm applies \textrm{sgn} in the feedback loop. Under the given assumptions, using a sliding mode argument, the feedback gain $k_p$ can be selected to guarantee zero error tracking in finite time provided that an upper bound $\gamma$ of the form $\sup_{\tau\in[0,\infty)}\|\dvectsf{u}(\tau)\|=\gamma<\infty$ is known~\cite{FC-YC-WR:12}. The dynamic consensus algorithm~\eqref{eq::alg_FC-YC-WR:12} can also be implemented without derivative information of the reference signals in an equivalent way as
\begin{subequations}\label{eq::alg_FC-YC-WR:12-2}
\begin{align}
\dot{p}^i &= k_p\sum_{j\in\mathcal{N}_\text{out}^i} \text{sgn}(x^i-x^j),\quad\quad \sum\nolimits_{j=1}^N p^j(0)=0,\\
x^i &= \mathsf{u}^i - p^i, \qquad \qquad\qquad \qquad i\in\until{N}.
\end{align}
\end{subequations}
The corresponding block diagram is shown in Figure~\ref{Fig:BlockDiagram_CT2}.

It is simple to see from the block diagram of Figure~\ref{Fig:BlockDiagram_CT1} why~\eqref{eq::alg_FC-YC-WR:12} is not robust to initial conditions; the integrator state is directly connected to the output and therefore affects the steady-state output in the consensus direction. This issue is addressed by the dynamic average consensus algorithm in Figure~\ref{Fig:BlockDiagram_CT3}, given by
\begin{subequations}\label{eq::alg_CT3}
\begin{align}
\dot{p}^i &= k_p\, \text{sgn}\bigl(\sum_{j\in\mathcal{N}_\text{out}^i} (x^i-x^j)\bigr),\quad\quad\quad p^i(0)\in\reals,\\
x^i &= \mathsf{u}^i - \sum_{j\in\mathcal{N}_\text{out}^i} (p^i-p^j), \qquad  \qquad i\in\until{N},
\end{align}
\end{subequations}
which moves the integrator before the Laplacian in the feedback loop. However, this dynamic average consensus algorithm has two Laplacian blocks directly connected which means that it requires two-hop communication to implement. In other words, two sequential rounds of communication are required at each time instant. In the time-domain, each agent must do the following (in order) at each time $t$: (1) communicate $p^i(t)$, (2) calculate $x^i(t)$, (3) communicate $x^i(t)$, and (4) update $p^i(t)$ using the derivative $\dot{p}^i(t)$. To only require one-hop communication, the dynamic average consensus algorithm in Figure~\ref{Fig:BlockDiagram_CT4}, given by
\begin{subequations}\label{eq::alg_CT4}
\begin{align}
\dot{q}^i &= -\alpha\,q^i + x^i \\
\dot{p}^i &= k_p\, \text{sgn}\bigl(\sum_{j\in\mathcal{N}_\text{out}^i} (q^i-q^j)\bigr),\\
x^i &= \mathsf{u}^i - \sum_{j\in\mathcal{N}_\text{out}^i} (p^i-p^j), \qquad p^i(0),q^i(0)\in\reals, \qquad i\in\until{N},
\end{align}
\end{subequations}
places a strictly proper transfer function in the path between the Laplacian blocks. The extra dynamics, however, cause the output to converge exponentially instead of in finite time~\cite{JG-RAF-KML:17}.

Alternatively, under the given assumptions, a sliding mode-based dynamic average consensus algorithm with zero error tracking, which can be  arbitrarily initialized is provided in~\cite{SR-WR:17} as
\begin{align*}
\dot{x}^i &= \dot{\mathsf{u}}^i-(x^i-\mathsf{u}^i)-k_p\sum_{j\in\mathcal{N}_\text{out}^i} \text{sgn}(x^i-x^j), \qquad x^i(0)\in\real,\quad i\in\until{N},
\end{align*}
or equivalently,
\begin{subequations}\label{eq::alg_FC-YC-WR:X-eq}
\begin{alignat}{2}
\dot{p}^i &= -p^i + k_p\sum_{j\in\mathcal{N}_\text{out}^i} \text{sgn}(x^i-x^j), \qquad &&p^i(0)\in\real, \\
x^i &= \mathsf{u}^i - p^i, &&i\in\until{N}.
\end{alignat}
\end{subequations}
However, this algorithm requires both the reference signals and their derivatives to be bounded with known values $\gamma_1$ and $\gamma_2$: $\sup_{\tau\in[0,\infty)}\|\vectsf{u}(\tau)\|=\gamma_1<\infty$ and $\sup_{\tau\in[0,\infty)}\|\dvectsf{u}(\tau)\|=\gamma_2<\infty$. These values are required to design the proper sliding mode gain $k_p$.

The continuous-time finite-time algorithms described in this section exhibit a sliding mode behavior: in fact, they converge
in finite time to the agreement manifold and then slide on it by switching continuously at an infinite frequency between two system structures. This phenomenon is called chattering. The reader may also recall that commutations at infinite frequency between two subsystems is termed as the
Zeno phenomenon in the literature of hybrid systems. The interested reader can find an enlightening discussion on the relation between first-order sliding mode chattering and Zeno phenomenon in~\cite{Yu2012}. From a practical perspective, chattering  is undesirable and leads to excessive control energy expenditure~\cite{SlotineLi:91}. 
A common approach to eliminate  chattering is to smooth out the control discontinuity in a thin boundary layer around the switching surface. However, this approach leads to a tracking error that is propositional to the thickness of the boundary layer. Another approach to address is the use of higher-order sliding mode control, see~\cite{LF-AL:02} for details. To the best of our knowledge, higher-order sliding mode control has not been used in the context of dynamic average consensus, although there exists results for other networked agreement problems~\cite{EGR-EJO-JGR-ESE-PZ-OG:17}.

\section{Conclusions}\label{sec::Conclusions}

This article has provided an overview of the state of the art on
  the available distributed algorithmic solutions to tackle the
  dynamic average consensus problem. We began by exploring several
  applications of dynamic average consensus in cyber-physical systems,
  including distributed formation control, state estimation, convex
  optimization, and optimal resource allocation. Using dynamic average
  consensus as a backbone, these advanced distributed algorithms
  enable groups of agents to coordinate in order to solve complex
  problems. Starting from the static consensus problem, we
  then derived dynamic average consensus algorithms for various
  scenarios. We first introduced continuous-time algorithms along with
  simple techniques for analyzing them, using block diagrams to
  provide intuition for the algorithmic structure. To reduce the
  communication bandwidth, we then showed how to choose the stepsizes
  to optimize the convergence rate when implemented in discrete time,
  along with how to accelerate the convergence rate by introducing
  extra dynamics. Finally, we showed how to use \textit{a priori}
  information about the reference signals to design algorithms with
  improved tracking performance.

  We hope that the article helps the reader obtain an overview on the
  progress and intricacies of this topic, and appreciate the design
  trade-offs faced when balancing desirable properties for large-scale
  interconnected systems such as convergence rate, steady-state error,
  robustness to initial conditions, internal stability, amount of
  memory required on each agent, and amount of communication between
  neighboring agents.  Given the importance of the ability to track
  the average of time-varying reference signals in network systems, we
  expect the number and breadth of applications for dynamic average
  consensus algorithms to continue to increase in areas such as the
  smart grid, autonomous vehicles, and distributed robotics.

  Many interesting questions and avenues for further research remain
  open. For instance, the emergence of opportunistic
  state-triggered ideas in the control and coordination of networked
  cyber-physical systems presents exciting opportunities for the
  development of novel solutions to the dynamic average consensus
  problem.  The underlying theme of this effort is to abandon the
  paradigm of periodic or continuous sampling/control in exchange for
  deliberate, opportunistic aperiodic sampling/control to improve
  efficiency.  Beyond the brief incursion on this topic in ``Dynamic
  Average Consensus Algorithms with Continuous-Time Evolution and
  Discrete-Time Communication'', further research is needed in
  synthesizing triggering criteria for individual agents that
  prescribe when information is to be shared with or acquired from
  neighbors, that lead to convergence guarantees, and that are
  amenable to the characterization of performance improvements over
  periodic discrete-time implementations. 
  The use of event-triggering   also opens up the way to employing other interesting forms of  communication and computation among the agents when solving the
  dynamic average consensus problem such as, for instance, the
  cloud. In cloud-based coordination, instead of direct peer-to-peer
  communication, agents interact indirectly by opportunistically
  communicating with the cloud to leave messages for other
  agents. These messages can contain information about their current
  estimates, future plans, or fallback strategies.  The use of the
  cloud also opens the possibility of network agents with limited
  capabilities taking advantage of high-performance computation
  capabilities to deal with complex processes.  The time-varying
  nature of the signals available at the individual agents in the
  dynamic average consensus problem raises many interesting challenges
  that need to be addressed to take advantage of this approach. Related to the focus of this effort on the communication aspects, the development of initialization-free  dynamic average consensus algorithms over directed graphs is also another important line of research.

  We believe that the interconnection of dynamic average consensus
  algorithms with other coordination layers in network systems is a
  fertile area for both research and applications.  Dynamic average
  consensus algorithms are a versatile tool in interconnected scenarios
  where it is necessary to compute changing estimates of quantities
  that are employed by other coordination algorithms, and whose
  execution in turn affects the time-varying signals available to the
  individual agents.  We have illustrated this in the ``Applications
  of Dynamic Average Consensus in Network Systems'' section, where we
  described how in resource allocation problems, a group of
  distributed energy resources can collectively estimate the mismatch
  between the aggregated power injection and the desired load using
  dynamic average consensus. The computed mismatch in turn informs the
  distributed energy resources in their decision making process
  seeking to determine the power injections that optimize their
  generation cost, which in turn changes the mismatch computed by the
  dynamic average consensus algorithm.  The fact that the time-varying
  nature of the signals is driven by a dynamic process that itself
  uses the output of the dynamic average consensus algorithms opens
  the way for the use of many concepts germane to systems and control,
  including stable interconnections, input-to-state stability, and
  passivity. Along these lines, we could also think of self-tuning
  mechanisms embedded within dynamic average consensus algorithmic
  solutions that tune the algorithm execution based on the evolution
  of the time-varying signals.

  Another interesting topic for future research is the privacy
  preservation of the signals available to the agents in the dynamic
  average consensus problem.  Protecting privacy and confidentiality
  of data is a critical issue in emerging distributed automated
  systems deployed in a variety of scenarios, including power
  networks, smart transportation, the Internet of Things, and
  manufacturing systems.  In such scenarios, the ability of a network
  system to optimize its operation, fuse information, compute common
  estimates of unknown quantities, and agree on a common worldview
  while protecting sensitive information is crucial.  In this respect,
  the design of privacy-preserving dynamic average consensus
  algorithms is in its infancy. Interestingly, the dynamic nature of
  the problem might offer advantages in this regard with respect to
  the static average consensus problem. For instance, in differential
  privacy, where the designer makes provably difficult for an
  adversary to make inferences about individual records from published
  outputs, or even detect the presence of an individual in the
  dataset, it is known that privacy guarantees weaken as more queries are made to the same database. However, if the database is changing, this limitation no longer applies, and this opens the way to studying how privacy guarantees change with the rate of variation of the time-varying signals in the dynamic average consensus problem.

\bibliographystyle{IEEEtran}
\bibliography{body}

\clearpage


\setcounter{equation}{0}
\renewcommand{\theequation}{S\arabic{equation}}
\setcounter{table}{0} \renewcommand{\thetable}{S\arabic{table}}
\setcounter{figure}{0} \renewcommand{\thefigure}{S\arabic{figure}}

\noindent \textbf{Sidebar: Article Summary}\\
\noindent
This article deals with the dynamic average consensus problem and the distributed coordination algorithms available to solve it.  Such problem arises in scenarios with multiple agents, where each one has access to a time-varying signal of interest (for example, a robot sensor sampling the position of a mobile target of interest or a distributed energy resource taking a sequence of frequency measurements in a microgrid).  The dynamic average consensus problem consists of having the multi-agent network collectively compute the average of the set of time-varying signals. Reasons for pursuing this objective are numerous, and include data fusion, refinement of uncertainty guarantees, and computation of higher-accuracy estimates, all enabling local decision making with network-wide aggregated information. Solving this problem is challenging because the local interactions among agents only involve partial information and the quantity that the network seeks to compute is changing as the agents run their routines.  The article provides a tutorial introduction to distributed methods that solve the dynamic average consensus problems, paying
special attention to the role of network connectivity, incorporating information about the nature of the time-varying signals, the performance trade-offs regarding convergence rate, steady-state error and memory and communication requirements, and algorithm robustness against initialization errors.


\clearpage
\noindent \textbf{Sidebar: Basic Notions from Graph Theory}\\
\noindent
The communication network of a multi-agent cooperative system can be modeled by a \emph{directed graph}, or \emph{digraph}. Here, we briefly review some basic concepts from graph theory following~\cite{FB-JC-SM:09-S}. A digraph is a pair $\GG = (\VV ,\EE )$, where
$\VV=\{1,\dots,N\}$ is the \emph{node set} and $\EE \subseteq
\VV\times \VV$ is the \emph{edge set}.  An edge from $i$ to $j$,
denoted by $(i,j)$, means that agent $j$ can send information to agent
$i$. For an edge $(i,j) \in\EE$, $i$ is called an \emph{in-neighbor}
of $j$, and $j$ is called an \emph{out-neighbor} of $i$. We denote the set of out-neighbors of each agent $i$ by $\NN_{\text{out}}^i$.
A graph is \emph{undirected} if
$(i,j) \in \EE$ anytime $(j,i)\in\EE$.  

A \emph{weighted digraph} is a triplet $\GG = (\VV
,\EE,\vect{\mathsf{A}})$, where $(\VV ,\EE )$ is a digraph and
$\vect{\mathsf{A}}\in\real^{N\times N}$ is a weighted
\emph{adjacency} matrix with the property that $a_{ij} >0$ if $(i, j)
\in\EE$ and $a_{ij} = 0$, otherwise. 
A weighted digraph is
\emph{undirected} if $a_{ij} = a_{ji}$ for all $i,j\in\VV$.  The
\emph{weighted out-degree} and \emph{weighted in-degree} of a node
$i$, are respectively, $\text{d}^{\text{out}}(i) =\sum^N_{j=1} a_{ji}$
and $\text{d}^{\text{in}}(i) = \sum^N_{j=1} a_{ij}$. We let
$\text{d}_{\max}^{\text{out}} = \underset{i \in \until{N}}{\max}
\text{d}^{\text{out}} (i)$  denote the maximum weighted out-degree.
A digraph is \emph{weight-balanced} if at each node $i\in\VV$, the weighted
out-degree and weighted in-degree coincide (although they might be
different across different nodes).  
The out-degree matrix $\vect{\mathsf{D}}^{\text{out}}$ is the diagonal matrix
with entries $\vect{\mathsf{D}}^{\text{out}}_{ii} = \text{d}^{\text{out}}(i)$,
for all $i\in\VV$. The \emph{(out-) Laplacian} matrix is $\lL =
\vect{\mathsf{D}}^{\text{out}} -\vect{\mathsf{A}}$. Note that
$\lL\vect{1}_N=0$. A weighted digraph $\GG$ is weight-balanced if
and only if $\vect{1}_N^\top\lL=0$. Based on the structure of
$\lL$, at least one of the eigenvalues of $\lL$ is zero and
the rest of them have nonnegative real parts. 
We denote the eigenvalues of $\lL$ by $\lambda_i$, $i \in
\until{N}$, where $\lambda_1=0$ and
$\Re(\lambda_i)\leq\Re(\lambda_j)$, for $i<j$. 
For strongly connected digraphs, $\rank(\lL)=N-1$. For strongly connected and weight-balanced digraphs, we denote the
eigenvalues of $\Sym{\lL}=(\lL+\lL^\top)/2$  by $\hat{\lambda}_1, \dots,\hat{\lambda}_N$, where
$\hat{\lambda}_1=0$ and $\hat{\lambda}_i\leq \hat{\lambda}_j$, for $i<j$. 
For strongly connected and weight-balanced digraphs, we have
\begin{equation}\label{eq::RLR}
    0<\hat{\lambda}_2 \vect{I}\leq \rR^\top\Sym{\lL}\rR\leq \hat{\lambda}_N \vect{I},
\end{equation}
where $\rR \in \real^{N \times (N-1)}$ satisfies $[\frac{1}{N}\vect{1}_N~~\rR][\frac{1}{N}\vect{1}_N~~\rR]^\top=[\frac{1}{N}\vect{1}_N~~\rR]^\top[\frac{1}{N}\vect{1}_N~~\rR]=\vect{I}_N$. Notice that for connected graphs, $\Sym{\lL}=\lL$, and consequently $\lambda_i=\hat{\lambda}_i$, for all $i\in\VV$.
\begin{figure}[h]
\begin{subfigure}{0.5\textwidth}
\centering  \begin{tikzpicture}[auto,thick,scale=1.1, every node/.style={scale=1.1}]
     \node (1) at (-0.50,-0.1) [ draw, minimum size=10pt,color=blue, circle, very thick, fill=blue!10] {{\scriptsize 1}};
      \node (2) at (0.5,1.1) [ draw, minimum size=10pt,color=blue, circle,very thick, fill=blue!10] {{\scriptsize 2}};
       \node (3) at (1.6,0.1) [ draw, minimum size=10pt,color=blue, circle, very thick, fill=blue!10] {{\scriptsize 3}};
    \node (4) at (2.8,1.2) [ draw, minimum size=10pt,color=blue, circle, very thick, fill=blue!10] {{\scriptsize 4}};
     \path[draw, -latex']  (2)--node[above]{{\scriptsize 1}}(1) ;
           \path[draw, -latex']  (1)--node[below]{{\scriptsize 1}}(3);
      \path[draw, -latex']  (3)--node[above]{{\scriptsize 2}}(2) ;
     \path[draw, -latex']  (2)--node[above]{{\scriptsize 1}}(4) ;
     \path[draw, -latex']  (4)--node[below]{{\scriptsize 1}}(3) ;
\end{tikzpicture}
\caption{Strongly connected, weight-balanced digraph:
\\ $\vectsf{A}=\left[\begin{smallmatrix}0\,&0\,&1\,&0\\
1\,&0\,&0\,&1\\
0\,&2\,&0\,&0\\
0\,&0\,&1\,&0\end{smallmatrix}\right]$,~ $\vectsf{L}=\left[\begin{smallmatrix}1&0&-1&0\\
-1&2&0&-1\\
0&-2&2&0\\
0&0&-1&1\end{smallmatrix}\right]$.}
\end{subfigure}%
\qquad
\begin{subfigure}{0.5\textwidth}
\centering  \begin{tikzpicture}[auto,thick,scale=1.1, every node/.style={scale=1.1}]
      \node (1) at (-0.50,-0.1) [ draw, minimum size=10pt,color=blue, circle, very thick, fill=blue!10] {{\scriptsize 1}};
      \node (2) at (0.5,1.1) [ draw, minimum size=10pt,color=blue, circle,very thick, fill=blue!10] {{\scriptsize 2}};
       \node (3) at (1.6,0.1)[ draw, minimum size=10pt,color=blue, circle, very thick, fill=blue!10] {{\scriptsize 3}};
    \node (4) at (2.8,1.2) [ draw, minimum size=10pt,color=blue, circle, very thick, fill=blue!10] {{\scriptsize 4}};
     \path[draw, -latex']  (2)--(1) ;
           \path[draw, -latex']  (1)--(3) ;
      \path[draw, -latex']  (3)--(2) ;
     \path[draw, -latex']  (2)--(4) ;
     \path[draw, -latex']  (4)--(3) ;
      \path[draw, -latex']  (1)--(2) ;
           \path[draw, -latex']  (3)--(1) ;
      \path[draw, -latex']  (2)--(3) ;
     \path[draw, -latex']  (4)--(2) ;
     \path[draw, -latex']  (3)--(4) ;
\end{tikzpicture}
\caption{Connected graph with unit edge weights:
\\ 
$ \vectsf{A}=\left[\begin{smallmatrix}
0\,&1\,&1\,&0\\
1\,&0\,&1\,&1\\
1\,&1\,&0\,&1\\
0\,&1\,&1\,&0\end{smallmatrix}\right]$,~ $\vectsf{L}=\left[\begin{smallmatrix}
2&-1&-1&0\\
-1&3&-1&-1\\
-1&-1&3&-1\\
0&-1&-1&2\end{smallmatrix}\right]$.}
\end{subfigure}
\caption{Examples of directed and undirected graphs.}
\label{Fig::graphs-directed-undirected}
\end{figure}

Intuitively, the Laplacian matrix can be viewed as a diffusion operator over the graph. To illustrate this, suppose each agent $i\in\VV$ has a scalar variable $x_i\in\real$. Stacking the variables into a vector $\vect{x}$, multiplication by the Laplacian matrix gives the weighted sum
\begin{equation}
    [\lL\vect{x}]_i = \sum_{j\in\VV} a_{ij}\,(x_i-x_j)
\end{equation}
where $a_{ij}$ is the weight of the link between agents $i$ and $j$.


\clearpage

\noindent \textbf{Sidebar: Input-to-State Stability of LTI Systems}\\
\noindent
For a linear time-invariant system
\begin{align}
\dvect{x}=\vect{A}\vect{x}+\vect{B}\vect{u},\quad\vect{x}\in\real^n,~\vect{u}\in\real^m,
\end{align}
we can write the solution  for $t\in\realnonnegative$ as 
\begin{align}\label{eq::LTI-sys}
\vect{x}(t)=\text{e}^{\vect{A}\,t}\,\vect{x}(0)+\int_{0}^t\text{e}^{\vect{A}(t-\tau)}\,\vect{B}\,\vect{u}(\tau)\text{d}\tau.
\end{align}
For a Hurwitz matrix $\vect{A}$, by using the bound 
\begin{align}\label{eq::exp_bound_LTI}
\|\text{e}^{\vect{A}\,t}\|\leq \kappa\, \text{e}^{-\underline{\lambda}\,t}, \quad t\in\realnonnegative, 
\end{align}
for some $\kappa,\underline{\lambda}\in\realpositive$, we can establish an upper bound on the norm of the trajectories of~\eqref{eq::LTI-sys} as follows
\begin{align}
\|\vect{x}(t)\|&\leq \kappa\, \text{e}^{-\underline{\lambda}\,t}\,\|\vect{x}(0)\|+\int_{0}^t\kappa\, \text{e}^{-\underline{\lambda}\,(t-\tau)}\|\vect{B}\|\,\|\vect{u}(\tau)\|\text{d}\tau,\nonumber\\
&\leq \kappa\, \text{e}^{-\underline{\lambda}\,t}\,\|\vect{x}(0)\|+\frac{\kappa\, \|\vect{B}\|}{\underline{\lambda}} \sup_{0\leq \tau\leq t}\|\vect{u}(\tau)\|,\quad\quad\forall t\in\realnonnegative.
\end{align}
The bound here shows that the zero-input response decays to zero exponentially fast, while the zero-state response is bounded for every bounded input, indicating an input-to-state stability (ISS) behavior. It is worth noticing that the ultimate bound on the system state is proportional to the bound on the input. 

Next, we comment on how to compute the parameters $\kappa,\underline{\lambda}\in\realpositive$ in~\eqref{eq::exp_bound_LTI}. Recall that~\cite[Fact 11.15.5]{DSB:05-S} for any matrix $\vect{A}\in\real^{n\times n}$, we can write 
\begin{align}\label{eq::matrix_exp_norm}
\|\text{e}^{\vect{A}\,t}\|\leq \text{e}^{\lambda_{\max}(\Sym{ \vect{A}})\,t}.\quad \forall \,t\in\realnonnegative
\end{align}
where $\Sym{\vect{A}}=\frac{1}{2}(\vect{A}+\vect{A}^\top)$. Therefore, for a Hurwitz matrix $\vect{A}$ whose $\Sym{\vect{A}}$ is also Hurwitz, the exponential bound parameters can be set to
\begin{align}\label{eq::LTI_exp_k_lam_sym}
    \underline{\lambda}=-\lambda_{\max}(\Sym{ \vect{A}}),\quad \kappa=1.
\end{align}
A tighter exponential bound of 
\begin{align}\label{eq::LTI_exp_k_lam_tight}
  \underline{\lambda}=\lambda^\star,\quad \kappa=\sqrt{\sigma_{\max}(\vect{P}^\star)/\sigma_{\min}(\vect{P}^\star)},
\end{align}
can also be obtained for any Hurwitz, according to~\cite[Proposition~5.5.33]{DH-AJP:05-S}, from the convex linear matrix inequality optimization problem 
\begin{subequations}\label{eq::matrix_exp_norm-tighter-Lya}
\begin{align}(\mathsf{\lambda}^\star,\vect{P}^\star)=&\argmin{\mathsf{\lambda}}\quad \text{s.t.}\\
&\vect{P}\vect{A}+\vect{A}^\top\vect{P}\leq -2\,{\mathsf{\lambda}}\,\vect{P},\quad \vect{P}>0,\quad\mathsf{\lambda}>0.\end{align}
\end{subequations}

Similarly, the state of the discrete-time linear time-invariant system
\begin{align}
\vect{x}_{k+1} = \vect{A} \vect{x}_k + \vect{B} \vect{u}_k, \quad \vect{x}_k\in\real^n, ~ \vect{u}_k\in\real^m
\end{align}
with initial condition $\vect{x}_0\in\real^n$ satisfies the bound
\begin{align}
\|\vect{x}_k\| &\leq \sqrt{\frac{\sigma_\text{max}(\vect{P})}{\sigma_\text{min}(\vect{P})}} \Bigl(\rho^k \|\vect{x}_0\| + \frac{1-\rho^k}{1-\rho} \|\vect{B}\|\sup_{0\leq j<k} \|\vect{u}_j\|\Bigr)
\end{align}
where $\vect{P}\in\real^{n\times n}$ and $\rho\in\real$ satisfy
\begin{align}
\vect{A}^\top \vect{P} \vect{A} - \rho^2\,\vect{P}\leq 0, \quad \vect{P}>0, \quad \rho\geq 0.
\end{align}



\clearpage
\noindent \textbf{Sidebar: Further Reading}\\
\noindent

Numerous works have studied the robustness of dynamic average consensus algorithms against a variety of disturbances and sources of error present in practical scenarios. These include fixed communication delays~\cite{HM-SSK:17-S}, additive input disturbances~\cite{GS-KHJ:13-S}, time-varying communication graphs~\cite{BV-RAF-KML:14-S}, and driving command saturation~\cite{SSK-JC-SM:15-ijrnc}. Variations of the dynamic average consensus problems explore scenarios where the algorithm design depends on the specific agent dynamics~\cite{FC-GF-LL-WR:15-S,SG-WR-FC-YS:17-S,SR-WR:17} or incorporates different agent roles, such as in leader-follower networks of mobile agents~\cite{WR:07-S,GS-YH-KHJ:12-S,ZM-DVD-KHJ:14-S}.

When dealing with directed agent interactions, a common assumption in solving the average consensus problem is that the communication graph is \emph{weight-balanced}, which is equivalent to the graph consensus matrix $\vect{W}\defeq \vect{I}-\lL$ being doubly stochastic. In~\cite{JMH-JNT:15}, it is shown in fact that calculating an average over a network requires either explicit or implicit use of either (1) the out-degree of each agent, (2) global node identifiers, (3) randomization, or (4) asynchronous updates with specific properties. In particular, the balanced assumption is necessary for scalable, deterministic, synchronous algorithms. In general, agents may not have access to their out-degree (for example, agents which use local broadcast communication). If each agent knows its out-degree, however, then distributed algorithms may be used to generate weight-balanced and doubly stochastic digraphs~\cite{BG-JC:09c-S,AR-TC-CNH:14-S}. Another approach is to explicitly use the out-degree in the algorithm by having agents share their out-weights and use them to adjust for the imbalances in the graph; this approach is referred to as the \emph{push-sum} protocol and has been applied to the static average consensus problem (see~\cite{FB-VB-PT-JT-MV:10-S,ADD-CNH:13-S,AN-AO:15-S,PR-BG-TL-BT:17-S}). Both of these approaches of dealing with unbalanced graphs require each agent to know its out-degree.

Furthermore, when communication links are time-varying, these approaches only work if the time-varying graph remains weight-balanced, see~\cite{SSK-JC-SM:15-ijrnc,SSK-JC-SM:15-auto2-S}. If communication failures caused by limited communication ranges or external events such as obstacle blocking destroy the weight-balanced character of the graph, it is still possible to solve the dynamic average consensus problem if the expected graph is balanced~\cite{BV-RAF-KML:14-S}.
Another set of works have explored the question of how to optimize the graph topology to endow consensus algorithms with better properties. These include designing the network topology in the presence of random link failures~\cite{SK-JMFM:08} and optimizing the edge weights for fast consensus~\cite{LX-SB:03,PY-RAF-KML:06}.


\clearpage

\noindent \textbf{Sidebar: Euler Discretizations of Continuous-Time Dynamic Average Consensus Algorithms}\\
\noindent
The continuous-time algorithms described in the article can also give rise to discrete-time strategies. Here we describe how to discretize them so that they are implementable over wireless communication channels. This can be done by using the (forward) Euler discretization of the derivatives,
\begin{align*}
  \dot{\vect{x}}(t)\approx\frac{\vect{x}(k+1)-\vect{x}(k)}{\delta},
\end{align*}
where $\delta\in\realpositive$ is the stepsize. To illustrate the discussion, we develop this approach for the  algorithm~\eqref{eq::alg-SSK-JC-SM:15-ijrnc-ct} over a connected graph topology. The discussion below can also be extended to include iterative forms of the other continuous-time algorithms studied in the paper.  Using the Euler discretization in the algorithm~\eqref{eq::alg-SSK-JC-SM:15-ijrnc-ct} leads to 
\begin{subequations}\label{eq::DCDisc_smlr}
  \begin{align}
   v^i(k+1) &=v^i(k)+\delta\alpha\beta
 \sum\nolimits_{j=1}^{N}{a}_{ij}({x}^i(k)\!-\!{x}^j(k)), \label{eq::SSK-JC-SM:15-ijrnc-dt-alg-vvi}\\
     x^i(k+1) &= x^i(k)+\Delta \mathsf{u}^i(k)-\delta\alpha
    (x^i(k)-\mathsf{u}^i(k))-\delta\beta\sum\nolimits_{j=1}^{N}\!\!{a}_{ij}({x}^i(k)-{x}^j(k))-\delta 
    v^i(k),\label{eq::SSK-JC-SM:15-ijrnc-dt-alg-xxi}
  \end{align}
\end{subequations}
where $\Delta \mathsf{u}^i(k) =\mathsf{u}^i(k+1)-\mathsf{u}^i(k)$. To implement this iterative form at each timestep $k$ we need to have access to the future value of the reference input at timestep $k+1$. Such a requirement is not practical when the reference input is sampled from a physical process or is a result of another on-line algorithm. We could circumvent this requirement using a backward Euler discretization, but the resulting algorithm tracks the reference dynamic average with one-step delay. A practical solution which avoids requiring the future values of the reference input is obtained by introducing an intermediate variable $z^i(k)=x^i(k)-\mathsf{u}^i(k)$ and representing the iterative algorithm~\eqref{eq::DCDisc_smlr} in the form 
\begin{subequations}\label{eq::SSK-JC-SM:15-ijrnc-dt-alg}
  \begin{align}
   v^i(k+1) &=v^i(k)+\delta\alpha\beta
 \sum\nolimits_{j=1}^{N}{a}_{ij}({x}^i(k)\!-\!{x}^j(k)), \label{eq::SSK-JC-SM:15-ijrnc-dt-alg-vi}\\
     z^i(k+1) &= z^i(k)-\delta\alpha
    z^i(k)-\delta\beta\sum\nolimits_{j=1}^{N}\!\!{a}_{ij}({x}^i(k)-{x}^j(k))-\delta
    v^i(k),\label{eq::SSK-JC-SM:15-ijrnc-dt-alg-zi}
    \\
    x^i(k)&= z^i(k)+\mathsf{u}^i(k), \label{eq::SSK-JC-SM:15-ijrnc-dt-alg-xi}
  \end{align}
\end{subequations}
for $i\in\until{N}$. Algorithm~\eqref{eq::SSK-JC-SM:15-ijrnc-dt-alg} is then implementable without the use of future inputs.

The question then is to characterize the adequate stepsizes that guarantee that the convergence properties of the continuous-time algorithm are retained by its discrete implementation. Intuitively, the smaller the stepsize, the better for this purpose. However, this also requires more communication. To ascertain this issue, the following result is particularly useful.

\begin{lem}[Admissible stepsize for Euler discretized form of LTI systems and a bound on their trajectories]\label{lem::discret-lemma}
Consider
\begin{align*}
  \dvect{x}=\vect{A}\vect{x}+\vect{B}\vect{u},\quad t\in\real_{\geq0},
\end{align*}
and its Euler discretized iterative form
\begin{align}\label{eq::Euler_A}
\vect{x}(k+1)=(\vect{I}+\delta\vect{A})\,\vect{x}(k)+\delta\vect{B}\,\vect{u}(k),\quad k\in\mathbb{Z}_{\geq0},
\end{align}
where $\vect{x}\in\real^n$ and $\vect{u}\in\real^m$ are, respectively,  state and input vectors, and $\delta\in\real_{>0}$ is the discretization stepsize. Let the system matrix $\vect{A}=[a_{ij}]\in\real^{n\times n}$ be a Hurwitz matrix with eigenvalues $\{\mu_i\}_{i=1}^n$, and the difference of input signal be  bounded, $\|\Delta\vect{u}\|<\rho<\infty$.
Then, for any $\delta\in(0,\bar{d})$ where 
\begin{align}\label{eq::d_bound}
\bar{d}=\min\Big\{-2\,\frac{\re{\mu_i}}{|\mu_i|^2}\Big\}_{i=1}^n
\end{align}
the eigenvalues of $(\vect{I}+\delta\vect{A})$ are all located inside the until circle in complex plane. Moreover, starting from any $\vect{x}(0)\in\real^n$, the trajectories of \eqref{eq::Euler_A} satisfy
\begin{align}
\lim_{k\to\infty}\|\vect{x}(k+1)\|\leq \frac{\kappa\,\rho\,\|\vect{B}\|}{1-\omega},
\end{align}
where $\omega\in(0,1)$, and $\kappa\in\real>0$ such that $\|\vect{I}+\delta\vect{A}\|^k\leq\kappa\,\omega^k$.
\end{lem}
%

The bounds $\omega\in(0,1)$ and $\kappa\in\real_{>0}$ in   $\|\vect{I}+\delta\vect{A}\|^k\leq \kappa\,\omega^k$ when all the eigenvalues of  $\vect{I}+\delta\vect{A}$ are located in the unit circle of the complex plane can be obtained from the following LMI optimization problem (see~\cite[Theorem 23.3]{WJR:93-S} for details)
  \begin{align}\label{eq::LMI-disc-bound}
  ({\omega},\kappa,\vect{Q})=&\argmin\omega^2,\quad\text{subject to }\\
&\frac{1}{{\kappa}} \,\vect{I}\leq \vect{Q}\leq \vect{I},\quad 0<\omega^2<1,\quad \kappa>1,\nonumber\\
&(\vect{I}+\delta\vect{A})^\top\vect{Q}\,(\vect{I}+\delta\vect{A})-\vect{Q}\leq -(1-\omega^2)\,\vect{I}.\nonumber
\end{align}

Building on Lemma~\ref{lem::discret-lemma}, the next result characterizes the admissible discretization stepsize for the algorithm~\eqref{eq::SSK-JC-SM:15-ijrnc-dt-alg} and its ultimate tracking behavior.

\begin{theorem}[Convergence of~\eqref{eq::SSK-JC-SM:15-ijrnc-dt-alg} over connected graphs~\cite{SSK-JC-SM:15-ijrnc}]\label{thm::DCDiscProb1Sol}
  Let $\GG$ be a connected, undirected graph. Assume that
  the differences of the inputs of the network satisfy
  $\max_{k\in\mathbb{Z}_{\geq0}}\|(\vect{I}-\frac{1}{N}\vect{1}_N\vect{1}_N^\top)\,\Delta\vectsf{u}(k)\| = \gamma<\infty$. Then,
  for any $\alpha,\beta>0$, the algorithm~\eqref{eq::SSK-JC-SM:15-ijrnc-dt-alg}
  over~$\GG$ initialized at $z^i(0)\in\real$ and $v^i(0)\in \real$
  such that $\SUM{i=1}{N}v^i(0)=0$ has bounded trajectories that satisfy 
  \begin{align}\label{eq::SSK-JC-SM:15-ijrnc-dt-alg-disc-bound}
    \lim_{k\to\infty} \left| x^i(k)-\mathsf{u}^\text{avg}(k)\right| \leq
  \frac{\delta \,\kappa\,\gamma }{1-\omega},\quad i \in \until{N}
  \end{align}
  provided $\delta \in(0,
  \min\{\alpha^{-1},2\,\beta^{-1}(\lambda_N)^{-1}\})$. Here, $\lambda_N$ is the largest eigenvalue of the Laplacian, and $\omega\in(0,1)$ and ${\kappa}\in\real_{>0}$ satisfy  $\|\vect{I}-\delta\,\beta\,\rR^\top\lL\rR\|^k\leq{\kappa}\,{\omega}^k$, $k\in\mathbb{Z}_{\geq0}$. 
\end{theorem}

Note that the characterization of the stepsize requires knowledge of the largest eigenvalue $\lambda_N$ of the Laplacian. Since such knowledge is not readily available to the network unless dedicated distributed algorithms are introduced to compute it,~\cite{SSK-JC-SM:15-ijrnc} provides the sufficient characterization $\delta \in(0,
  \min\{\alpha^{-1},\beta^{-1}(\dout^\text{max})^{-1}\})$ along with the ultimate tracking bound
\begin{equation*}
    \lim_{k\to\infty} \left| x^i(k)-\mathsf{u}^\text{avg}(k)\right| \leq
  \frac{\delta \gamma }{\beta\,\lambda_2},\quad i \in \until{N}.
\end{equation*}

\clearpage

%
\noindent \textbf{Sidebar: Dynamic Average Consensus Algorithms with Continuous-Time Evolution and Discrete-Time Communication}\\
\noindent

We discuss here an alternative to the discretization route explained in ``Euler Discretizations of Continuous-Time Dynamic Average Consensus Algorithms" to produce implementable strategies from the continuous-time algorithms described in the paper. This approach is based on the observation that, when implementing the algorithms over digital platforms, computation can still be reasonably approximated by continuous-time evolution (given the every growing computing capabilities of modern embedded processors and computers) whereas communication is a process that still requires proper acknowledgment of its discrete-time nature. Our basic idea is to opportunistically trigger, based on the network state, the times for information sharing among agents to take place, and to allow individual agents to determine these autonomously. This has the potential to result in more efficient algorithm implementations as performing communication usually requires more energy than computation~\cite{HK-AW:05-S}.  In addition, the use of fixed communication stepsizes can lead to a wasteful use of the network resources because of the need to select it taking into account worst-case scenarios. These observations are aligned with the ongoing research activity~\cite{WPMHH-KHJ-PT:12-S,LH-CF-HO-AS-EF-JR-SIN:17-S} on  event-triggered control and aperiodic sampling for controlled dynamical systems that seeks to trade computation and decision
making for less communication,
sensing or actuator effort while still guaranteeing a desired
level of performance. The surveys~\cite{CN-EG-JC:17-S,LD-QLH-XG-XMZ:18-S} describe how these ideas  
can be employed to design event-triggered communication laws for static average consensus.

Motivated by these observations,~\cite{SSK-JC-SM:15-auto2-S} investigates a discrete-time communication implementation of the continuous-time algorithm~\eqref{eq::alg-SSK-JC-SM:15-ijrnc-ct} for dynamic average consensus. Under this strategy the algorithm becomes
\begin{subequations}\label{eq::EventTrig_Alg}
  \begin{align}
    \dot{v}^i & =\alpha\beta\sum\nolimits_{j=1}^N
    \mathsf{a}_{ij}(\hat{x}^i-\hat{x}^j), \label{eq::EventTrig_Alg-a}
    \\
    \dot{x}^i &
    =\dot{\mathsf{u}}^i\!-\!\alpha(x^i-\mathsf{u}^i)\!-\!\beta\sum\nolimits_{j=1}^N
    \mathsf{a}_{ij}(\hat{x}^i\!-\!\hat{x}^j)\!-\!v^i,
   \label{eq::EventTrig_Alg-b}
  \end{align}
\end{subequations}
for each $i\in\until{N}$, where $\hat{x}^i (t) = x^i(t^i_{k})$ for
$t\!\in\![t^i_{k},t^i_{k+1})$, with $\{t^i_{k}\} \subset \realnonnegative$ denoting the sequence of times at which agent $i$ communicates with its in-neighbors. The basic idea is that agents share their information with neighbors when the uncertainty in the outdated information is such that the monotonic convergent behavior of the overall network can no longer be guaranteed. The design of such triggers is challenging because of the following requirements:
(a) triggers need to be distributed, so that agents can check them with the information available to them from their out-neighbors, (b) they must guarantee the absence of Zeno behavior (the undesirable situation where an infinite number of communication rounds are triggered in a finite amount of time), and (c) they have to ensure the network achieves dynamic average consensus even though agents operate with outdated information while inputs are changing with~time.

Consider the following event-triggered communication law~\cite{SSK-JC-SM:15-auto2-S}: each agent is to communicate with its in-neighbors at times
$\{t^i_{k}\}_{k \in \naturals} \subset \realnonnegative$, starting at $t^i_1=0$, determined by
\begin{align}\label{eq::TrigLaw_Distributed_Directed}
    t^i_{k+1} \!=\! \argmax \setdef{t \in [t^i_{k},\infty)\!}{\!
      |{x}^i(t^i_{k})-x^i(t)|\leq \eps_i }.
\end{align}
Here, $\eps_i\in\realpositive$ is a constant value which each agent chooses locally to control its inter-event times and avoid Zeno behavior. Specifically, the inter-execution times of each agent $i\in\until{N}$ employing~\eqref{eq::TrigLaw_Distributed_Directed} are lower bounded by
\begin{equation}\label{eq::dist_event_time_lw_bnd}
    \tau^i = \frac{1}{\alpha} \ln \Big(1+\frac{\alpha
      \eps_i}{c^i} \Big) ,
\end{equation}
where $c^i$ and $\eta$ are positive real numbers that depend on the initial conditions and network parameters (we omit here for simplicity their specific form, but the interested reader is referred to~\cite{SSK-JC-SM:15-auto2-S} for explicit expressions).
The lower bound~\eqref{eq::dist_event_time_lw_bnd} shows that for a positive nonzero $\eps^i$, the inter-execution times are bounded away from zero and we have the guarantees that for networks with finite number of agents the implementation of the algorithm~\eqref{eq::EventTrig_Alg} with the communication trigger law~\eqref{eq::TrigLaw_Distributed_Directed} is Zeno free. The following result formally describes the convergence behavior of the algorithm~\eqref{eq::EventTrig_Alg} under~\eqref{eq::TrigLaw_Distributed_Directed}  When the interaction topology is modeled by a strongly connected and weight-balanced digraph.

\begin{theorem}[Convergence of~\eqref{eq::EventTrig_Alg} over
   strongly connected and weight-balanced digraph
  with asynchronous distributed event-triggered
  communication~\cite{SSK-JC-SM:15-auto2-S}]\label{thm::Alg_D_Event}
  Let $\GG$ be a strongly connected and weight-balanced digraph.
  Assume the reference signals satisfy
  $\sup_{t\in[0,\infty)}|\dot{\mathsf{u}}^i(t)|= \kappa^i<\infty$, for  $i\in\until{N}$, and  $\sup_{t\in[0,\infty)}\|\vect{\Pi}_N\dvectsf{u}(t)\| =
  \gamma<\infty$.  For any $\alpha, \beta \in \realpositive$, the algorithm~\eqref{eq::EventTrig_Alg} over $\GG$ starting from $x^i(0)\in\reals$ and $v^i(0)\in\reals$ with $\sum_{i=1}^Nv^i(0)=0$, where each agent $i \in\until{N}$ communicates with its neighbors at times $\{t^i_{k}\}_{k \in \naturals} \subset \realnonnegative$,  starting at $t^i_1=0$,  determined by~\eqref{eq::TrigLaw_Distributed_Directed} with  $\vect{\eps} \in
  \realpositive^N$, 
  satisfies
  \begin{align}\label{eq::Alg_D_Event_ultimate_bound}
     \limsup_ {t\to\infty}\! \Big|x^i(t)\!-\!\mathsf{u}^{\text{avg}}(t) \Big|
    \!\leq \!\frac{\gamma\!+\!\beta
      \|\lL\|\,\|\vect{\eps}\|}{\beta{\Hlambda}_{2}} ,
  \end{align}
  for $ i \in\until{N}$ with an exponential rate of convergence of
  $\min\{\alpha,\beta{\Hlambda}_2\}$. Furthermore, the
  inter-execution times of agent $i\in\until{N}$ are lower bounded by~\eqref{eq::dist_event_time_lw_bnd}.
\end{theorem}

The expected trade-off between the desire for longer inter-event time and the adverse effect on systems convergence and performance is captured in~\eqref{eq::dist_event_time_lw_bnd} and~\eqref{eq::Alg_D_Event_ultimate_bound}. The lower bound $\tau^i$ in~\eqref{eq::dist_event_time_lw_bnd} 
on the inter-event times allows a designer to compute bounds on the maximum number of communication rounds (and associated energy
spent) by each agent $i\in \until{N}$ (and hence the network) during
any given time interval. It is interesting to analyze how this
lower bound depends on the various problem ingredients: $\tau^i$
is an increasing function of $\eps_i$ and a decreasing function of
$\alpha$ and $c^i$. Through the latter variable, the bound also depends on the graph topology and the design parameter $\beta$. Given the definition of $c^i$, we can deduce that the faster an
input of an agent is changing (larger $\kappa^i$) or the farther the agent initially starts from the average of the inputs, the more often that agent would need to trigger communication.  The connection between the network performance and the communication overhead can also be observed here. Increasing $\beta$ or
decreasing $\eps_i$ to improve the ultimate tracking error
bound~\eqref{eq::Alg_D_Event_ultimate_bound} results in smaller
inter-event times.  Given that the rate of convergence
of~\eqref{eq::EventTrig_Alg} under~\eqref{eq::TrigLaw_Distributed_Directed} is~$\min\{\alpha,\beta\Hlambda_{2}\}$, decreasing $\alpha$ to increase the inter-event times slows down the convergence.

\begin{figure}[ht]
\begin{subfigure}{0.45\textwidth}
\centering  \includegraphics[trim=0mm 0mm 0mm 0mm,clip,width=3.2in, height=1.3in]{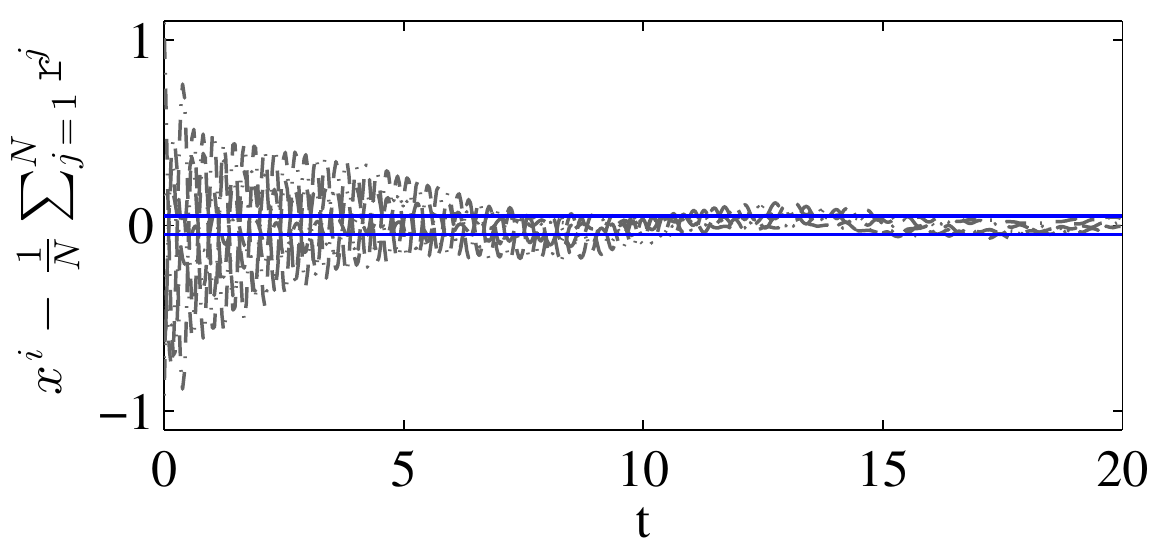}
\caption{}
\end{subfigure}\quad\quad
\begin{subfigure}{0.5\textwidth}
\centering  \includegraphics[trim=0mm 0mm 0mm 0mm,clip,width=3.2in, height=1.3in]{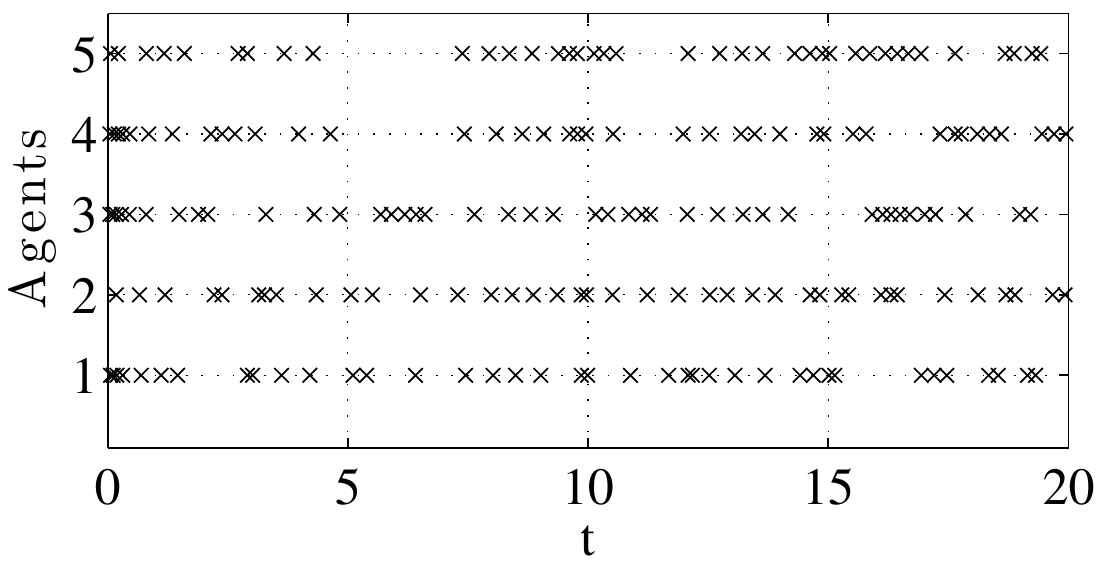}
\caption{}
\end{subfigure}
\caption{Comparison between the event-triggered  algorithm~\eqref{eq::EventTrig_Alg}
    employing the event-triggered communication law~\eqref{eq::TrigLaw_Distributed_UD} and the Euler
    discretized implementation of  algorithm~\eqref{eq::alg-SSK-JC-SM:15-ijrnc-ct} as described in~\eqref{eq::SSK-JC-SM:15-ijrnc-dt-alg} with fixed stepsize~\cite{SSK-JC-SM:15-auto2-S}. Both of these algorithms use~$\alpha=1$ and $\beta=4$. The network is a
    weight-balanced digraph of $5$ agents with unit
    weights. The inputs are $\mathsf{r}^1(t)\!=\!0.5\sin(0.8t)$,
    $\mathsf{r}^2(t)\!=\!0.5\sin(0.7t)\!+\!0.5\cos(0.6t)$,
    $\mathsf{r}^3(t)=\sin(0.2t)\!+\!1$,
    $\mathsf{r}^4(t)\!=\!\text{atan}(0.5t)$,
    $\mathsf{r}^5(t)\!=\!0.1\cos(2t)$. In plot (a),
    the black (resp. gray) lines correspond to the tracking error of the event-triggered algorithm~\eqref{eq::EventTrig_Alg} employing event-triggered law~\eqref{eq::TrigLaw_Distributed_UD} with $\eps_i/(2\sqrt{\dout^i})=0.1$ (resp. the Euler discretized algorithm~\eqref{eq::SSK-JC-SM:15-ijrnc-dt-alg} with fixed stepsize $\delta=0.12$). Recall from ``Euler Discretizations of Continuous-Time Dynamic Average Consensus Algorithms" that convergence for algorithm~\eqref{eq::SSK-JC-SM:15-ijrnc-dt-alg} is guaranteed if $\delta \in (0,\min\{\alpha^{-1},\beta^{-1}(\text{d}_{\max}^{\text{out}})^{-1}\})$, which for this example results in $\delta\in(0,0.125)$. The horizontal blue lines show the $\pm 0.05$ error bound for reference. Plot (b) shows the communication times of each agent using the event-triggered strategy. As seen in plot (a), both these algorithms exhibit comparable tracking performance. The number of times that agents $\{1,2,3,4,5\}$ communicate in the time interval $[0,20]$ is $(39, 40,42, 40,39)$, respectively, when implementing event-triggered communication~\eqref{eq::TrigLaw_Distributed_UD}. These numbers are significantly smaller than the communication rounds required by each agent in the Euler discretized algorithm~\eqref{eq::SSK-JC-SM:15-ijrnc-dt-alg} ($20/0.12 \simeq 166$ rounds).}
\label{Event-triger-Examples}
\end{figure}

When the interaction topology is a connected graph, the properties of the  Laplacian allow us to identify an alternative event-triggered communication law which, compared to~\eqref{eq::TrigLaw_Distributed_Directed}, has a longer inter-event time but similar dynamic average tracking performance. Now, consider the sequence of communication times $\{t^i_{k}\}_{k \in \naturals}$ determined by
 \begin{align}\label{eq::TrigLaw_Distributed_UD}
    t^i_{k+1}  = \argmax \setdef{&t \in
      [t^i_{k},\infty)}{|\hat{x}^i(t) -x^i(t) |^2
   \leq\\
      & \quad\frac{1}{4 \dout^i } \sum_{j=1}^N
      {\mathsf{a}}_{ij}(t)|\hat{x}^i(t) - \hat{x}^j(t)|^2 + \frac{1}{4\dout^i }\eps_i^2}, \nonumber
\end{align}
Compared to~\eqref{eq::TrigLaw_Distributed_Directed}, the extra term $\frac{1}{4 \dout^i } \sum_{j=1}^N {\mathsf{a}}_{ij}(t)|\hat{x}^i(t) - \hat{x}^j(t)|^2$ in the communication law~\eqref{eq::TrigLaw_Distributed_UD} allows agents to have longer inter-event times.
Formally, the inter-execution times of agent $i\in\until{N}$ implementing~\eqref{eq::TrigLaw_Distributed_UD} are lower bounded by 
\begin{equation}\label{eq::dist_event_time_lw_bnd_undirected}
    \tau^i = \frac{1}{\alpha} \ln \Big(1+\frac{\alpha
      \eps_i}{2\bar{c}^i\sqrt{\dout^i}} \Big) ,
\end{equation}
for positive constants $\bar{c}^i$, see~\cite{SSK-JC-SM:15-auto2-S} for explicit expressions.
Numerical examples in~\cite{SSK-JC-SM:15-auto2-S} show that the implementation of~\eqref{eq::TrigLaw_Distributed_UD} for connected graphs results in inter-event times longer than the ones of the event-triggered law~\eqref{eq::TrigLaw_Distributed_Directed}. 
Figure~\ref{Event-triger-Examples} shows one of those examples.  Similar results can also be derived for time-varying, jointly connected graphs, see~\cite{SSK-JC-SM:15-auto2-S} for a complete exposition.



\clearpage

\subsection{Authors Information} 

\emph{\textbf{Solmaz S. Kia}} is an Assistant Professor of Mechanical
and Aerospace Engineering at the University of California, Irvine, CA,
USA. She obtained her Ph.D. degree in Mechanical and Aerospace
Engineering from UCI, in 2009, and her M.Sc. and B.Sc. in Aerospace
Engineering from the Sharif University of Technology, Iran, in 2004
and 2001, respectively. She was a senior research engineer at SySense
Inc., El Segundo, CA from Jun. 2009 to Sep. 2010. She held
postdoctoral positions in the Department of Mechanical and Aerospace
Engineering at the UC San Diego and UCI. She was a recipient of UC
president's postdoctoral fellowship in 2012-2014 and NSF CAREER award
in 2017. Her main research interests, in a broad sense, include
distributed optimization/coordination/estimation, nonlinear control
theory, and probabilistic robotics.

\emph{\textbf{Bryan Van Scoy}} is a postdoctoral researcher at the
University of Wisconsin--Madison, WI, USA. He received the Ph.D. degree
in Electrical Engineering and Computer Science from Northwestern University
in 2017, and B.S. and M.S. degrees in Applied Mathematics along with a B.S.E.
in Electrical Engineering from the University of Akron in 2012. His
research interests include distributed algorithms for multi-agent
systems, and the analysis and design of optimization algorithms.

\emph{\textbf{Jorge Cort\'es}} (M'02-SM'06-F'14) is a Professor of
Mechanical and Aerospace Engineering at the University of California,
San Diego, CA, USA. He received the Licenciatura degree in mathematics
from Universidad de Zaragoza, Zaragoza, Spain, in 1997, and the
Ph.D. degree in engineering mathematics from Universidad Carlos III de
Madrid, Madrid, Spain, in 2001. He held postdoctoral positions with
the University of Twente, Twente, The Netherlands, and the University
of Illinois at Urbana-Champaign, Urbana, IL, USA. He was an Assistant
Professor with the Department of Applied Mathematics and Statistics,
University of California, Santa Cruz, CA, USA, from 2004 to 2007.  He
is the author of Geometric, Control and Numerical Aspects of
Nonholonomic Systems (Springer-Verlag, 2002) and co-author (together
with F. Bullo and S. Mart{\'\i}nez) of Distributed Control of Robotic
Networks (Princeton University Press, 2009). He was an IEEE Control
Systems Society Distinguished Lecturer (2010-2014) and is an IEEE
Fellow. His current research interests include cooperative control,
network science, game theory, multi-agent coordination in robotics,
power systems, and neuroscience, geometric and distributed
optimization, nonsmooth analysis, and geometric mechanics and control.

\emph{\textbf{Randy Freeman}} is a Professor of Electrical Engineering
and Computer Science at Northwestern University, Evanston, IL, USA. He
received the Ph.D. degree in electrical engineering from the
University of California at Santa Barbara, Santa Barbara, CA, USA, in
1995.  He has been Associate Editor of the IEEE Transactions on
Automatic Control. His research interests include nonlinear systems,
distributed control, multi-agent systems, robust control, optimal
control, and oscillator synchronization. Prof. Freeman received the
National Science Foundation CAREER Award in 1997. He has been a member
of the IEEE Control System Society Conference Editorial Board since
1997, and has served on program and operating committees for the
American Control Conference and the IEEE Conference on Decision and
Control.

\emph{\textbf{Kevin Lynch}} (S'90-M'96-SM'05-F'10) is a Professor and
the Chair of the Mechanical Engineering Department, Northwestern
University, Evanston, IL, USA.  He received the B.S.E. degree in
electrical engineering from Princeton University, Princeton, NJ, USA,
in 1989, and the Ph.D. degree in robotics from Carnegie Mellon
University, Pittsburgh, PA, USA, in 1996.  He is a member of the
Neuroscience and Robotics Laboratory and the Northwestern Institute on
Complex Systems. His research interests include dynamics, motion
planning, and control for robot manipulation and locomotion;
self-organizing multi-agent systems; and functional electrical
stimulation for restoration of human function. He is a coauthor of the
textbooks Principles of Robot Motion (MIT Press, 2005), Embedded
Computing and Mechatronics (Elsevier, 2015), and Modern Robotics:
Mechanics, Planning, and Control (Cambridge University Press, 2017).

\emph{\textbf{Sonia Mart{\'\i}nez}} (M'02-SM'07-F'18) is a Professor
of Mechanical and Aerospace Engineering at the University of
California, San Diego, CA, USA. She received the Ph.D. degree in
Engineering Mathematics from the Universidad Carlos III de Madrid,
Spain, in May 2002. Following a year as a Visiting Assistant Professor
of Applied Mathematics at the Technical University of Catalonia,
Spain, she obtained a Postdoctoral Fulbright Fellowship and held
appointments at the Coordinated Science Laboratory of the University
of Illinois, Urbana-Champaign during 2004, and at the Center for
Control, Dynamical systems and Computation (CCDC) of the University of
California, Santa Barbara during 2005.  In a broad sense, her main
research interests include the control of network systems,
multi-agent systems, nonlinear control theory, and robotics.  For her
work on the control of underactuated mechanical systems she received
the Best Student Paper award at the 2002 IEEE Conference on Decision
and Control. She was the recipient of a NSF CAREER Award in 2007. For
the paper ``Motion coordination with Distributed Information,"
co-authored with Jorge Cort\'es and Francesco Bullo, she received the
2008 Control Systems Magazine Outstanding Paper Award. She has served
on the editorial boards of the European Journal of Control
(2011-2013), and currently serves on the editorial board of the
Journal of Geometric Mechanics and IEEE Transactions on Control of
Network Systems.

\end{document}